\documentclass[11pt]{article}
\usepackage[utf8]{inputenc} 
\usepackage{geometry}
\usepackage[english]{babel}
\usepackage{graphicx}
\usepackage[mathscr]{euscript}
\usepackage{amsmath}
\usepackage{slashed}
\usepackage{graphicx}
\usepackage{amssymb}
\usepackage{cite}
\usepackage{amsfonts} 
\usepackage{verbatim} 
\usepackage{subfig} 
\usepackage{amsmath,amscd}
\usepackage{tikz}
\usepackage{tabu,multirow}
\usepackage{tensor}
\usepackage{xfrac}
\usepackage{lscape}
\usepackage{cancel}
\usepackage{MnSymbol}
\usepackage{rotating}
\usepackage{dsfont}
\usepackage[T1]{fontenc}
\usepackage{titlesec, blindtext, color}
\definecolor{gray75}{gray}{0.75}

\usetikzlibrary{matrix,arrows,decorations.pathmorphing}

\usepackage{setspace}

\usepackage{amssymb}
\usepackage{bbm}

\usepackage{hyperref}
\begin{document}

\pagestyle{empty}
\rightline{LTH 1265} 
\vskip 1.5 true cm  
\begin{center}  
{\Large Supersymmetry algebras in arbitrary signature and their R-symmetry groups}\\[.5em]
\vskip 1.0 true cm   
{L.~Gall and T.~Mohaupt} \\[5pt] 
{Department of Mathematical Sciences\\ 
University of Liverpool\\
Peach Street \\
Liverpool L69 7ZL, UK\\[2ex]  
Louis.Gall@astrazeneca.com,
Thomas.Mohaupt@liv.ac.uk \\[1em]
}
August 11, 2021
\end{center}
\vskip 1.0 true cm  

\baselineskip=18pt  

\begin{abstract}  
\noindent  
String theory, specifically type-II superstring theory, can be formulated in any ten-dimensional signature. To facilitate the
study of supergravity and superstring theories in this setting, we present  a uniform construction of supersymmetry algebras in arbitrary dimension and signature, which generalizes
the ideas underlying symplectic Majorana spinors. In our formalism R-symmetry
acts on an auxiliary multiplicity space which makes its action manifest. This allows
us to provide extensive tables which list the R-symmetry groups of extended supersymmetry algebras
for all signatures together with other useful information. Twisted (`type-*')
supersymmetry algebras in Lorentz signature with non-compact R-symmetry groups are shown to 
be part of a general pattern resulting from the interplay between complex superbrackets
and reality conditions.  As an application we show how the relations between type-II string theories in ten and nine
dimensions can be extracted from their supersymmetry algebras. We also use our results to determine
the special geometry of vector and hypermultiplet scalar manifolds of four-dimensional $\mathcal{N}=2$
and three-dimensional $\mathcal{N}=4$ supergravity theories for all signatures. 

\end{abstract}

\newpage

\tableofcontents

\newpage

\pagestyle{plain}

\pagenumbering{arabic}

\section{Introduction}

String theory extends our concepts of space-time geometry and 
symmetry in various directions. Besides the well known and widely 
explored concepts of T-duality, mirror symmetry and the AdS/CFT correspondence,
there is the less explored idea of timelike T-duality, which has far reaching 
consequences. Firstly, in Minkowski signature, it relates the type-II string theories 
to the so-called type-II* string theories, which realize `twisted' versions of the
standard $\mathcal{N}=2$ supersymmetry algebras \cite{Hull:1998vg}. Secondly, together with 
standard (i.e. spacelike) T-duality and with S-duality, it creates a web of type-II
string theories which covers all possible space-time signatures $(t,s)$, $t+s=10$ 
in ten dimensions \cite{Hull:1998ym}. This type-II network is related to three versions of eleven-dimensional
M-theory with space-time signature $(1,10)$, $(2,9)$, $(5,6)$. While the interpretation
of theories with multiple time dimensions is not obvious, these exotic theories
are arguably part of the configuration space of string theory, and therefore
their properties deserve detailed investigation.  The effective field theories
of type-* theories contain fields with negative kinetic energy, but are, as full
string theories, equivalent to standard type-II theories, at least as 
long as the timelike circle has finite radius. Type-* theories admit  de Sitter solutions,
and theories with multiple times may admit interesting brane world models. 
Both formal and phenomenological aspects of exotic string theories, by which 
we refer to type-* as well as non-Minkowksi signature theories, have been
investigated in more detail in \cite{Dijkgraaf:2016lym,Blumenhagen:2020xpq}.

Supersymmetry and supergravity in Euclidean and other non-Lorentzian
signatures have been studied to some extent in the literature.  
Supersymmetry algebras in arbitrary signature have been discussed in 
\cite{Ferrara:2001sx}. 
Lower-dimensional supergravity theories in non-standard signatures have been 
constructed using dimensional reduction in 
\cite{Cremmer:1998em,Hull:1998br,Gutowski:2012yb,Sabra:2016abd,Sabra:2015tsa,Sabra:2017xvx}.
A Euclidean version of the special geometry of $\mathcal{N}=2$ vector and hypermultiplets 
has been developed in \cite{Cortes:2003zd,Cortes:2005uq,Cortes:2009cs,Cortes:2015wca},
while $\mathcal{N}=2$ vector multiplets in arbitrary signature were constructed in \cite{Gall:2018ogw,Cortes:2019mfa}.
Four-dimensional supersymmetric solutions 
in neutral signature have been investigated in \cite{Klemm:2015mga,Gutowski:2019hnl},
brane-like solutions in arbitrary dimension and signature have been 
constructed  in \cite{Sabra:2020gio} and supersymmetric solutions of five-dimensional
vector multiplets coupled to supergravity have recently been studied for
arbitrary signature in \cite{Sabra:2021omz}.

The concrete form of the supersymmetry algebra varies from 
dimension to dimension, and from signature to signature, depending on 
whether the supercharges are Dirac, Majorana, Weyl or Majorana-Weyl spinors. 
The main result of this paper is a universal construction, which generalizes the idea
underlying symplectic Majorana spinors, and applies it to any dimension and
signature. The general idea is to start with a complex, hence signature 
independent  supersymmetry algebra 
and then to impose reality conditions which select a space-time signature and
reduce the complex R-symmetry group to one of its real forms. As we will 
see the possible reality conditions include, besides standard Majorana
and symplectic Majorana conditions, the `twisted' or $\mathrm{O}(p,q)$ 
Majorana conditions which were used in \cite{Hull:1998ym} to describe
the supersymmetry algebras of ten-dimensional type-II string theories in
general signature. Our formalism provides a systematic way of identifying
reality conditions that define real supersymmetry algebras by selecting 
real forms of the complex R-symmetry group.
Such a uniform approach useful if one wants to explore the web of string 
dualities across dimensions and signatures, as we illustrate using type-II string
theories and their compactifications as an example. 
One advantage of our formalism is that it disentangles
the actions of the spin and R-symmetry groups, making the R-symmetry
group manifest. This allows one to easily distinguish between non-isomorphic
supersymmetry algebras which have the same number of supercharges, 
and to identify Lorentz signatures  where twisted (type-*) supersymmetry 
algebras exist. For example the ten-dimensional type-IIB and type-IIB$^*$
algebras have R-symmetry groups $\mathrm{O}(2)$ and $\mathrm{O}(1,1)$, 
while the standard and twisted four-dimensional $\mathcal{N}=2$ 
supersymmetry algebra have R-symmetry groups $\mathrm{U}(2)$ and 
$\mathrm{U}(1,1)$, respectively. We present tables where we classify
the possible R-symmetry groups appearing in our construction up to dimension
12 for all signatures.

\subsection{Background}

Before we give an overview of our construction, we need to provide some
background. The following section is partly based on  \cite{Alekseevsky:1997},
whose method, notation and terminology we have adopted.
Given a real vector space $V\cong \mathbb{R}^{t,s}$,
equipped with a scalar product (non-degenerate real bilinear form) of signature
$(t,s)$, the associated Poincar\'e Lie algebra is 
\begin{equation}
 \mathfrak{g}_0 =
 \mathfrak{so}(V) + V \;,
\end{equation}
where $\mathfrak{so}(V) \cong \mathfrak{so}(t,s)$ is the `Lorentz Lie algebra'
(Lie algebra of infinitesimal isometries of the scalar product), and where $V$ 
is the Lie algebra of translations.\footnote{Depending on context, 
$V\cong \mathbb{R}^{t,s}$ denotes either a vector space, or the affine space modelled on 
this vector space (interpreted as a flat spacetime), or the Lie algebra of translations acting on the affine space.}
To extend this to a {\em Poincar\'e Lie superalgebra} $\mathfrak{g} = \mathfrak{g}_0 + \mathfrak{g}_1$ 
one adds a real $\mathrm{Spin}(V)$ module
$\mathfrak{g}_1$,
which may be reducible, and introduces a $\mathbb{Z}_2$-grading under which
$\mathfrak{g}_0$ is even, while $\mathfrak{g}_1$ is odd,
\begin{equation}
\mathfrak{g} = \mathfrak{g}_0 + \mathfrak{g}_1 = \mathfrak{so}(V) + V + \mathfrak{g}_1 \;.
\end{equation}
We will use the terms $\mathfrak{g}$-module, $G$-module, or module for short, 
for a vector space
carrying a representation of a Lie algebra $\mathfrak{g}$ or of a Lie group $G$.
The group
$\mathrm{Spin}(V)=\mathrm{Spin}(t,s)$, which is a two-fold cover of special orthogonal group
$\mathrm{SO}(V)=\mathrm{SO}(t,s) $, is contained in the real Clifford algebra, $\mathrm{Spin}(V) \subset
Cl(V)=Cl_{t,s}$. The same applies to its Lie algebra 
$\mathfrak{spin}(t,s) \cong \mathfrak{so}(t,s)$.\footnote{All standard facts about Clifford
algebras and their relation to spin groups used in this paper can be found in
\cite{SpinGeometry}.} 
The vector space $\mathfrak{g}_1$ is a finite sum of irreducible real spinor representations
which, depending on signature, can be Dirac spinors, Majorana spinors, Weyl spinors, or Majorana-Weyl spinors. Note that complex representations
can be regarded as real representations by `forgetting the complex structure,'
and that it is this underlying real representation that is relevant for the construction and
classification of real supersymmetry algebras. 
In signatures where  Majorana spinors do not exist, the elements of the irreducible 
real spinor representations are Dirac or Weyl spinors.

 While
the action of $\mathfrak{so}(V)$ on $\mathfrak{g}_1$ is determined
by the spinor representations we have chosen, it can be shown that the
translation algebra $V$ must operate trivially on $\mathfrak{g}_1$, and that 
the only freedom besides the choice of $\mathfrak{g}_1$ is the choice
of a superbracket, that is of a symmetric bracket 
\begin{equation}
\Pi\;: \;\;\;\mathfrak{g}_1 \times \mathfrak{g}_1 \rightarrow V \;,\;\;\;
(\lambda, \chi) \mapsto \Pi(\lambda, \chi) =  \{ \lambda , \chi \} \;,
\end{equation}
which is covariant with respect to the action of $\mathfrak{so}(V)$.
Mathematically, such a bracket corresponds to a real, symmetric, 
non-degenerate, spin-equivariant, vector-valued bilinear form $\Pi$
on the spinor module $\mathfrak{g}_1$.  
In the physics literature the superbracket is usually defined by writing
down the anti-commutation relations of the
supercharges $Q_{i\alpha}$, where $\alpha$ is a spinor index corresponding
to an irreducible representation, while $i=1\ldots, \mathcal{N}$ labels 
copies of the irreducible real spinor representation.
Let us illustrate  this using the case 
where the irreducible real spinor representation is given by Majorana 
spinors.
Then the $\mathcal{N}$-extended supersymmetry algebra\footnote{We
use the term `supersymmetry algebra' for the supertranslation algebra $V + \mathfrak{g}_1$, whose only non-trivial 
algebraic relation is the $Q$-$Q$ anti-commutator. As mentioned before, this
is the only `moving part' in our analysis once $\mathfrak{g}_1$ has been chosen.} 
takes the form
\begin{equation}
\label{QQ_basic}
\{ Q_{i \alpha} \;, Q_{j\beta} \}  = M_{ij} (\gamma^\mu C^{-1})_{\alpha \beta} P_\mu \;.
\end{equation}
Note that we need to include the inverse $C^{-1}$ of the charge conjugation matrix
$C$ in order to lower one index of the matrix $\gamma^\mu = (\gamma^{\mu\;\;\beta}_{\;\;\alpha})$,
see Appendices \ref{app:spinor_indices} -- \ref{secpoincare}
for our index conventions. Since the bracket is symmetric,
the right-hand side must be symmetric in the multi-indices $(i \alpha), (j\beta)$. 

We will prefer to work with vector-valued bilinear forms on $\mathfrak{g}_1$, which allows us
to suppress spinor indices. In this language the relation \eqref{QQ_basic} is expressed
using a bilinear form 
\begin{equation}
\beta\;: \;\;\; ( S \otimes \mathbb{R}^{\mathcal{N}}) \times (S \otimes \mathbb{R}^{\mathcal{N}})
\rightarrow \mathbb{R} \;,\;\;\;  \beta(\lambda, \chi) = (\lambda^i)^T  C \chi^j M_{ji} \;,
\end{equation}
where we have written out the spinor module $\mathfrak{g}_1$ in terms of irreducible real
$\mathrm{Spin}(t,s)$ modules $S$, which in our case are Majorana representations:
\begin{equation} 
\mathfrak{g}_1 = S \otimes \mathbb{R}^{\mathcal{N}} \cong S \oplus  \cdots \oplus  S \;\;\;\ (\mathcal{N}-\text{times})\;.
\end{equation}
Since we take a sum of  isomorphic modules, we can 
rewrite the $\mathcal{N}$-fold sum as a tensor product with 
an internal `multiplicity space' $\mathbb{R}^{\mathcal{N}}$. 
Then the 
spin group only acts on the first factor but not on the internal space.
Our strategy for disentangling the spin group and R-symmetry group will 
be to have the R-symmetry group acting only (or almost only) on the internal space, 
as we will discuss in more detail below.

The bilinear form $\beta$ is scalar-valued. Vector-valued
bilinear forms, and, more generally,  bilinear forms valued in antisymmetric tensors 
are obtained by substituting antisymmetrized products 
$\gamma^{(p)} = \gamma^{\mu_1 \cdots \mu_p} =\gamma^{[\mu_1} \cdots 
\gamma^{\mu_p]}$ of $\gamma$-matrices into the first argument. 
In particular, the vector-valued bilinear form
$\beta^{(1)} = \beta(\gamma^\mu \cdot, \cdot)$ allows us to express the superbracket
\eqref{QQ_basic} as
\begin{equation}
\label{QQ_vectorform}
\beta(\gamma^\mu \lambda, \chi) = (\gamma^\mu \lambda^i)^T C \chi^j M_{ji} \;.
\end{equation}
To recover the superbracket \eqref{QQ_basic} from \eqref{QQ_vectorform} one
expands the spinors $\lambda, \chi$ in a basis given by the supercharges, 
$\lambda = \lambda^{i\alpha} Q_{i\alpha}$,
$\chi = \chi^{j \beta} Q_{j \beta}$. We refer to Appendix \ref{secpoincare}
for details. Up to isomorphism the superbracket only 
depends on a few invariants of the bilinear forms defined by the matrices 
$C=(C^{\alpha \beta})$ and $M=(M_{ij})$.

To define a superbracket, the vector-valued bilinear form $\beta(\gamma^\mu \cdot,
\cdot)$ must be symmetric and equivariant (covariant) with respect to the spin group. As shown in \cite{Alekseevsky:1997}
this is achieved by using {\em admissible} bilinear forms $\beta$ which 
are characterized by 
having a definite
{\em symmetry} $\sigma_\beta\in \{ 1,-1 \} $ and {\em type} $\tau_\beta \in \{ 1, -1 \}$, 
where\footnote{With regard to the symmetry, note
that we work with commuting spinors in this paper. The translation to a 
formalism with anti-commuting spinors is straightforward and only
introduces an additional sign. Using anti-commuting spinors is required when 
discussing properties of spinor bilinears in Lagrangians, a subject that we will
not discuss in this paper. See, however,  \cite{Gall:2018ogw} where our formalism has
been used to construct five-dimensional vector multiplets for arbitrary signature. }

\begin{equation}
\label{symmetry+type}
\beta(\lambda, \chi) = \sigma_\beta \beta(\chi, \lambda) \;,\;\;
\beta(\gamma^\mu \lambda, \chi)  = \tau_\beta 
\beta(\lambda, \gamma^\mu \chi)  \;.
\end{equation}
The vector-valued bilinear form $\beta(\gamma^\mu\cdot, \cdot)$ is symmetric if 
$\sigma_\beta \tau_\beta =1$. Therefore such brackets $\beta$ will be called
{\em super-admissible}. It can be shown that any super-admissible bracket $\beta$ on 
a spinor module $\mathfrak{g}_1$ defines a supersymmetry algebra. In particular the vector-valued
bracket is automatically spin-equivariant, and the super-Jacobi identity required
to make $\mathfrak{so}(V) + V + \mathfrak{g}_1$ a Lie superalgebra holds automatically. 
Conversely, the space of real, symmetric, spin-equivariant vector-valued bilinear forms,
and, hence, the space of real Poincar\'e Lie superalgebra structures related to 
a given spinor module $\mathfrak{g}_1$ is  a finite-dimensional real vector space which admits
a basis given by forms of the type $\beta(\gamma^\mu\cdot, \cdot)$, where
$\beta$ is super-admissible. In this sense all extended Poincar\'e Lie
superalgebras are known for all dimensions and signatures. 
A basis of super-admissible bilinear forms has been constructed in
\cite{Alekseevsky:1997}.\footnote{One can also include poly-vector charges (BPS charges), 
that is additional terms in the supersymmetry anti-commutator
which transform as antisymmetric Lorentz tensors \cite{Alekseevsky:2003vw}. The inclusion of such charges 
in our formalism will be left for future work.}
Since any linear combination of super-admissible bilinear forms defines a Poincar\'e Lie superalgebra
(as long as it is non-degenerate, which is the generic case), superbrackets form continuous families. This raises the question of classification, that is
to decide which brackets define non-isomorphic Poincar\'e Lie superalgebras. To address
this question one needs to study the 
{\em Schur group}  ${\cal C}^*(\mathfrak{g}_1)$, which is defined as 
the subgroup of automorphism group $\mbox{Aut}(\mathfrak{g}_1) = \mathrm{GL}(\mathfrak{g}_1)$ 
of the real vector space $\mathfrak{g}_1$ whose action commutes with the action of $\mathfrak{spin}(V)$, 
\begin{equation}
\mathcal{C}^*(\mathfrak{g}_1) = \{ Z \in \mbox{Aut}({\mathfrak{g}_1}) | 
[Z, \mathfrak{spin}(V)]=0 \} 
= Z_{\mathrm{GL}(\mathfrak{g}_1)}(\mathfrak{spin}(V))
\end{equation} 
that is, the centralizer of $\mathfrak{spin}(V) \subset \mathrm{GL}(\mathfrak{g}_1)$.  
Any two superbrackets which are in the same orbit of the action of the Schur group
on the space of super-admissible bilinear forms are isomorphic. For later use
we also define the {\em Schur algebra}  
$\mathcal{C}(\mathfrak{g}_1) = Z_{\mathrm{End}(\mathfrak{g}_1)}(\mathfrak{spin}(V))$, which 
is the centralizer of $\mathfrak{spin}(V)$ in  the algebra $\mbox{End}(\mathfrak{g}_1)$ of endomorphisms
of $\mathfrak{g}_1$. 
\begin{equation}
\mathcal{C}(\mathfrak{g}_1) = \{ Z \in \mbox{End}({\mathfrak{g}_1}) | 
[Z, \mathfrak{spin}(V)]=0 \} 
= Z_{\mathrm{GL}(\mathfrak{g}_1)}(\mathfrak{spin}(V))
\end{equation} 
The Schur group is the group of invertible elements of the Schur algebra.

The classification problem for Poincar\'e Lie superalgebras with odd part $\mathfrak{g}_1$
is almost, but not quite,
equivalent to the problem of classifying the orbits of the Schur group on the
space of superbrackets on $\mathfrak{g}_1$.\footnote{We will use the terms
super-admissible bilinear form and superbracket interchangeably.} The reason is that elements of the
pin group $\mathrm{Pin}(V)$
which are not contained in the spin group $\mathrm{Spin}(V)$ may lead to 
isomorphisms between brackets which belong to different orbits. Recall
that $\mathrm{Pin}(V)$ is a double cover of the full orthogonal group
$\mathrm{O}(V)$, while $\mathrm{Spin}(V)$ is a double cover of the 
special orthogonal group $\mathrm{SO}(V)$. Both groups are contained
in the Clifford algebra $Cl(V)$. A precise criterion for two Poincar\'e
Lie superalgebras to be isomorphic is given by Theorem 1 of  \cite{Cortes:2019mfa}.
As illustrated in  \cite{Cortes:2019mfa} by the classification of four-dimensional supersymmetry algebras
with eight real supercharges for arbitrary signature,
this classification can be done case by case but requires some work. 
However there is a sufficient condition for two supersymmetry algebras
to be non-isomorphic which is easier to check, namely that
their R-symmetry groups are different. We define the {\em R-symmetry
group} of a superbacket as the subgroup of the Schur group under which 
this superbracket invariant (its stabilizer group):
\begin{equation}
G_R = \{ R \in \mathcal{C}^*(\mathfrak{g}_1) | \beta( \gamma^\mu R \cdot, 
R \cdot) = \beta(\gamma^\mu \cdot, \cdot) \} \;.
\end{equation}
Note that our
definition does not depend on how the R-symmetry group acts on 
the fields in particular field theoretic realizations of the algebra. 
Moreover, with our definition the R-symmetry
group is not necessarily a connected group. This may lead to slight
differences when comparing our tables to the literature. 

We will also need to consider the complexification $\mathfrak{g}^\mathbb{C}$ 
of a Poincar\'e Lie superalgebra. In this context one can define complex versions 
of the Schur algebra, 
Schur group and R-symmetry group, which will be denoted $\mathcal{C}_\mathbb{C}(\mathfrak{g}_1^\mathbb{C})$, $\mathcal{C}^*_\mathbb{C}(\mathfrak{g}_1^\mathbb{C})$, and 
$G_R^\mathbb{C}$.

\subsection{Overview of the construction}

To explain the main idea of our construction, recall how supersymmetry 
can be formulated in terms of symplectic Majorana spinors. In signature
$(1,3)$ the unique (up to isomorphism) irreducible real spinor representation 
is the Majorana representation,
and the smallest or $\mathcal{N}=1$ supersymmetry algebra is based on 
supercharges $Q_\alpha$ which are Majorana spinors. The standard 
$\mathcal{N}=2$ algebra can be written in terms of two Majorana spinors,
or of a single Dirac spinor, but there also is a third option, namely 
to take two Dirac spinors $Q_{i\alpha}$, and to impose a reality condition. 
The supersymmetry algebra is then defined by a complex superbracket, 
subject to a reality condition:\footnote{When formulating this algebra using Majorana 
spinors, one uses a different charge conjugation matrix $C'$, which is symmetric,
and a different matrix $B'$, which satisfies $B'B'^* = + \mathrm{Id}$. See Tables
\ref{Table:epsilon-structures}  and
\ref{bilinears} for information about the properties
of the matrices
$C$ and $B$ in arbitrary dimension and signature.}
\begin{equation}
\label{susy_symp_Maj}
\{ Q_{i \alpha} \;, Q_{j\beta} \}  = M_{ij} (\gamma^\mu C^{-1})_{\alpha \beta} P_\mu \;, 
\;\;\; (Q^i)^* = \alpha B Q^j  L_{ji} \;,\;\;\;i,j=1, 2\;,
\end{equation}
where in our specific case $M_{ij} = L_{ij}= \epsilon_{ij}$, the charge conjugation matrix $C$
is antisymmetric, and $B$ satisfies $BB^* = - \mathrm{Id}$, indices $i,j$ are
raised and lowered using $\epsilon_{ij}$, $\epsilon^{ij}$, and $\alpha$ is a conventional
phase.  A pair of Dirac spinors
$\lambda^i$ which satisfies 
\[
(\lambda^i)^* = \alpha B \lambda^j \epsilon_{ji}
\]
is called a pair of {\em symplectic Majorana spinors}. 

The same construction
can be applied in signature $(1,4)$, where Majorana spinors do not exist
and Dirac spinors are irreducible, so that  the smallest
supersymmetry algebra has eight real supercharges. As in signature (1,3) 
one can replace a Dirac spinor by a pair of symplectic Majorana spinors, with
the supersymmetry algebra taking the form  (\ref{susy_symp_Maj}).
While one also can express the supersymmetry algebra in terms of Dirac 
spinors (see for example \cite{Cortes:2003zd}), the formulation using symplectic Majorana
spinors is the standard one in five dimensions. One of its advantages is
its manifest R-symmetry: the group which leaves both the complex superbracket and
the reality condition invariant is 
$\mathrm{USp}(2,\mathbb{R}) \cong \mathrm{SU}(2)$.

Formuling the supersymmetry algebra in terms of 
symplectic Majorana spinors can be interpreted as 
complexifying the space of Dirac spinors
$\mathbb{S}$, and then imposing a reality condition. The space $\mathbb{S}$ 
of Dirac spinors, which is also called the {\em complex spinor module}, is obtained
by restricting an irreducible representation of the complex Clifford algebra 
$Cl(V^\mathbb{C})$, where $V^\mathbb{C} = V \otimes_\mathbb{R} \mathbb{C}$, 
to the real spin group $\mathrm{Spin}(V)$. 
The complexification of $\mathbb{S}$, regarded as a real spin module,
is $\mathbb{S}^{\mathbb{C}} =
\mathbb{S} \otimes_{\mathbb{R}} \mathbb{C}$. Since $\mathbb{S}$ admits
a spin invariant complex structure, its complexification is self-conjugate, 
$\mathbb{S}^{\mathbb{C}} = \mathbb{S} \oplus \bar{\mathbb{S}}$, where
$\bar{\mathbb{S}}$  is the complex-conjugate module.
Moreover, in any dimension and signature one can
find a $\mathrm{Spin}(t,s)$ invariant matrix $B$ acting on $\mathbb{S}$ which satisfies
$BB^*= \pm \mathrm{Id}$.
Therefore the complex spinor module $\mathbb{S}$ always carries either an invariant
real structure or an invariant quaternionic structure, and thus is self-conjugate, 
$\mathbb{S} \cong \bar{\mathbb{S}}$ as $\mathrm{Spin}(t,s)$ as module.
Therefore
\[
\mathbb{S}^{\mathbb{C}} = \mathbb{S} \otimes_{\mathbb{R}}\mathbb{C} = \mathbb{S} 
\oplus \overline{\mathbb{S}} \cong \mathbb{S} \oplus \mathbb{S} \cong 
\mathbb{S}\otimes_{\mathbb{C}} \mathbb{C}^2 \;.
\]
In the last step we have rewritten the sum $\mathbb{S} \oplus \mathbb{S}$ as 
a complex tensor product, 
where the second
factor is an auxiliary multiplicity space which encodes that we have two copies
of $\mathbb{S}$. This corresponds to the expression for the complex superbracket in 
(\ref{susy_symp_Maj}) where we use a pair of Dirac spinors $Q_{i\alpha}$, with the
index $i=1,2$ referring to the multiplicity space 
$\mathbb{C}^2$. 

The complex superbracket (\ref{susy_symp_Maj}) defines a 
complex Poincar\'e Lie superalgebra $\mathfrak{so}(V^\mathbb{C}) \oplus {V}^\mathbb{C}
\oplus \mathfrak{g}_1^\mathbb{C}$, where $V^\mathbb{C} = V \otimes_\mathbb{R} \mathbb{C}$
is the complexified translation algebra, $\mathfrak{so}(V^\mathbb{C})$ the complexified
Lorentz Lie algebra and  $\mathfrak{g}_1^\mathbb{C} = \mathbb{S}^\mathbb{C}$ 
the complexified spinor module. By imposing the reality condition (\ref{susy_symp_Maj}) 
we recover the real supersymmetry algebra as a real form of this complex algebra.

We already mentioned that the multiplicity space $\mathbb{C}^K = \mathbb{C}^2$ is useful
in making explicit the action of the R-symmetry group. 
In five dimensions, where $\mathbb{S}$ is irreducible, the Schur group
acts trivially on the factor $\mathbb{S}$ in $\mathbb{S}\otimes_\mathbb{C} \mathbb{C}^2$
and therefore the complex Schur group is $\mbox{GL}(2,\mathbb{C})$.\footnote{To be precise, by 
Schur's lemma the R-symmetry group acts on $\mathbb{S}$ by scalar multiplication, 
which we can absorb
into its action on $\mathbb{C}^2$.} The R-symmetry group 
of the complex supersymmetry algebra is the subgroup which in addition preserves the bilinear form
defined by $M_{ij} = \epsilon_{ij}$ on $\mathbb{C}^2$, that is 
$G_R^\mathbb{C} = \mbox{Sp}(2,\mathbb{C})$. The R-symmetry group $G_R$ of the 
real supersymmetry algebra is the real form of $G_R^\mathbb{C}$ which also preserves
the reality condition in (\ref{susy_symp_Maj}), that is $G_R = \mathrm{USp}(2,\mathbb{R})
\cong \mathrm{SU}(2)$. In four dimension, $\mathbb{S}$ becomes reducible, 
which enlarges the R-symmetry group to $\mathrm{U}(2)$.\footnote{We will explain how to deal 
with this in due course. In Appendix \ref{secmatrixnot} we present
two methods which allow to disentangle the spin and R-symmetry groups.}

The observation that we will expand on in this paper is that the construction 
of real supersymmetry algebras based on symplectic Majorana spinors 
can be adapted to arbitrary dimension and signature. This requires of course
to also consider other types of reality conditions. Extended real supersymmetry algebras 
are included by enlarging the internal multiplicity space.

Thus, schematically, the idea is:
\begin{equation}
\text{Real supersymmetry algebra} \leftrightarrow 
\left\{ \begin{array}{l}
\text{Complex bilinear form (superbracket)}, \\
\text{Reality condition}.
\end{array}
\right. 
\end{equation}

Depending on the signature, there either is a unique (up to isomorphism) 
irreducible real spinor representation given by Dirac or Majorana spinors,
or there are two inequivalent ones, given by Weyl spinors or 
Majorana-Weyl spinors. For simplicity we first explain how our construction
works when the irreducible real spinor module is unique. 
\begin{enumerate}
\item
We start with a complex supersymmetry algebra with spinor module
given by $K$ copies of the complex spinor module,
$\mathfrak{g}_1^\mathbb{C} = \mathbb{S} \otimes \mathbb{C}^K$. 
We will sometimes refer to this as the $K$-extended spinor module, 
or extended spinor module for short.
To define a complex supersymmetry algebra, we need to 
specify a complex super-admissible bilinear form $\beta$. 
Since the Gram matrix of such a bilinear form has precisely 
the properties of a charge conjugation matrix $C$, this amounts
to choosing a charge conjugation matrix. 
If the bilinear form is super-admissible
we can extend it to $\mathbb{S} \otimes \mathbb{C}^K$ as
$\beta = C \otimes \delta$, where $\delta$ is the complex bilinear form 
on $\mathbb{C}^K$ defined by the unit matrix. If $C$ is not
super-admissible, we can still obtain a super-admissible bilinear form
on $\mathbb{S} \otimes \mathbb{C}^K$ for $K$ even, if we extend
$C$ as $\beta = C\otimes \epsilon$, where $\epsilon$ defines 
a non-degenerate anti-symmetric bilinear form. 
Charge conjugation matrices are classified by their symmetry $\sigma$ 
and type $\tau$. In odd dimensions charge conjugation matrices are 
unique up to isomorphism, while in even dimensions there are two
inequivalent ones, denoted $C_\pm=C_{- \tau}$, which are of opposite type $\tau$,
see Table \ref{bilinears}.
We can now write down the possible complex supersymmetry algebras
in any dimension, and for any value of $K$, and determine their Schur groups
and R-symmetry groups.
\item
Then we choose a signature $(t,s)$ and a $\mathrm{Spin}(t,s)$ invariant
reality condition on $\mathfrak{g}_1^\mathbb{C} = \mathbb{S} \otimes \mathbb{C}^K$. 
In odd dimensions $\mathbb{S}$ admits either a spin invariant real
structure or a spin invariant quaternionic structure, defined using 
the matrix $B$ which relates the $\gamma$-matrices to their complex 
conjugates. In even dimensions there are two non-equivalent such 
matrices, $B_\pm$, and, depending on signature,  $\mathbb{S}$
admits two inequivalent real structures or two inequivalent quaternionic
structures, or one of each type. 
We emphasize that while
$B_\pm$ can be constructed out of $C_\pm$, the choice of a reality
condition is independent of the choice of the bilinear form on $\mathbb{S}$,
that is, there are four possible combinations of $C_\pm$ with $B_{\pm}$
in our formalism. To define a real supersymmetry algebra we need a 
real structure on $\mathbb{S} \otimes \mathbb{C}^K$, which can 
be obtained either as a product of two real structures or of two
quaternionic structures on the factors.  The real or
quaternionic structure on $\mathbb{C}^K$ is chosen such that it
defines a real form $G_R$ of the complex R-symmetry group $G_R^\mathbb{C}$, 
that is, we construct it using an involutive automorphism of the Lie 
algebra of $G_R^\mathbb{C}$. We can then list, for all dimensions
and signatures, the real supersymmetry algebras that result from a
consistent pairing of complex superbrackets with reality conditions, 
and determine their R-symmetry groups.
\end{enumerate}
There are two additional issues in even dimensions, 
where Dirac spinors are reducible and 
decompose into Weyl spinors, $\mathbb{S}\cong \mathbb{S}_+ \oplus \mathbb{S}_-$.
\begin{itemize}
\item
Firstly, there are signatures with two inequivalent irreducible real spinor representations,
which can be either Weyl spinors or Majorana-Weyl spinors. If this happens there
potentially are supersymmetry algebras based on  spinor modules of the form 
$\mathfrak{g}_1^\mathbb{C} = \mathbb{S}_+ \otimes \mathbb{C}^{K_+} 
\oplus \mathbb{S}_- \otimes \mathbb{C}^{K_-}$, where $K_\pm$ can
be chosen independently.\footnote{As it is clear from context that $\mathbb{S}_+$ 
and $\mathbb{C}^{K_+}$ are considered as complex modules, we have written 
$\otimes$ rather than $\otimes_\mathbb{C}$. Similarly, we will simply write $\otimes$ in the future
when the tensor product is between a complex spin representation and an auxiliarly 
complex multiplicity space, which encodes copies of equivalent representations.}
 The existence of such chiral supersymmetry
algebras requires more than the existence of inequivalent irreducible real spinor modules. 
Firstly, there must exist a superbracket which pairs $\mathbb{S}_+$ with 
$\mathbb{S}_+$ and $\mathbb{S}_-$ with 
$\mathbb{S}_-$, while vanishing between $\mathbb{S}_+$ and $\mathbb{S}_-$.
This can be decided using an invariant of the charge conjugation matrix $C$,
its isotropy $\iota$. Superbrackets are either orthogonal, $\iota=1$, which 
means that they are non-degenerate on $\mathbb{S}_+ \times \mathbb{S}_+$ 
and $\mathbb{S}_- \times \mathbb{S}_-$ but vanish on 
$\mathbb{S}_+ \times \mathbb{S}_-$ and $\mathbb{S}_- \times \mathbb{S}_+$,
or they are isotropic, $\iota=-1$, which 
means that they are non-degenerate on $\mathbb{S}_+ \times \mathbb{S}_-$ 
and $\mathbb{S}_- \times \mathbb{S}_+$ but vanish on 
$\mathbb{S}_+ \times \mathbb{S}_+$ and $\mathbb{S}_- \times \mathbb{S}_-$.
The isotropy of a superbracket only depends on the dimension, and we will see
that superbrackets are orthogonal in dimensions $2,6,10,\ldots$. 
An orthogonal complex superbracket
is necessary, but not sufficient for a chiral supersymmetry algebra to exist, 
because in order to obtain a real superbracket we
also need to impose a reality condition. A
reality condition can either respect chirality, in which case we call it `Weyl compatible'
or it can flip it, in which case we call it `Weyl incompatible'. Weyl compatibility is a signature
dependent property, and therefore the existence of chiral supersymmetry algebras is
signature dependent.

Chiral supersymmetry algebras
require both an orthogonal complex bilinear form and a Weyl compatible
reality condition, and therefore Weyl (in-)compatibility is a signature dependent property.
\item
Secondly, since $\mathbb{S}$ is reducible the R-symmetry acts non-trivially on 
$\mathbb{S}$. However it acts irreducibly, and therefore as a multiple of  the 
identity $\mathrm{Id}$
on the complex irreducible Weyl spinor modules
$\mathbb{S}_+$ and $\mathbb{S}_-$.  This can be used to determine 
the R-symmetry group. 
The details depend on whether the complex
bilinear form is orthogonal or isotropic, and on whether the reality condition 
is Weyl compatible or Weyl incompatible.
Since there are four cases to consider,
the classification of R-symmetry is somewhat involved in even dimensions. 
Using a certain block matrix notation one can
manifestly disentangle the actions of the spin and R-symmetry groups.
\end{itemize}

\subsection{Organisation of the paper}

The program described above is carried out as follows in the bulk of
this paper.
\begin{itemize}
\item
In Section \ref{secspinmod} we provide the necessary background
on Clifford and spin representations, with further details relegated
to Appendices A,B and C.
\item
In Section 3 we obtain complex supersymmetry algebras by 
constructing super-admissible bilinear forms on the
complexified extended spinor module
$\mathfrak{g}_1^\mathbb{C}$, where 
$\mathfrak{g}_1^\mathbb{C} = \mathbb{S} \otimes \mathbb{C}^K$ or
$\mathfrak{g}_1^\mathbb{C} = \mathbb{S}_+ \otimes \mathbb{C}^{K_+}
\oplus \mathbb{S}_- \otimes \mathbb{C}^{K_-}$. 
To do so we first have to study admissible bilinear forms on the spaces
$\mathbb{S}$ and $\mathbb{S}_\pm$ of Dirac and Weyl spinors, 
and bilinear forms on $\mathbb{C}^K$. The resulting complex 
supersymmetry algebras are classified by their R-symmetry 
groups which are listed in Table \ref{bilinears}. This table 
contains all relevant information about the superbrackets that
we use in our construction. 
\item
In Section 4 we systematically construct spin invariant reality conditions
on $\mathfrak{g}_1^\mathbb{C}$. These are built out of real and 
quaternionic structures on  $\mathbb{S}$, $\mathbb{S}_\pm$ and  
$\mathbb{C}^K$. To discuss real and quaternionic structures in 
parallel, we use the unifying concept of an $\epsilon$-quaternionic structure.
For dimensions up to twelve, and for all signatures, Table \ref{Table:epsilon-structures} lists
all inequivalent real and quaternionic structures on the complex spinor
module $\mathbb{S}$, and their properties. This table encodes
all relevant information about reality conditions used in our formalism, and, in particular,
allows to read off the the 
Schur algebras $\mathcal{C}(\mathbb{S})$. 
\item
In Section 5 we analyze how reality conditions can be imposed consistently
on complex superbrackets in order to define real supersymmetry algebras. 
\item
In Section 6 we classify the real R-symmetry groups which arise in our 
construction.  Table \ref{tableoddrealforms} shows how the real forms 
of the complex R-symmetry groups listed in Table \ref{bilinears}
arise from reality conditions. Table \ref{tableoddR} lists all possible real R-symmetry groups in
odd dimensions up to 11. In even dimensions the possible supersymmetry algebras
and R-symmetry groups fall into four classes, depending on whether 
the bilinear form is orthogonal or isotropic, and whether the reality condition 
is Weyl-compatible or Weyl-incompatible. In orthogonal Weyl-incompatible
signatures the real R-symmetry group only depends on the dimension, while
the real R-symmetry groups for orthogonal, Weyl-compatible signatures are listed
in Table \ref{tableorthwcrealforms} and for  isotropic signatures with either reality condition
in Table \ref{tableRisotropic}. For reference and convenience, we also provide a master
table, Table  \ref{mastertableeven},  for general even 
dimensions and a master table, Table \ref{mastertable}, for  all dimensions. 
\item
In Section 7 we explain why real supersymmetry algebras constructed by our method
are classified by their R-symmetry group, together with the choice of a relative sign
for chiral supersymmetry algebras. Details about the isomorphisms required to show
this have been relegated to Appendix \ref{Sec:Details_Isomorphisms}. 
\item
In Section 8 we apply our results to study supersymmetry algebras of 
type-II string theories in dimensions 10 and 9 and of their Calabi-Yau 
compactifications to dimensions 4 and 3. We show how our formalism
allows one to identify which theories exist in a given signature, and how to
determine their mutual relations by spacelike and timelike reduction and T-duality. 
For four-dimensional $\mathcal{N}=2$ theories 
and three-dimensional $\mathcal{N}=4$ theories 
we explain how the geometry of the scalar manifold can be read off from the R-symmetry 
group. Tables \ref{10d-9d}, \ref{Tab:4d_susy} and \ref{Tab:3d_susy} provide summaries.
\item
In Section 9 we make some concluding remarks and provide an outlook onto open questions and future 
work.
\end{itemize}

In order to keep the bulk of this paper as short as possible, some details 
have been relegated to appendices, together
with additional background information that is helpful but would have 
interrupted the flow of the main narrative. In Appendix A we summarize
our conventions and notation, and in particular explain how the
(partially) index free notation based on bilinear forms that we use in the
main part of the paper relates to the standard notation using anti-commutators
of supercharges. We also list various formulae which are used in the 
main part of this paper. In Appendix B we explain the relation between 
real spinors and real semi-spinors, as defined in the mathematics literature,
to Majorana, Weyl and Majorana-Weyl spinors, as defined in the physics
literature. In Appendix C we present details on the complexification of
spinor modules, which are relevant for understanding the precise relation 
between the odd parts $\mathfrak{g}_1$ and $\mathfrak{g}_1^\mathbb{C}$ 
of real and complex supersymmetry algebras. In Appendix D we
review the matrix notation for Weyl spinors introduced in \cite{Cortes:2019mfa}, which
is used in the main part to disentangle the actions of the spin and R-symmetry 
group in even dimensions by doubling the auxiliary multiplicity space of our
construction. Alternatively one can work without doubling, in which case the
R-symmetry group operates on spinor indices, in a way that we also explain
in Appendix D.
In Appendix E we provide the details of several isomorphisms which are needed
to show that the supersymmetry algebras that we have constructed are 
classified, essentially, by their R-symmetry groups. In Appendix F we 
collect formulae which allow one to carry out the spacelike and timelike dimensional
reduction of spinors, superbrackets and reality conditions.

\section{Clifford and spinor representations} \label{secspinmod}

In this section we review the necessary background on Clifford algebras, spin groups 
and their representations for arbitrary signature.
Our presentation is partially based on  \cite{VanProeyen:1999ni}, whose
conventions we follow except for some tweaks where we 
follow \cite{Alekseevsky:1997,Alekseevsky:2003vw,Cortes:2003zd}. We will also
use certain facts about Clifford algebras and
real associative algebras, see  for example \cite{SpinGeometry,Lang:Algebra}.

\subsection{Clifford and spinor modules}

We consider flat space-times $V\cong \mathbb{R}^{t,s}$ 
of arbitrary signature $(t,s)$ and dimension $D=t+s$. Our convention 
for the metric is 
$\eta = \mbox{diag}(-1, \ldots, -1, 1, \ldots, 1)$ with $t$ entries $-1$. 
The associated Clifford algebras $Cl_{t,s}$ are the real algebras
with generators $\gamma^\mu$, $\mu=1, \ldots, n$ and relations
\begin{equation}
\label{Clifford_algebra}
\{ \gamma^\mu, \gamma^\nu \}  = 2 \eta^{\mu \nu} \mathbbm{1} \;.
\end{equation}

Clifford algebras are real, associative algebras with a unit, and isomorphic
to real, complex or quaternionic matrix algebras of the form
$\mathbb{R}(n), \mathbb{C}(n), \mathbb{H}(n), 2 \mathbb{R}(n), 2 \mathbb{H}(n)$,
where $\mathbb{K}(n)$ is the algebra of $n\times n$ matrices over 
$\mathbb{K}=\mathbb{R}, \mathbb{C}, \mathbb{H}$, where 
$2 \mathbb{K}(n) := \mathbb{K}(n) \oplus \mathbb{K}(n)$, 
and where $n=2^{[\frac{D}{2}]}$. By allowing complex linear combinations 
of the generators $\gamma^\mu$ 
one obtains the complex Clifford algebras $\mathbb{C}l_{t+s} =
Cl_{t,s} \otimes_\mathbb{R} \mathbb{C}$, which are isomorphic 
to matrix algebras of the form $\mathbb{C}(n)$ in even dimensions
and  to matrix algebras of the form $2\mathbb{C}(n) = \mathbb{C}(n) 
\oplus \mathbb{C}(n)$ in odd dimensions.\footnote{See for example 
Tables I and II of \cite{SpinGeometry} for a complete list of Clifford algebras
as matrix algebras.}

Algebras of the form $m\mathbb{K}(n)$ have $m$ inequivalent irreducible 
representations, where one of the summands $\mathbb{K}(n)$ acts on 
$\mathbb{K}^n$ by matrix multiplication, while the other summands
act trivially. Therefore Clifford algebras have either one to two
inequivalent irreducible representations. Clifford representations
give rise to spinor representations by restriction, because the 
real and complex spin groups are naturally embedded into 
the real and complex Clifford algebras. 
The even subalgebra $Cl^0_{t,s}\subset Cl_{t,s}$ of the real Clifford
algebra is the subalgebra which is 
generated by even products of the generators $\gamma^\mu$. This 
subalgebra is itself a Clifford algebra. The spin group $\mathrm{Spin}(t,s)$
is the subgroup of (the group of invertible elements of)
$Cl^0_{t,s}$  which is generated by unit norm elements. 
Therefore irreducible $Cl^0_{t,s}$ modules become irreducible
$\mathrm{Spin}(t,s)$ modules by restriction. The {\em real spinor module}
$S$ is the $\mathrm{Spin}(t,s)$ module obtained by restricting an
irreducible $Cl_{t,s}$ module. As a real spin representation, $S$ is either irreducible, or decomposes
into two irreducible {\em real semi-spinor modules} $S_\pm$, 
$S = S_+ + S_-$. The real semi-spinor modules can be 
isomorphic to each other, $S_+ \cong S_-$ or non-isomorphic, $S_+\not{\cong} S_-$.
One can decide which of these three cases is realised by
comparing the matrix algebras realizing $Cl_{t,s}$ and $Cl^0_{t,s}$,
and using that $\mathbb{K}(n)$ has one,  while $2\mathbb{K}(n)$
has two inequivalent irreducible representations. 
We can also determine
the Schur algebras
$\mathcal{C}(S)$ and $\mathcal{C}(S_\pm)$ by using corrolaries
to Schur's Lemma, see \cite{Lang:Algebra}. For the case at hand,
the relevant statements are:
\begin{itemize}
\item
If $\Sigma = \mathbb{K}^n$
is an irreducible module of the real spin group
 $\mathrm{Spin}(t,s) \subset Cl^0_{t,s}$ ,  then its Schur algebra
is $\mathbb{K}$.\footnote{The cases where $\mathbb{K}=\mathbb{R}, \mathbb{C}, \mathbb{H}$ 
correspond to what is usually called real, complex and quaternionic representations, respectively. 
In all cases we are interested in the underlying real representation. Complex and quaternionic 
representations are thus viewed as real representations with additional invariant structures
that are encoded by the Schur algebra.}
\item
If $\Sigma_1 \oplus \Sigma_2$ is the sum of
two irreducible modules of the real spin group $\mathrm{Spin}(t,s) \subset Cl^0_{t,s}$, 
then the Schur algebra is
$2\mathbb{K}$ if $\Sigma_1\not\cong \Sigma_2$ and 
$\mathbb{K}(2)$ if $\Sigma_1 \cong \Sigma_2$. 
\end{itemize}

The complex spin group $\mathrm{Spin}(t+s, \mathbb{C})$ is the subgroup
of unit norm elements of the even subalgebra $\mathbb{C}l^0_{t+s} \subset
\mathbb{C}l_{t+s}$. The {\em complex spinor module} $\mathbb{S}$ 
is the $\mathrm{Spin}(t,s)$-module obtained by restricting an irreducible 
$\mathbb{C}l_{t+s}$-module. Its elements are the {\em Dirac spinors}.
Since $\mathbb{C}l^0_{t+s} \cong \mathbb{C}l_{t+s-1}$, it follows that
$\mathbb{S}$ is irreducible in odd dimensions, but decomposes into
two irreducible complex semi-spinor modules $\mathbb{S}_\pm$ 
in even dimensions, $\mathbb{S} = \mathbb{S}_+ + \mathbb{S}_-$. 
The elements of $\mathbb{S}_\pm$ are also called {\em Weyl spinors}. 
Note that while $\mathbb{S}_\pm$ are non-isomorphic as $\mathbb{C}l^0_{t+s}$ modules, 
they may or may not be isomorphic
as real $\mathrm{Spin}(t,s)$-modules. 
The decomposition of $\mathbb{S}$ into irreducible $\mathrm{Spin}(t,s)$-modules
can be obtained by comparing the matrix algebras realizing
$\mathbb{C}l_{t+s}$, $Cl_{t,s}$ and $Cl^0_{t,s}$. This also allows one to 
determine the complex Schur algebras 
$\mathcal{C}_\mathbb{C}(\mathbb{S})$, $\mathcal{C}_\mathbb{C}(\mathbb{S}_\pm)$,
and the real Schur algebras $\mathcal{C}(\mathbb{S})$, $\mathcal{C}(\mathbb{S}_\pm)$.
The complex Schur algebra of an irreducible representation of $\mathrm{Spin}(t+s,\mathbb{C}) \subset
\mathbb{C}l_{t+s}^0$ is $\mathbb{C}$, while the complex Schur algebra of the 
sum of two irreducible complex representations is $2\mathbb{C}$ if the representations are 
not equivalent and $\mathbb{C}(2)$ if they are equivalent. 

As an example, consider the case where $(t,s)=(1,3)$. Then 
$\mathbb{C}l_4 = \mathbb{C}(4)$ and $\mathbb{C}l_4^0 = 2 \mathbb{C}(2)$ 
which implies $\mathbb{S}  = \mathbb{C}^4$ and $\mathbb{S}_\pm =
\mathbb{C}^2$ which are the Dirac and Weyl spinors respectively.
Since 
$\mathbb{S}_\pm$ are complex irreducible, their complex Schur algebras are 
$\mathcal{C}_\mathbb{C}(\mathbb{S}_\pm)
\cong \mathbb{C}$, and since they are inequivalent as complex modules,
$\mathcal{C}_\mathbb{C} (\mathbb{S}) = 2 \mathbb{C}$. Since $Cl_{1,3} = \mathbb{R}(4)$, the
real spinor module is $S=\mathbb{R}^4$. This implies that 
$\mathbb{S}= S \otimes_\mathbb{R} \mathbb{C}$ and shows that in this case real spinors are Majorana spinors,
that is, they arise by imposing a reality condition on Dirac spinors.  
The even part of the real Clifford algebra is $Cl_{1,3}^0 = \mathbb{C}(2)$ and since
$\mathbb{C}^2 \cong \mathbb{R}^4$, real spinors are irreducible.  
There are no real semi-spinors (which would be Majorana-Weyl
spinors), and Majorana spinors are equivalent, as real spin representations, to Weyl 
spinors $S\cong \mathbb{S}_\pm$.\footnote{This is reflected by the familiar fact that
in this case the 
$\mathcal{N}=1$ supersymmetry algebra can be equivalently expressed
using Majorana spinor or Weyl spinors.}
The Schur algebra of the real spinor module 
$S=\mathbb{C}^2$ 
is $\mathcal{C}_{1,3}(S) =
\mathbb{C}$. 
Since $\mathbb{S}_\pm$ are equivalent as real spin modules, 
the Schur algebra of $\mathbb{S}$, considered as a real spin module, 
is $\mathcal{C}_{1,3}(\mathbb{S}) =
\mathbb{C}(2)$. 

It is instructive to compare this to the case $(t,s)=(3,1)$ which is
Minkowski space with a mostly minus convention for the metric. In this case
the real Clifford algebra is $Cl_{3,1} = \mathbb{H}(2)$ and therefore the real
spinor module is $S = \mathbb{H}^2 \cong \mathbb{C}^4 = \mathbb{S}$. Now
real spinors are Dirac spinors, and because $Cl_{3,1}^0=\mathbb{C}(2)$ they are reducible
and decompose into real semi-spinors, $S = S_+ \oplus S_-$, where $S_\pm = 
\mathbb{C}^2\cong \mathbb{R}^4$. For dimensional reasons $\mathbb{S}_\pm \cong S_\pm$.
Since $Cl_{3,1}^0\cong \mathbb{C}(2)$ only has one inequivalent irreducible representation, 
$S_\pm$ are isomorphic as real
spin modules. Therefore the  Schur algebra of $\mathbb{S}=S$ is 
$\mathcal{C}_{3,1}(\mathbb{S}) = \mathcal{C}_{3,1}(S) = \mathbb{C}(2)$. 
In summary,
real semi-spinors and complex semi-spinors are equivalent to each other and correspond
to Weyl spinors of either chirality, which are equivalent, as real spin representations,  
to Majorana spinors.\footnote{We
do not distinguish between Majorana and Pseudo-Majorana spinors, but see for 
example \cite{Hull:1998ym}.}

 For more examples we refer to
\cite{Gall:2018ogw,Cortes:2019mfa} where the spinor modules and Schur algebras
have been worked for $D=4,5$ for all signatures. This already provides examples of
all possible cases.

\subsection{Spinor representations in arbitrary signature  \label{Sec:Explicit_spinor_modules}}

The Clifford generators can be realized explicity as 
$\gamma$-matrices, which like the abstract generators we denote by  
$\gamma^\mu$. These 
are complex $2^{[\frac{D}{2}]} \times 2^{[\frac{D}{2}]}$ matrices and 
can be constructed
as tensor products of the Pauli matrices and the $2\times 2$ identity
matrix, see for example \cite{VanProeyen:1999ni}.  
The $\gamma$-matrices generate a representation of the real Clifford algebra 
$Cl_{t,s}$ 
on $\mathbb{C}^{[\frac{D}{2}]}$, which extends to an irreducible representation of its
 complexification $\mathbb{C}l_{t+s}$. By restriction to $\mbox{Spin}(t,s)$, this
 becomes the complex spinor module $\mathbb{S}$ introduced above.
Its elements are the {\em Dirac spinors}.

The $\gamma$-matrices can be changed by equivalence transformations, which
can be used to select representations which are convenient for performing
computations. 
We impose that timelike $\gamma$-matrices are anti-Hermitian, while
spacelike $\gamma$-matrices are Hermitian:
\begin{equation}
(\gamma^\mu)^\dagger = \left\{ \begin{array}{cl}
- \gamma^\mu \;, & \mu = 1, \ldots, t  \;, \\
\gamma^\mu \;, & \mu=t+1, \ldots t+s = D \;. \\
\end{array} \right.
\end{equation} 
The remaining freedom of
performing unitary transformation will be used later to impose further conditions. 
In odd dimensions the product $\omega = \gamma_1 \cdots \gamma_D$
commutes with all generators $\gamma_\mu$ and distinguishes between
the two inequivalent representations of $Cl_{t,s}$ (which define equivalent
spin representations). In even dimensions $\omega$ anti-commutes with 
all generators, and commutes with the spin generators $\gamma^{\mu \nu}=\frac{1}{2}
[\gamma^\mu, \gamma^\nu]$.
Therefore $\omega$ can be used to decompose the complex spinor module $\mathbb{S}$ into
{\em complex semi-spinor modules} $\mathbb{S}_\pm$. Depending on signature,
$\omega^2 = (-1)^t \mathbbm{1}$ or $\omega^2 = (-1)^{t+1} \mathbbm{1}$. 
Setting $\gamma_* = \pm \omega$ or $\gamma_* = \pm i \omega$ 
one obtains an operator which satisfies $\gamma_*^2=1$ and therefore
can be used to define projectors $\frac{1}{2} (1 \pm \gamma_*)$ onto 
the complex semi-spinor modules. For Lorentz signature $(1,D-1)$ 
(or, using a `mostly minus' convention $(D-1,1)$)  $\gamma_*$ is the
chirality operator which defines Weyl spinors. We will therefore refer
to complex semi-spinors as {\em Weyl spinors}. For computational purposes
it is convenient to fix the relation between $\gamma_*$ and $\omega$.
We choose
\begin{equation}
\label{gamma_star}
\gamma_* = (-i)^{\frac{D}{2} + t} \gamma_1 ... \gamma_D \;.
\end{equation}

One can always construct matrices $A,B,C$ which relate the $\gamma$-matrices
to their Hermitian conjugates, complex conjugates, and transposed, respectively:
\begin{eqnarray}
(\gamma^\mu)^\dagger &=& (-1)^t A \gamma^\mu A^{-1}\;, \label{A}\\
(\gamma^\mu)^*  &=& (-1)^t  \tau B \gamma^\mu B^{-1} \;, \\
(\gamma^\mu)^T &=& \tau  C \gamma^\mu C^{-1} \;, \label{C}
\end{eqnarray}
where $\tau=\pm 1$, with the allowed values depending on the dimension 
$D$.\footnote{$\tau$ is related to the 
parameter $\eta$ in \cite{VanProeyen:1999ni} by $\tau=-\eta$. Note that some authors,
for example \cite{Hull:1998ym}
use a definition of $\eta$ which is signature dependent. In our convention 
and the one of  \cite{VanProeyen:1999ni}, $\tau=-\eta$ is signature independent, and all
signature dependent factors in the relations \eqref{gamma_star} are explicit factors $(-1)^t$. 
This is natural, because,  as we will see later, 
$\tau$ and another parameter $\sigma$ are invariants which characterize the
properties of a complex bilinear form with Gram matrix $C$, for which signature
does not have an invariant meaning. }
One possible choice for $A$ is the product $\gamma_1 \cdots \gamma_t$ of
all timelike $\gamma$-matrices. In Lorentz signature $(1,D-1)$, this gives
the usual 
$A=\gamma_0$, where we have shifted the range of Lorentz indices 
to  $\mu = 0, \ldots, D-1$.\footnote{In general we use
$\mu=1,\ldots, D$, but for signature $(1,D-1)$ we may shift this to the conventional
range. We also shift the index range when performing a dimensional reduction, see
Appendix \ref{App:Dim_Red}.}
We remark that our choice for $A$ is not unique, and for Lorentz
signature there are conventions which differ from ours by a factor $-1$
or $\pm i$.

The matrix $C$ is the {\em charge conjugation matrix}.
One can always choose a representation where $C$ is Hermitian and
unitary,
\begin{equation}
C^\dagger = C^{-1} = C \;.
\end{equation}
$C$ is either symmetric or antisymmetric 
\begin{equation}
C^T = \sigma C \;,
\end{equation}
where $\sigma=\pm 1$.\footnote{$\sigma$ is related to the parameter $\epsilon$ used
in  \cite{VanProeyen:1999ni}
 by $\sigma = - \epsilon$.} 
Which values of $\tau$ and $\sigma$ are possible depends on the dimension $D$, 
and we have listed these values in  Table \ref{bilinears}. The well known periodicity
modulo 8 of the classification of real Clifford algebras implies that this table is
periodic modulo 8, so that it encodes the values for arbitrary dimension. 
While in odd dimensions there is only one charge conjugation matrix up to
equivalence, there are two inequivalent choices in even dimensions, which
are distinguished by the corresponding value of $\tau$,
\begin{equation}
C_\pm := C_{-\tau} \;.
\end{equation}
Both charge conjugation matrices are related through multiplication by the chirality
matrix $\gamma_*$:
\begin{equation}
C_\pm = \gamma_* C_\mp \;.
\end{equation}
With our choice \eqref{gamma_star} for the chirality matrix $\gamma_*$, this
implies
\begin{align}
    &C_\pm \gamma_* = \begin{cases} \pm i C_\mp \;,\qquad &\text{for D = 2, 6, 10, \ldots }\\ C_\mp  \;,\qquad &\text{for D = 4, 8, 12, \ldots} \end{cases} 
\end{align}
see Appendix \ref{Sect_Useful_Formulae} for a collection of useful relations.

Given a choice of $A$ and $C$, we can take $B:= (CA^{-1})^T$. The matrix $B$
satisfies
\begin{equation}
BB^\dagger = \mathbbm{1} \;,\;\;\; BB^* = \epsilon \mathbbm{1} \;,
\end{equation}
where $\epsilon=\pm 1$, depending on signature.
$B$ is not completely fixed by our choice of $A,C$ since 
$\alpha B$, where $\alpha$ is a phase factor, has the same properties. 
The matrix $B$ can be used to define the (family of) complex-antilinear
maps
\begin{equation}
J \;: \psi \mapsto \alpha^* B^* \psi^*  \;.
\end{equation}
Since 
\begin{equation}
J^2 (\psi) = J(\alpha^* B^* \psi^*) = \alpha^* B^* \alpha B \psi = \epsilon \psi \;,
\end{equation}
the matrix $B$ either defines a real structure ($\epsilon=1$) or a 
quaternionic structure ($\epsilon=-1$) 
on $\mathbb{S}$, which is $\mathrm{Spin}(t,s)$-invariant. 
$\mathbb{S}$ carries a 
natural spin invariant complex structure
\[
I\;:  \psi \mapsto i \psi \;.
\]
Since $J$ is complex antilinear, $I$ and $J$ anticommute with each other and
with their product $K := IJ$. Since in addition
\begin{equation}
I^2 = - \mathrm{Id} \;,\;\;\;
J^2 = K^2 = \epsilon \mathrm{Id} \;,
\end{equation}
$I,J$ generate a real four-dimensional algebra, which for $\epsilon=-1$ is the
algebra $\mathbb{H}$ of quaternions. For $\epsilon=1$, the resulting
algebra $\mathbb{H}'$ is called the algebra of {\em para-quaternions}
(or split-quaternions). The algebra $\mathbb{H}'$ 
is isomorphic to the algebra of real $2\times 2$ matrices, 
$\mathbb{H}' \cong \mathbb{R}(2)$. Compared to the quaternions,
the two generators $J,K$ are not complex structures (which square 
to minus the identity), but {\em para-complex} structures (which square
to the identity and have an equal number of eigenvalues $\pm 1$). 
To be able to discuss both cases in parallel, we will use the terms
$\epsilon$-{\em complex} and $\epsilon$-{\em quaternionic structure}, and use
the notation $J^{(\epsilon)}=J^{(\pm)}$, if $(J^{(\epsilon)})^2 = \epsilon \mbox{Id}$.
We will also use 
the notation $\mathbb{H}_\epsilon$, where $\mathbb{H}_{-1} = \mathbb{H}$
and $\mathbb{H}_1 = \mathbb{H}'$. 

The algebra $\mathbb{H}_\epsilon$ is a subalgebra of the Schur algebra $\mathcal{C}(\mathbb{S})$
of the complex spinor module. 
By comparison to the classification of Clifford algebras one can verify that
this is in fact the full Schur algebra in odd dimensions. In even dimensions
one has two charge conjugation matrices $C_\pm$ which can be used
to define two $B$-matrices $B_\pm$, which satisfy 
$B_\pm B^*_\pm = \epsilon_\pm \mathrm{Id}$. The corresponding 
$\epsilon$-quaternionic structures are denoted $J_{\pm}^{(\epsilon_\pm)}$, where
the lower index encodes the corresponding $B$-matrix $B_\pm$, while 
the upper index encodes whether the $\epsilon_\pm$-quaternionic structure
is a real or quaternionic structure.

The signs $(\epsilon_+, \epsilon_-)$ 
depend on the signature and are listed in Table  \ref{Table:epsilon-structures}, 
which again is periodic modulo 8 in dimension.
Note that all three inequivalent combinations occur, that is,
in even dimensions $\mathbb{S}$ either carries two spin invariant real
structures, or two spin invariant quaternionic structures, or one structure of 
each type. The algebras generated by these structures are $2\mathbb{H}' = \mathbb{H}' \oplus
\mathbb{H}'$ for $(\epsilon_+, \epsilon_-) = (1,1)$ (two real structures), 
$2\mathbb{H} = \mathbb{H} \oplus \mathbb{H}$ for $(\epsilon_+, \epsilon_-) = (-1,-1)$ (two quaternionic structures), 
and the algebra $\mathbb{C}(2)$ of complex $2\times 2$ matrices 
for $(\epsilon_+, \epsilon_-) = (\pm 1,\mp 1)$ (one real and one quaternionic structure). 
Note that in the third case $\mathbb{H} \subset \mathbb{C}(2)$ and 
$\mathbb{H}' \subset \mathbb{C}(2)$, but $\mathbb{C}(2) \not\cong
\mathbb{H} \oplus \mathbb{H}'$. By comparison to the classification 
of Clifford algebras one can verify that this is the full Schur algebra
$\mathcal{C}(\mathbb{S})$ in even dimensions. We refer to 
\cite{Cortes:2019mfa} for more details. Note that the entries in 
Table  \ref{Table:epsilon-structures} allow one to read off the Schur algebras
$\mathcal{C}(\mathbb{S})=\mathcal{C}_{t,s}(\mathbb{S})$ for all signatures $(t,s)$.

In signatures where there exists a $B$-matrix which defines a real structure one can impose
the reality condition 
\begin{equation}
\psi^* = \alpha B \psi
\end{equation}
on a Dirac spinor. We will refer to such spinors as {\em Majorana spinors}. Note
that we do not require that the $\gamma$-matrices have real entries, and that we 
do not distinguish between Majorana spinors and pseudo-Majorana spinors.\footnote{This
distinction may be relevant for deciding which type of terms, for example mass terms,
can appear in a supersymmetry Langrangian. This is beyond the scope of our paper,
see for example \cite{Hull:1998ym} for a discussion.}

\begin{table*}[th!]
\begin{tabu}{|[1.5pt] c|c|c|c|c|c|c|c|[1.5pt]}
\tabucline[1.5pt]{1-8} $D$ & $(0, D)$ & $(1, D - 1)$ & $(2, D - 2)$ & $(3, D-3)$ & $(4, D-4)$ & $(5, D-5)$ & $(6, D- 6)$ \\
\tabucline[1.5pt]{1-8} 1 & $+1$ & $+1$ & & & & & \\
\hline 2 & $-1_+,+1_-$ & $+1_++1_-$ & $+1_+,-1_-$ & & & & \\
\hline 3 & $-1$ & $+1$ & $+1$ & $-1$ & & & \\
\hline 4 & $-1_+,-1_-$ & $+1_+,-1_-$ & $+1_+,+1_-$ & $-1_+,+1_-$ & $-1_+,-1_-$ & & \\
\hline 5 & $-1$ & $-1$ & $+1$ & $+1$ & $-1$ & $-1$ & \\
\hline 6 & $+1_+,-1_-$ & $-1_+,-1_-$ & $-1_+,+1_-$ & $+1_+,+1_-$ & $+1_+,-1_- $ & $-1_+,-1_-$ & $-1_+,+1_-$ \\
\hline 7 & $+1$ & $-1$ & $-1$ & $+1$ & $+1$ & $-1$ & $-1$ \\
\hline 8 & $+1_+,+1_-$ & $-1_+,+1_-$ & $-1_+,-1_-$ & $+1_+,-1_-$ & $+1_+,+1_-$ & $-1_+,+1_-$ & $-1_+,-1_-$ \\
\hline 9 & $+1$ & $+1$ & $-1$ & $-1$ & $+1$ & $+1$ & $-1$ \\
\hline 10 & $-1_+,+1_-$ & $+1_+,+1_-$ & $+1_+,-1_- $ & $-1_+,-1_-$ & $-1_+,+1_-$ & $+1_+,+1_-$ & $+1_+,-1_- $ \\
\hline 11 & $-1$ & $+1$ & $+1$ & $-1$ & $-1$ & $+1$ & $+1$ \\
\hline 12 & $-1_+,-1_- $ & $+1_+,-1_- $ & $+1_+,+1_-$ & $-1_+,+1_-$ & $-1_+,-1_-$ & $+1_+,-1_-$ & $+1_+,+1_-$\\ \tabucline[1.5pt]{1-8}
\end{tabu}
\caption{This table lists the spin invariant real and quaternionic structures on the complex spinor
module $\mathbb{S}$ for all signatures $(t,s)$ in dimensions up to twelve. Note that the table is periodic modulo
8 in dimension and therefore covers all possible dimensions and signatures. 
It is also invariant under $(t,s)\leftrightarrow (s,t)$, provided that in even dimension
 one exchanges $J_\pm^{(\epsilon)}$ with $J_\mp^{(\epsilon)}$. 
The entries 
in the table are the values of $\epsilon=\pm 1$ which tell us whether 
the  $\epsilon$-quaternionic structure  $J^{(\epsilon)}$ is a real structure, $\epsilon=1$, or
a quaternionic structure, $\epsilon=-1$. 
In even dimensions the sign subscript
on $\pm 1_\pm$ indicates whether the corresponding $\epsilon$-quaternionic structure  
$J^{(\epsilon)}_\pm$ has been constructed using $B_+$ or $B_-$. Majorana spinors
exists in signatures where at least one of the entries is $+1$, while symplectic Majorana
spinors exist whenever at least one entry is $-1$. 
In even dimensions, if both entries are equal to each other, reality conditions are `Weyl-compatible,' that is
they respect chirality, whereas if they are different from each other, reality conditions
are `Weyl-incompatible', that is, they flip chirality. Majorana-Weyl spinors exist when 
both entries are equal to $+1$. If both entries are equal to $-1$ one has a quaternionic
structure compatible with chirality and can define `symplectic Majorana-Weyl spinors.'
If  the signs are not equal, one has one real and one quaternionic structures, which 
both flip chirality. 
The Schur algebra of the complex spinor module $\mathbb{S}$ 
is $\mathbb{H}_\epsilon$ in odd dimensions and either $\mathbb{H}_\epsilon \oplus
\mathbb{H}_\epsilon$ (for $\epsilon_+=\epsilon_-$) or $\mathbb{C}(2)$ 
(for $\epsilon_+=-\epsilon_-$)
in even
dimensions. Some of the above statements will only be proved in later sections.
\label{Table:epsilon-structures}}
\label{structures}
\end{table*}
Note that Majorana spinors are not the same as real spinors, that is elements
of the real spinor module $S$. Depending on signature, either $\mathbb{S}\cong S$,
and real spinors are Dirac spinors, 
or $\mathbb{S} \cong S \otimes_\mathbb{R}  \mathbb{C}$, and real spinors are 
Majorana spinors. We remark that if $\mathbb{S}\cong S$, Majorana spinors
may still exist. In this case real spinors are reducible, $S\cong S_+ \oplus S_-$, 
and Majorana spinors correspond real semi-spinors. Note that  in contrast to complex
semi-spinors, real semi-spinors can exist in odd dimensions. The interested
reader is referred to Appendix \ref{App:Real_vs_Majorana} for details.

\subsection{Complexification of spinor modules \label{complexification}}

The odd part $\mathfrak{g}_1$ of a Poincar\'e Lie superalgebra 
consists of copies of irreducible real spinor representations. 
If the real spinor module $S$ is the unique irreducible real spinor 
representation, then the only choice we have is the number $N$ of copies, 
$\mathfrak{g}_1 = S^{\oplus  N} \cong S \otimes \mathbb{R}^N$. 
If $S$ is reducible, but the real semi-spinor
modules are isomorphic, $S_+\cong S_-$, we can take 
$\mathfrak{g}_1 =  S_+^{\oplus N}  \cong S_+ \otimes \mathbb{R}^{N}$
without loss of generality.
 If the real semi-spinors are not equivalent $S_+ \not\cong S_-$,
we can choose their multiplicity independently, 
$\mathfrak{g}_1 = S^{\oplus N_+} _+ \oplus  S_-^{\oplus N_-}  \cong
S_+ \otimes \mathbb{R}^{N_+} \oplus 
S_- \otimes \mathbb{R}^{N_-}$.

The complexification $\mathfrak{g}_1^\mathbb{C} \otimes_\mathbb{R} \mathbb{C}$
of $\mathfrak{g}_1$ takes the form 
$\mathbb{S} \otimes \mathbb{C}^K$ or $\mathbb{S}_+ \otimes \mathbb{C}^{K_+}
\oplus \mathbb{S}_- \otimes \mathbb{C}^{K_-}$, where the relation between
$K,K_+,K_-$ and ${N}, {N}_+, {N}_-$ depends on 
whether $\mathbb{S}$, $\mathbb{S}_\pm$ carry a $\mathrm{Spin}(t,s)$ invariant real
structure, or not.  Here we summarize these relations, while details are given in 
Appendix \ref{App:Details_Complexification}. There are three cases, depending 
on the properties of the real spinor module $S$:
\begin{enumerate}
\item
$S$ irreducible:
\[
\mathfrak{g}_1 \otimes_{\mathbb{R}} \mathbb{C} = 
S^{\oplus N} \otimes_{\mathbb{R}} \mathbb{C} = \mathbb{S} \otimes_{\mathbb{C}} \mathbb{C}^K\;,
\left\{ \begin{array}{ll}
K=N \;, & \mbox{if} \;\; \mathbb{S} \cong  S \otimes_{\mathbb{R}} \mathbb{C} \;, \\
K = 2N \;, & \mbox{if}\;\; \mathbb{S} \cong S \;. \end{array} \right.
\]
\item
$S=S_+ + S_-$ and $S_+ \cong S_-$:
\[
\mathfrak{g}_1 \otimes_{\mathbb{R}} \mathbb{C} = 
S_+^{\oplus N} \otimes_{\mathbb{R}} \mathbb{C} = \mathbb{S} \otimes_{\mathbb{C}} \mathbb{C}^K\;,
\]
where $K=N$, since this only occurs when $\mathbb{S} \cong S_\pm \otimes_{\mathbb{R}} 
\mathbb{C}$. 
\item
$S=S_+ + S_-$ and $S_+ \not{\cong} S_-$:
\begin{eqnarray*}
 && \mathfrak{g}_1 \otimes_{\mathbb{R}} \mathbb{C} = 
\left( S_+^{\oplus N_+} \oplus S_-^{\oplus N_-} \right) \otimes_{\mathbb{R}} \mathbb{C}  \\
&& 
=
\mathbb{S}_+ \otimes_{\mathbb{C}} \mathbb{C}^{K_+}  \oplus
\mathbb{S}_- \otimes_{\mathbb{C}} \mathbb{C}^{K_-}  \;,
\left\{ \begin{array}{ll}
K_\pm = N_\pm \;,& \mbox{if}\;\; \mathbb{S}_\pm \cong S_\pm \otimes_{\mathbb{R}} \mathbb{C}\;, \\
K_\pm = 2 N_\pm \;, & \mbox{if}\;\; \mathbb{S}_\pm \cong S_\pm  \;.\\
\end{array} \right. 
\end{eqnarray*}
\end{enumerate}
Note that the relation between $K, K_\pm$ and $N, N_\pm$ is completely determined
by the information  whether 
$\mathbb{S}, \mathbb{S}_\pm$ admit invariant real structures or not. 
We remark that $N,N_\pm$ do not count supersymmetries 
in multiples of the minimal supersymmetry algebra 
in a given signature. The reason is that the definition of a supersymmetry
algebra requires the existence of  a non-degenerate bracket 
$\mathfrak{g}_1 \times \mathfrak{g}_1 \rightarrow V$, which is not guaranteed.
The convention 
that we find convenient for labelling supersymmetry algebras in arbitrary signature
is that  $\mathcal{N}=K$ in cases 1 and 2 
and $\mathcal{N}_\pm = K_\pm$ in case 3. 
This means that we count in units of Majorana and Majorana-Weyl spinors, irrespective
of whether these exist in the given signature. Thus in the non-chiral case the
smallest supersymmetry algebra is labeled $\mathcal{N}=1$ if the supercharges form 
a single Majorana spinors and $\mathcal{N}=2$ if they form a Dirac spinor. 
In the chiral case algebras based on a single Majorana-Weyl spinor are denoted
$(1,0)$ and $(0,1)$ while algebras based on a single Weyl spinor are 
denoted $(2,0)$ or $(0,2)$. 
We will illustrate how our conventions compare to standard $(1,D-1)$ signature
conventions using explicit examples later, see in particular Section \ref{sect:R_even}.

\section{Bilinear forms and complex supersymmetry algebras \label{secbilform}}

The first step in our programme is to construct complex supersymmetry
algebras with odd part $\mathfrak{g}_1^\mathbb{C}$ equal to
\begin{align}
    \mathbb{S} \otimes \mathbb{C}^K \qquad \text{or} \qquad \mathbb{S}_+ \otimes \mathbb{C}^{K_+} \oplus \mathbb{S}_- \otimes \mathbb{C}^{K_-}.
\end{align}
This requires us to specify a complex superbracket, 
$\mathfrak{g}_1^\mathbb{C} \times \mathfrak{g}_1^{\mathbb{C}} \rightarrow V^\mathbb{C}$,
where $V^\mathbb{C} = \mathbb{C}^{t+s}  \cong \mathbb{R}^{t,s} \otimes_\mathbb{R} \mathbb{C}$. 
As explained in the introduction, this is equivalent to defining 
a complex symmetric, spin-equivariant vector-valued bilinear form $\Pi_\beta$ on 
$\mathfrak{g}_1^\mathbb{C}$, which  in turn can be defined using a super-admissible 
complex bilinear form $\beta$ on $\mathfrak{g}_1^\mathbb{C}$. In this section we show
how such forms are constructed out of bilinear forms on the complex spinor
modules $\mathbb{S}$ and $\mathbb{S}_\pm$, 
and on the auxiliary spaces $\mathbb{C}^K$ and $\mathbb{C}^{K_\pm}$. 

\subsection{Bilinear forms on the complex spinor module $\mathbb{S}$} \label{subsecbilformons}

The definitions of symmetry $\sigma_\beta$ and type $\tau_\beta$ of a bilinear
form $\beta$ were given (\ref{symmetry+type}). Bilinear forms of definite type and
symmetry are called admissible and are automatically spin invariant. The choice
of an admissible complex bilinear form on $\mathbb{S}$ is equivalent to the choice
of a charge conjugation matrix $C$, 
\begin{equation}
C: \mathbb{S} \times \mathbb{S} \rightarrow \mathbb{C} \;, \;\;
    C(\lambda, \chi) = \lambda^T C \chi = \lambda_\alpha C^{\alpha \beta} \chi_\beta  \;.
\end{equation}
We will refer to such bilinear forms as  {\em Majorana bilinear forms.}  The symmetry $\sigma_C$ 
and type $\tau_C$ of the bilinear form $C$ are equal to the parameters $\sigma,\tau$ 
associated with a charge conjugation matrix, $\sigma=\sigma_C, \tau = \tau_C$.
The values of $\sigma, \tau$ are listed in Table \ref{bilinears}.

Given an admissible bilinear form on $\mathbb{S}$, we can define a spin equivariant
vector valued form 
\begin{equation}
\beta^{(1)} \;: \mathbb{S} \times \mathbb{S} \rightarrow V^\mathbb{C} \;,\;\;\;
(s,t) \mapsto \beta(\gamma^\mu s, t) e_\mu \;,
\end{equation}
which is symmetric for $\sigma \tau=1$ and anti-symmetric for $\sigma \tau =-1$. 
Symmetric vector valued bilinear forms on $\mathbb{S}$ define a Poincar\'e
Lie superalgebra with $\mathfrak{g}_1^\mathbb{C} = \mathbb{S}$, and the associated 
scalar bilinear forms are called super-admissible. Since we are interested 
in defining supersymmetry algebras with $\mathfrak{g}_1^\mathbb{C} = \mathbb{S} \otimes
\mathbb{C}^K$, we are not restricted to super-admissible bilinear forms, since
we have the additional freedom of choosing a bilinear form on $\mathbb{C}^K$. Therefore all
admissible bilinear forms on $\mathbb{S}$ qualify as building blocks for superbrackets.

\subsection{Bilinear forms on the Weyl Spinor modules $\mathbb{S}_\pm$} \label{secbilformweyl}

In even dimensions the complex spinor module decomposes into the complex
semi-spinor modules $\mathbb{S}_\pm$, and we can define two inequivalent
complex bilinear forms $C_{-\tau}= C_\pm$ using the two available charge conjugation matrices. 
Let us consider the case where one argument of the bilinear form $C_{-\tau}$ is a Weyl spinor, while
the other is arbitrary. 
Using $\gamma_* \lambda_\pm = \pm \lambda_\pm$ and \eqref{cgammastar}
we find
\begin{align}
    C_{-\tau } (\cdot, \lambda_\pm) = C_{-\tau} (\cdot, \pm \gamma_* \lambda_\pm) =
    \begin{cases} i C_{\tau} (\cdot, \lambda_\pm) \;,\qquad &\text{D = 2, 6, 10},  \ldots\\ \pm C_{\tau} (\cdot, \lambda_\pm)\;, \qquad &\text{D = 4, 8, 12}, \ldots .\end{cases}   
\end{align}
We observe that the two bilinear forms become proportional when we restrict one argument
to be a Weyl spinor. This shows that the two charge conjugation matrices and bilinear forms only 
differ by a relative sign or factor of $i$ between their restrictions to the complex semi-spinor modules. 
For any admissible bilinear form $\beta$ on $\mathbb{S} = \mathbb{S}_+  \oplus
\mathbb{S}_-$ we can define an associate admissible bilinear form $\beta' = \beta(\cdot, \gamma_* \cdot)$ 
which has opposite type and is proportional to $\beta$ when restricted to a definite
chirality in one argument.

Admissible bilinear forms on $(\mathbb{S}_+ \oplus \mathbb{S}_-) \times
(\mathbb{S}_+ \oplus \mathbb{S}_-)$ have a third invariant besides
the symmetry $\sigma_\beta$ and the type $\tau_\beta$. An admissible bilinear
form $\beta$ has isotropy $\iota_{\beta,0}=1$ and
is called {\em orthogonal} if its
restrictions to $\mathbb{S}_+ \times \mathbb{S}_-$ and 
$\mathbb{S}_- \times \mathbb{S}_+$
are identically zero, and it has isotropy $\iota_{\beta,0}=-1$ and
 is called {\em isotropic} if its restrictions
to $\mathbb{S}_+ \times \mathbb{S}_+$ and
$\mathbb{S}_- \times \mathbb{S}_-$
are identically zero.
To determine the isotropy $\iota_{\beta,0}$ we use 
that bilinear
forms which differ by the insertion of $\gamma_*$ in one argument are 
proportional to one another when restricted to a fixed chirality in one argument. 
We compute:
\begin{eqnarray*}
&& C_+ (\lambda_\pm, \chi_\pm) = k  C_-(\lambda_\pm, \chi_\pm )  \;,\mbox{where}\;\;\;
k = \left\{  \begin{array}{ll} i \;, &   D=2,6,10,\ldots\\  1 \;, & D=4,8,12, \ldots \\
\end{array} \right. \\
&\Rightarrow& \sigma_+ C_+ (\lambda_\pm, \chi_\pm) = k \sigma_- C_-(\lambda_\pm, \chi_\pm) 
= k k^{-1}  \sigma_- C_+ (\lambda_\pm, \chi_\pm)
= \sigma_- C_+ (\lambda_\pm, \chi_\pm) \;,
\end{eqnarray*}
where $\sigma_\pm = \sigma_{C_\pm}$. 
Therefore either $\sigma_+= \sigma_-$, or $C_+$ and $C_-$ are completely degenerate 
when restricted to semi-spinors of the same chirality. 
On the other hand
\begin{eqnarray*} 
&&  C_+ (\lambda_\pm, \chi_\mp) = k' C_-(\lambda_\pm, \chi_\mp ) \;,\mbox{where}\;\;\;
k' = \left\{ \begin{array}{ll}  i \;, &   D=2,6,10,\ldots\\  -1 \;, & D=4,8,12, \ldots \\
\end{array} \right. \\
&\Rightarrow& \sigma_+ C_+ (\lambda_\mp, \chi_\pm) = k  \sigma_- C_-(\lambda_\mp, \chi_\pm) 
= k k' \sigma_- C_+ (\lambda_\mp, \chi_\pm)
=- \sigma_- C_+ (\lambda_\mp, \chi_\pm) \;.
\end{eqnarray*}
Note that compared to the previous case we have obtained a minus sign, because
the chirality of the second argument has changed. In this case either
$\sigma_+ = - \sigma_-$, 
or the bilinear forms $C_\pm$ are completely degenerate on semi-spinors of opposite
chirality. Thus any admissible bilinear form is
either orthogonal or isotropic, and its isotropy is given by 
\begin{equation}
\iota_{\beta, 0} = \sigma_+ \sigma_- \;.
\end{equation}
If we insert a $\gamma$-matrix into the first argument, this flips the isotropy, since
$\gamma^\mu$ anti-commutes with $\gamma_*$ in even dimensions. More generally,
if we substitute $p$-fold anti-symmetrized products of $\gamma$-matrices into 
$\beta$ the isotropy is given by
\begin{equation}
\iota_{\beta,p} = (-1)^p \sigma_+ \sigma_-  \;,\;\;\;p=0,1, \ldots \;.
\end{equation}
Since superbrackets are defined by vector-valued bilinear forms, the relevant
isotropy is $\iota_{\beta,1}$, which we will therefore denote by $\iota = \iota_\beta:= \iota_{\beta, 1}$
in the following. By inspection of Table \ref{bilinears}, we see that 
vector-valued bilinear forms $C_\pm(\gamma^\mu\cdot, \cdot)$ 
are orthogonal in dimension $D=2,6,10,\ldots$ and isotropic in dimension 
$D=4,8,12,\ldots$. This is important because `chiral' supersymmetry algebras which 
only involve supersymmetry generators of one chirality (or which, more generally, 
have different anti-commutation relations depending on chirality) can only exist in dimensions where the vector-valued
bilinear form is of orthogonal type.

\subsection{Bilinear forms on the auxiliary space $\mathbb{C}^{K}$}

To define a complex bilinear form on the extended spinor module $\mathbb{S} \otimes \mathbb{C}^K$,
we also need to choose a bilinear form $M$ on $\mathbb{C}^K$, 
\begin{equation}
    M: \mathbb{C}^{K} \times \mathbb{C}^{K} \rightarrow \mathbb{C} \;,\;\;\;
    M(w,z) = w^i z^j M_{j i} \qquad i, j = 1,...,M \;.
\end{equation}
Our index convention is chosen to be consistent with the $NW-SE$ convention in the case
where $M$ is antisymmetric. 
Since the spin group does not
act on the multiplicity space $\mathbb{C}^K$, spin-equivariance is not an issue. If 
the bilinear form $\beta$ we have chosen on $\mathbb{S}$ is super-admissible, then
$M$ needs to be symmetric, if $\beta$ is not super-admissible, then 
$M$ needs to be antisymmetric, in order that $\beta \otimes M$
defines a superbracket on $\mathbb{S} \otimes \mathbb{C}^K$. 
The symmetry of $M$ is denoted $\sigma_M$,
\begin{align}
    M(w,z) = \sigma_M M(z,w), \qquad M_{i j} = \sigma_M M_{j i}, \qquad \sigma_M = \pm 1.
\end{align}
Non-degenerate
symmetric and
antisymmetric complex bilinear forms on  $\mathbb{C}^K$ are unique up to isomorphism.
In the symmetric case we will use the standard symmetric bilinear form $\delta(\cdot, \cdot)$
with Gram matrix given by the Kronecker-symbol, $M_{i j} = \delta_{i j}$.
In the anti-symmetric case the bilinear form is only non-degenerate for $K$ even. 
Since degenerate superbrackets effectively involve a smaller spinor module 
$\mathfrak{g}_1^\mathbb{C}$ with a non-degenerate superbracket, 
we don't need to consider them separately.\footnote{The vector
space of real superbrackets contains degenerate superbrackets, which correspond to 
higher co-dimension orbits of the action of the Schur group. In particular, the completely
degenerate bracket always forms a zero-dimensional orbit. 
If superbrackets exist which are degenerate, but not completely degenerate, they 
define smaller supersymmetry 
algebras with a lower number of supercharges. For example the $\mathcal{N}=1$ 
supersymmetry algebras in signatures $(1,3)$ and $(2,2)$ correspond to co-dimension one
orbits in the space of $\mathcal{N}=2$ superbrackets, see \cite{Cortes:2019mfa}.
Such real superbrackets can also be obtained directly by imposing a reality condition on a non-degenerate
complex superbracket and the restriction to even $K$ does not restrict the generality of our
method.}
 For $K$ even, we use the standard non-degenerate anti-symmetric bilinear form 
\begin{equation}
(J_K)_{i j} = \begin{pmatrix}
            0 & \mathds{1}_k \\ -\mathds{1}_k & 0
        \end{pmatrix} \qquad K = 2k \;.
\end{equation}
$(J_2)_{i j}$ is the Levi-Civita symbol $\epsilon_{i j}$. 
The bilinear form represented by the Gram matrix $(J_K)_{i j}$ will be denoted $J(\cdot, \cdot)$ or just $J$.
The $K$ subscript will be omitted when the context is clear.

The groups acting linearly on $\mathbb{C}^{K}$ under which these bilinear forms are invariant will be denoted $G_{\mathbb{C}^{K}}$. Depending on the symmetry of $M$ the invariance group is either the complex
orthogonal group or the complex symplectic group:
\begin{align}
    G_{\mathbb{C}^{K}} = \begin{cases} \mathrm{O}(K,\mathbb{C}) \;, \quad &\text{for   }M=\delta, \\
        \mathrm{Sp}(K,\mathbb{C}) \;,\quad &\text{for   }M=J.    \end{cases}
\end{align}
As we will see, the R-symmetry groups of the real supersymmetry algebras will be determined
by the groups $G_{\mathbb{C}^{K}}$, together with isotropy properties of the bilinear form and
the reality conditions.

 \subsection{Complex supersymmetry algebras}

By combining our previous results, we can now define complex superbrackets.

\subsubsection{Odd dimensions}

In odd dimensions, the complexified spinor module is always of the form 
$\mathbb{S} \otimes \mathbb{C}^K$, and bilinear forms can be built by
taking tensor products of bilinear forms on each factor. 
We will use the notation $\beta = C \otimes M$, where
\begin{equation}
    \beta: (\mathbb{S} \otimes \mathbb{C}^{K}) \times (\mathbb{S} \otimes \mathbb{C}^{K}) \rightarrow \mathbb{C} \;,\;\;\l
    \beta (\lambda, \chi) = (\lambda^i)^T C \chi^j M_{j i} \;.
\end{equation}
It will be useful to display indices referring to the multiplicity space $\mathbb{C}^K$ explicitly while
suppressing spinor indices. 

The symmetry of $\beta$ is the product of those of $C$ and $M$, while the type
is inherited from $C$ since the $\gamma$-matrices do not act on $\mathbb{C}^K$:
$\sigma_\beta = \sigma_C \sigma_M$, 
$\tau_\beta = \tau_C$. To define a superbracket we need $\sigma_\beta \tau_\beta =
\sigma_C \tau_C \sigma_M=1$. 

\subsubsection{Even Dimensions}

In even dimensions we have to distinguish two cases. If the real spinor module 
is irreducible then the extended spinor module has the same form 
$\mathbb{S} \otimes \mathbb{C}^K$ as in odd dimensions, and everything works like there.
In signatures where  the real spinor module decomposes into non-isomorphic real semi-spinor modules
the extended spinor module is  $\mathbb{S}_+ \otimes \mathbb{C}^{K_+} \oplus \mathbb{S}_- \otimes \mathbb{C}^{K_-}$. How we proceed depends on 
whether  the vector-valued Majorana bilinear forms are orthogonal or isotropic on $\mathbb{S} = \mathbb{S}_+ \oplus \mathbb{S}_-$, which is determined by $\iota=\iota_{\beta,1} = - \sigma_+ \sigma_-$. 
This only depends on the dimension and
 is unaffected by the $\mathbb{C}^K$ factor.

For orthogonal vector-valued bilinear forms we can have $K_+ \neq K_-$, and the corresponding
bilinear forms on $\mathbb{C}^{K_\pm}$ will be called $M_\pm$. We can choose bilinear forms on 
each Weyl spinor module individually:
\begin{align}
    &\beta^{(1)}_+: (\mathbb{S}_+ \otimes \mathbb{C}^{K_+}) \times (\mathbb{S}_+ \otimes \mathbb{C}^{K_+}) \rightarrow  V_\mathbb{C} \;, \\
    &\beta_+ (\gamma^{\mu} \lambda_+, \chi_+) = (\gamma^{\mu}\lambda^i_+)^T C \chi^j_+ M_{+ j i} \nonumber, \\
    &\beta^{(1)}_-: (\mathbb{S}_- \otimes \mathbb{C}^{K_-}) \times (\mathbb{S}_- \otimes \mathbb{C}^{K}_-) \rightarrow V_\mathbb{C}
    , \\
    &\beta_- (\gamma^{\mu} \lambda_-, \chi_-) = (\gamma^{\mu}\lambda^i_-)^T C \chi^j_- M_{- j i} \nonumber.
\end{align}
Note that we have suppressed an additional `$\pm$' related to the choice of the charge
conjugation matrix $C_\pm$ for notational simplicity. Also note that without loss of generality
we can choose the same Majorana bilinear form on $\mathbb{S}_+$ and $\mathbb{S}_-$,
since choosing different Majorana bilinear forms only changes the bilinear form on 
the extended spinor module $\mathbb{S}_\pm \otimes \mathbb{C}^{K_\pm}$ 
by an overall factor.

For an isotropic vector-valued Majorana bilinear form we necessarily need 
$K_+=K_-$ to define a non-degenerate bracket. 
The vector-valued bilinear form is 
\begin{align}
    &\beta^{(1)}: (\mathbb{S}_\pm \otimes \mathbb{C}^{K}) \times (\mathbb{S}_\mp \otimes \mathbb{C}^{K}) \rightarrow  V_\mathbb{C} , \\
    &\beta (\gamma^{\mu} \lambda_\pm , \chi_\mp) = (\gamma^{\mu} \lambda^i_\pm)^T C \chi^j_\mp M_{j i} \nonumber.
\end{align}

The extended spinor module is $\mathbb{S}_+ \otimes \mathbb{C}^{K} \oplus \mathbb{S}_- \otimes \mathbb{C}^{K}$, and it is natural 
to combine the Weyl spinors into Dirac spinors, $\lambda^i = \lambda^i_+ + \lambda^i_-$, so that one works with $\mathbb{S} \otimes \mathbb{C}^K$. For even dimensions with isotropic bilinear forms ($D=4,8,12,...$) we will therefore construct superalgebras with supercharges that are elements of the $K$-extended spinor modules $\mathbb{S}\otimes \mathbb{C}^K$ regardless of whether the real spinor module is reducible or irreducible. 

This completes our construction of complex supersymmetry algebras. The superbracket is
defined by the choice of a charge conjugation matrix. This is unique in odd dimensions, 
while in even isotropic dimensions there are two inequivalent choices. In even 
orthogonal dimensions we can choose complex superbrackets independently 
in each chiral sector. The resulting R-symmetry groups will be determined in Section 
 \ref{Sec:complex_R-symmetry}, and in Section \ref{Sec:Isomorphisms} we will 
 show that complex supersymmetry algebras are determined by their R-symmetry
 group.  See Table \ref{bilinears} for a list of superadmissible complex bilinear forms,
complex  supersymmetry algebras, and their R-symmetry groups.

\begin{table*}[h!]
\begin{center}
    \begin{tabu}{|[1.5pt]c c|c|c|c|c|c|c|[1.5pt]}
    \tabucline[1.5pt]{1-8} $D$ && $\sigma$ & $\tau$ & $\iota$ & $M$ & $G_{\mathbb{C}^{K}}$ & $G^\mathbb{C}_R$ \\
   \tabucline[1.5pt]{1-8} 1 & & $+1$ & $+1$ & N/A & $\delta$ & $\mathrm{O}(K,\mathbb{C})$  & $\mathrm{O}(K,\mathbb{C})$\\
    \hline 2 & $C_+$ & $-1$ & $-1$ & $+1$ & $J$ & $\mathrm{O}(K,\mathbb{C})$ & $\mathrm{O}(K_+, \mathbb{C}) \times \mathrm{O}(K_-, \mathbb{C})$ \\
          &  $C_-$ & $+1$ & $+1$ & $+1$ & $\delta$ & $\mathrm{O}(K,\mathbb{C})$ & \\
    \hline 3 & & $-1$ & $-1$ & N/A & $\delta$ & $\mathrm{O}(K,\mathbb{C})$ & $\mathrm{O}(K,\mathbb{C})$ \\
    \hline 4 & $C_+$& $-1$ & $-1$ & $-1$& $\delta$ & $\mathrm{O}(K,\mathbb{C})$ & $\mathrm{GL}(K,\mathbb{C})$ \\
         & $C_-$ & $-1$ & $+1$ & $-1$ & $J$ & $\mathrm{Sp}(K,\mathbb{C})$ & \\
    \hline 5 & & $-1$ & + 1 & N/A & $J$ & $\mathrm{Sp}(K,\mathbb{C})$ & $\mathrm{Sp}(K,\mathbb{C})$ \\
    \hline 6 & $C_+$ & $+1$ & $-1$ & $+1$ & $J$ & $\mathrm{Sp}(K,\mathbb{C})$ & $\mathrm{Sp}(K_+,\mathbb{C}) \times \mathrm{Sp}(K_-, \mathbb{C})$ \\
          &   $C_-$ & $-1$ & $+1$ & $+1$ & $J$ & $\mathrm{Sp}(K,\mathbb{C})$ & \\
    \hline 7 && $+1$ & $-1$ & N/A & $J$ & $\mathrm{Sp}(K,\mathbb{C})$ & $\mathrm{Sp}(K,\mathbb{C})$ \\
    \hline 8 & $C_+$ & $+1$ & $-1$ & $-1$ & $J$ & $\mathrm{Sp}(K,\mathbb{C})$ & $\mathrm{GL}(K,\mathbb{C})$ \\
            &$C_-$ & $+1$ & $+1$ & $-1$ & $\delta$ & $\mathrm{O}(K,\mathbb{C})$ & \\
    \hline 9 && $+1$ & $+1$ & N/A & $\delta$ & $\mathrm{O}(K,\mathbb{C})$  & $\mathrm{O}(K,\mathbb{C})$ \\
    \hline 10 & $C_+$ & $-1$ & $-1$ & $+1$ & $J$ & $\mathrm{O}(K,\mathbb{C})$ & $\mathrm{O}(K_+,\mathbb{C}) \times \mathrm{O}(K_-, \mathbb{C})$ \\
            &$C_-$ & $+1$ & $+1$ & $+1$ & $\delta$ & $\mathrm{O}(K,\mathbb{C})$ & \\
    \hline 11 & & $-1$ & $-1$ & N/A & $\delta$ & $\mathrm{O}(K,\mathbb{C})$ & $\mathrm{O}(K,\mathbb{C})$ \\
    \hline 12 & $C_+$ & $-1$ & $-1$ & $-1$ & $\delta$ & $\mathrm{O}(K,\mathbb{C})$ & $\mathrm{GL}(K,\mathbb{C})$ \\
              & $C_-$ & $-1$ & $+1$ & $-1$ & $J$ & $\mathrm{Sp}(K,\mathbb{C})$ &  \\
	\tabucline[1.5pt]{1-8}
    \end{tabu}
    \end{center}
    \caption{This table provides a list of complex supersymmetry algebras. 
       For each dimension we list the inequivalent Majorana bilinear forms (charge conjugation matrices) $C$ together with their invariants $\sigma$ (symmetry), $\tau$ (type), and, where applicable $\iota$ (isotropy).
   For each $C$ we list the corresponding choice of a bilinear form $M$ on the internal space 
   $\mathbb{C}^K$ which makes $\beta = C\otimes M$ super-admissible, thus defining a complex supersymmetry algebra.  $G_{\mathbb{C}^K}$ is the isometry group of $M$ and $G_R^\mathbb{C}$ the resulting complex R-symmetry group. The pattern in this table repeats modulo 8, and since $C_+$ and
   $C_-$ define isomorphic superbrackets, complex supersymmetry algebras are classified by their
   R-symmetry groups.
       \label{bilinears}}
\end{table*}

\subsection{Complex  R-symmetry groups \label{Sec:complex_R-symmetry}}

The R-symmetry group is defined as the subgroup of the 
invariance group of the vector-valued bilinear form (and hence of the associated superbracket) that commutes with the Lie algebra of the spin group. Equivalently, it is subgroup of the Schur group (centralizer
of the spin Lie algebra), which leaves the superbracket invariant.
The R-symmetry group
of the complex vector-valued form bilinear form on $\mathfrak{g}_1^{\mathbb{C}}$,
\begin{equation}
G_R^{\mathbb{C}} = \{ R \in \mathcal{C}^*( \mathfrak{g}_1^\mathbb{C}) | 
\beta(\gamma^{\mu} R \cdot, R \cdot) = \beta(\gamma^{\mu} \cdot, \cdot) \} \;,
\end{equation}
is called the {\em complex  R-symmetry group}. 
The complex R-symmetry group only depends on the dimension. There are 
three distinct cases: superalgebras in odd dimensions, in even orthogonal dimensions 
$(D=2, 6, 10, \ldots)$  and in even isotropic dimensions $(D=4,8,12, \ldots)$. 

\subsubsection{Odd Dimensions}

In odd dimensions the extended spinor module is $\mathbb{S} \otimes \mathbb{C}^K$, and 
the complex spinor module $\mathbb{S}$ is complex-irreducible. By Schur's lemma
the R-symmetry coincides with the invariance group $G_{\mathbb{C}^{K}}$
of the bilinear form $M$ on $\mathbb{C}^K$. The complex bilinear form $M$ is
symmetric (anti-symmetric) if the vector-valued Majorana bilinear form $C(\gamma^\mu \cdot, \cdot)$ 
on $\mathbb{S}$ is symmetric
(anti-symmetric). In odd dimensions there is only one inequivalent charge conjugation 
matrix $C$, whose symmetry $\sigma=\sigma_C$ and type $\tau=\tau_C$ 
therefore determines the complex
R-symmetry group:
\begin{align}
    G^\mathbb{C}_R = \begin{cases} \mathrm{O}(K, \mathbb{C})\;,  \qquad &D = 1, 3, 9, 11, \ldots\\
    \mathrm{Sp}(K, \mathbb{C}) \;,\qquad &D = 5, 7, \ldots  \end{cases}
\end{align}
We will use the following notation for R-symmetry transformations, which
reflects that $R$ does not act on spinor indices (which have been suppressed):
\begin{align}
    \lambda^i \rightarrow R\indices{^i _j} \lambda^j, \;,\;\;\; i,j=1, \ldots K \;.
\end{align}
The corresponding R-symmetry Lie algebra element is written $r\indices{^i _j}$ such that $R\indices{^i _j} = \exp(r\indices{^i _j})$.

\subsubsection{Even Dimensions, orthogonal bilinear form}

In even dimensions the complex spinor module is reducible, 
$\mathbb{S}=\mathbb{S}_+ \oplus \mathbb{S}_-$,
and the complex semi-spinor modules $\mathbb{S}_\pm$ are inequivalent as complex modules. 
Therefore R-symmetry transformations act block-diagonally on the associated 
multiplicity spaces $\mathbb{C}^{K_\pm}$. Further details depend on whether the 
bilinear form preserves or flips chirality.

In orthogonal dimensions $D=2,6,10,\ldots$, the vector-valued bilinear form preserves chirality, and therefore
the complex R-symmetry group acts independently on left- and right-handed spinors. 
Using the matrix notation explained in Appendix \ref{secmatrixnot}, they take the form
\begin{align}
    \begin{pmatrix} \underline{\lambda}_+ \\ \underline{\lambda}_- \end{pmatrix} \rightarrow R \begin{pmatrix} \underline{\lambda}_+ \\ \underline{\lambda}_- \end{pmatrix} = \begin{pmatrix} A & 0 \\ 0 & B \end{pmatrix} \begin{pmatrix} \underline{\lambda}_+ \\ \underline{\lambda}_- \end{pmatrix} = \begin{pmatrix} A\indices{^i _j} & 0 \\ 0 & B\indices{^{\hat{i}} _{\hat{j}}} \end{pmatrix} \begin{pmatrix} \lambda^j_+ \\ \lambda^{\hat{j}} _- \end{pmatrix}\;,
\end{align}

with $i = 1,..., K_+$ and $\hat{i} = 1, ..., K_-$. The matrices $A\indices{^i _j}$ and $B\indices{^{\hat{i}} _{\hat{j}}}$ act only on the internal spaces $\mathbb{C}^{K_+}$ and $\mathbb{C}^{K_-}$ because 
$\mathbb{S}_\pm$ are complex irreducible and by 
Schur's lemma R-symmetry transformations are inert on the spinor indices.\footnote{To be
precise they can act by multiplication with non-zero complex numbers, which we
absorb into the action on the auxiliary space.}

Invariance of the vector-valued bilinear form defined by the matrices $M,M'$ implies
\begin{align}
    R^T \begin{pmatrix} M & 0 \\ 0 & M' \end{pmatrix} R  = \begin{pmatrix} M & 0 \\ 0 & M' \end{pmatrix} \label{orthwc1},
\end{align}

which, after inserting the components of $R$, becomes
\begin{align}
    \begin{pmatrix} A^T M A & 0 \\ 0 & B^T M' B \end{pmatrix}  = \begin{pmatrix} M & 0 \\ 0 & M' \end{pmatrix}.
\end{align}

By inspection of Table \ref{bilinears} $\sigma_+ \tau_+ = \sigma_- \tau_-$ in 
orthogonal dimensions $D=2,6,10,\ldots$, that is, $M$ and $M'$ 
are either both symmetric or both anti-symmetric, though they can have different size. 
Since $\sigma_\pm \tau_\pm = 1$ for $D=2,10, \ldots$ and 
$\sigma_\pm \tau_\pm = -1$ for $D=6, \ldots$, 
the complex R-symmetry groups in orthogonal dimensions are:
\begin{align}
    G^\mathbb{C}_R = \begin{cases} \mathrm{O}(K_+, \mathbb{C}) \times \mathrm{O}(K_-, \mathbb{C})\;,\qquad &D = 2, 10 ,\\
    \mathrm{Sp}(K_+, \mathbb{C}) \times \mathrm{Sp}(K_-, \mathbb{C}) \;,\qquad &D = 6. \end{cases}
\end{align}

\subsubsection{Even Dimensions, Isotropic Bilinear Forms}

In isotropic dimensions $D=4,8,12,\ldots$  R-symmetry transformations still have to act block-diagonally, 
but since the vector-valued bilinear form flips chirality, the blocks 
$A$ and $B$ must have the same size $K_+ = K_- = K$,
\begin{align}
    \begin{pmatrix} \underline{\lambda}_+ \\ \underline{\lambda}_- \end{pmatrix} \rightarrow R \begin{pmatrix} \underline{\lambda}_+ \\ \underline{\lambda}_- \end{pmatrix} = \begin{pmatrix} A & 0 \\ 0 & B \end{pmatrix} \begin{pmatrix} \underline{\lambda}_+ \\ \underline{\lambda}_- \end{pmatrix} = \begin{pmatrix} A\indices{^i _j} & 0 \\ 0 & B\indices{^i _j} \end{pmatrix} \begin{pmatrix} \lambda^j_+ \\ \lambda^j _- \end{pmatrix}.
\end{align}

Invariance of the vector-valued bilinear form implies
\begin{align}
    R^T \begin{pmatrix} 0 & M \\ M & 0 \end{pmatrix} R  = \begin{pmatrix} 0 & M \\ M & 0 \end{pmatrix}.
\end{align}

This leads to
\begin{align}
    \begin{pmatrix} 0 & A^T M B \\ B^T M A & 0 \end{pmatrix}  = \begin{pmatrix} 0 & M \\ M & 0 \end{pmatrix}.
\end{align}

This is solved by $B = M^{-1} (A^T)^{-1} M$ and therefore
\begin{align}
    R = \begin{pmatrix} A & 0 \\ 0 &  M^{-1} (A^T)^{-1} M \end{pmatrix}.
\end{align}

$A$ must be invertible but is otherwise unconstrained, i.e. $A \in \mathrm{GL}(K, \mathbb{C})$. We observe
\begin{align}
     \big(M^{-1} (A^T)^{-1} M\big) \big(M^{-1} (A'^T)^{-1} M\big) = M^{-1} ((A A')^T)^{-1} M.
\end{align}
Therefore the complex R-symmetry group in isotropic dimensions is 
\begin{align}
    G^\mathbb{C} _R = \mathrm{GL}(K, \mathbb{C}) \;, \qquad D = 4, 8, 12,
\end{align}
and acts as the direct sum of the fundamental representation $A\rightarrow A$ with 
the representation $A\rightarrow M^{-1} (A^T)^{-1} M$ which is equivalent 
to the dual (contragradient) representation $A\rightarrow (A^T)^{-1}$.

\subsubsection{Summary Table}

Our results for complex R-symmetry groups are summarized 
in Table \ref{bilinears}, together with information about the bilinear form $M$,
the charge 
conjugation matrices and their invariants. The invariants were taken from 
\cite{VanProeyen:1999ni}, which uses the notation $\epsilon = - \sigma$ and $\eta = -\tau$.
While the pattern repeats modulo 8, we have included all dimensions up to 12
for convenience. In even dimensions there a two different choices 
$C_\pm$ 
for the charge conjugation matrix, but we will show in Section 
\ref{Sec:Isomorphisms}  that there is a map
which relates the superbrackets defined by $C_\pm$ to one another. Therefore
complex supersymmetry algebras are classified by their R-symmetry groups.

\section{Reality conditions and $\epsilon$-quaternionic structures} \label{secepstruc}

So far we have constructed complex Poincar\'e Lie superalgebras. To obtain
real supersymmetry algebras we need to impose reality conditions, which 
must be $\mathrm{Spin}(t,s)$ equivariant and compatible with the superbracket.
In this section we deal with the first condition, while the second condition will be the 
subject of the next section.

Since $\mathfrak{g}_1^\mathbb{C}$ is the product of a complex spinor or complex semi-spinor
module with an auxiliary complex vector space, the natural way to obtain reality conditions
is to either take the product of two real structures or of two quaternionic structures. It is
therefore convenient to use the terminology of $\epsilon$-quaternionic structures 
introduced in Section \ref{Sec:Explicit_spinor_modules}.

\subsection{$\epsilon$-quaternionic structures on the complex spinor module $\mathbb{S}$}

We have seen in Section \ref{Sec:Explicit_spinor_modules} 
that using the matrices $A$ and $C$ we can define a new matrix $B$
\begin{align}
    B = (C A^{-1})^T \;,
\end{align}
which satisfies $B^*B = \epsilon \mathbbm{1}$. It can be shown that 
\begin{align}
    B^*B=\sigma(-\tau)^t(-1)^{t(t+1)/2} \mathbbm{1}\;,
\end{align}
where $\sigma, \tau$ are the symmetry and type of $C$, and where $t$ is the number
of timelike dimensions. Through the dependence on $t$ the type of the structure
varies with the signature. 
While in odd dimensions $C$ and $B$ are unique up to equivalence, there are two 
inequivalent choices of $C$, and hence of $B$, in even dimensions, denoted
$B_{-\tau} = (C_{-\tau} A^{-1})^T$.
Given the matrix  $B$ we can define define 
a one-parameter family of $\mathrm{Spin}(t,s)$-invariant  
maps
\begin{align}
    J^{(\epsilon)(\alpha)} : \lambda \rightarrow \alpha^* B^* \lambda^*, \qquad |\alpha| = 1\;,
\end{align}
which are complex anti-linear and satisfy $(J^{(\epsilon)(\alpha)})^2 =\epsilon \mathrm{Id}$. 
The presence of the phase $\alpha$ reflects  that while we have conventionally fixed $B$ in terms of $A$ and $C$, 
there remains the freedom of multiplying $B$ by a phase. 
The freedom of choosing $\alpha$ is important
if one wants to impose that expressions which are obtained by imposing 
a reality condition on a complex expression are real-valued, rather than just
being restricted to a generic real subspace.
We will use this in Section \ref{secsuperalg} to insure that the 
vector-valued bilinear form obtained by imposing a reality condition on the complex
vector-valued form is real-valued. 
 See also \cite{Gall:2018ogw,Cortes:2019mfa}
for how this freedom is used when constructing supersymmetric theories. 
To simplify notation, we will omit the superscript $(\alpha)$ whenever the value
of the phase is unimportant. 


In even dimensions, we have two possible charge conjugation matrices and two corresponding $\mathrm{Spin}(t,s)$-invariant $\epsilon$-quaternionic structures
\begin{align}
    J^{(\epsilon)(\alpha)}_{\pm} : 
    \lambda \rightarrow \alpha^* B _\pm ^* \lambda^*, \qquad |\alpha| = 1.
\end{align}

The subscript on $J^{(\epsilon)(\alpha)}_{\pm}$ refers to $B_\pm$ being used to define the structure. Later, we will admit different numbers of copies of $\mathbb{S}_{\pm}$, that is, $K_+\not=K_-$, and
then the phase $\alpha$ will also acquire a subscript, $\alpha_\pm$, since we can use different
types of structures on each Weyl spinor module.

The values of $\epsilon$ depend on the signature and are listed in Table \ref{Table:epsilon-structures}
for both odd and even dimensions up to dimension 12.  In even dimensions a subscript $\pm$ indicates
whether the value $\pm 1$ of $\epsilon$ refers to $J_+$ or $J_-$. As discussed 
in Section \ref{Sec:Explicit_spinor_modules}, in even dimensions there can
either be two real structures, or two quaternionic structures, or one of each type.
We have made use of the natural $(t,s) \leftrightarrow (s,t)$ symmetry, though in even dimensions one must 
then also replace 
$J^{(\epsilon)} _{-\tau}$ with $J^{(\epsilon)} _{\tau}$. For example, if we 
have  $J^{(-1)}_-$ in signature $(t,s)$, then 
in signature $(s,t)$ there will be a quaternionic structure $J^{(-1)}_+$. See Appendix \ref{appproofsigflip} for details.

In even dimensions, the matrices $C_\pm$ and hence the matrices $B_\pm$ 
are related to each other through multiplication with the chirality operator $\gamma_*$, 
see Appendix \ref{Sect_Useful_Formulae} for explicit expressions which involve
factors $\pm 1, \pm i$. This means that the difference between $C_+, B_+$ and
$C_-, B_-$ lies in how they act on the complex semi-spinor modules
$\mathbb{S}_\pm$. We need to investigate this further in order to fully understand
reality conditions on extended spinor modules of the form $\mathbb{S}_+ \otimes
\mathbb{C}^{K_+} \oplus \mathbb{S}_- \otimes \mathbb{C}^{K_-}$. We start from 
the observation that 
the matrices $B_\pm$ either both commute or both anti-commute with $\gamma_*$, depending
on the signature. Combining \eqref{bgammastar} and \eqref{gammastarb} from 
appendix \ref{Sect_Useful_Formulae}, we find
\begin{equation}
\label{Bpm_comm_gamma_star}
B_{\pm} \gamma_* = \left\{ \begin{array}{ll}
(-1)^{t+1} \gamma_* B_\pm \;, & D=2,6, 10, \ldots \\
(-1)^t \gamma_* B_\pm \;, & D=4,8, 12, \ldots
\end{array} \right.
\end{equation}
Since $\gamma_*$ is real, the $\epsilon$-quaternionic structure $J^{(\epsilon)}_\pm$
(anti-)commutes with $\gamma_*$ if and only if $B_\pm$ does. 
Therefore $J^{(\epsilon)} _{\pm}$ 
either preserves chirality and restricts to  an $\epsilon$-quaternionic structures on 
the semi-spinor modules $\mathbb{S}_\pm$, or it flips chirality and maps the 
two semi-spinor modules to one another 
\begin{equation}
J^{(\epsilon)}_{\pm}\;: \mathbb{S}_\pm \rightarrow \mathbb{S}_\pm \;,\;\;\mbox{or}\;\;\;
J^{(\epsilon)}_{\pm}\;: \mathbb{S}_\pm \rightarrow \mathbb{S}_\mp \;.
\end{equation}
We will refer to $\epsilon$-quaternionic structures which preserve chirality 
as {\em Weyl-compatible} and to those which flip chirality 
 as {\em Weyl-incompatible}. Since 
Weyl compatibility only depends on the space-time signature, we will also 
use the terminology of {\em Weyl-compatible signatures}  and {\em 
Weyl-incompatible signatures.}

The Weyl  (in-)compatibility of $J^{(\epsilon)}_\pm$ is correlated with $J^{(\epsilon)}_+$
and $J^{(\epsilon)}_-$ being of the same or of the opposite type. Using $\eqref{gammastarb}$ and $\eqref{bstargammastar}$ from Appendix \ref{Sect_Useful_Formulae} 
we obtain
\begin{align}
    B^*_+ B_+ = \begin{cases} i B^*_- \gamma_* B_+\;,  &= (-1)^{t+1} B^*_- B_- \qquad D = 2, 6, 10, \\ B^*_- \gamma_* B_+ \;, &= (-1)^t B^*_- B_- \qquad D = 4, 8, 12.
    \end{cases}
\end{align}
which by comparison to \eqref{Bpm_comm_gamma_star} shows that two 
$\epsilon$-complex structures $J^{(\epsilon)}_\pm$ are either of the same
type and both Weyl-compatible, or of opposite type and both Weyl-incompatible. 
These properties alternate with signature, which is due to the fact that the
increase of timelike directions adds one $\gamma$-matrix to the matrix $A$, and hence
to $B_\pm$. The pattern is shifted between the orthogonal dimensions
$D=2,6,10$ and the isotropic dimensions $D = 4, 8, 12$. The values of $\epsilon$ 
for $J_\pm^{(\epsilon)}$ have been listed in Table \ref{Table:epsilon-structures}.

We also note that
in Weyl-compatible signatures the Schur algebra of $\mathbb{S}$ is 
semi-simple, while in Weyl-incompatible signatures it is simple. 
This reflects that $C_\pm$ and, hence, $B_\pm$ and $J^{(\epsilon)}_\pm$ are related 
through multiplication by $\gamma_*$. In the case of semi-simple Schur algebra $\mathbb{H}_\epsilon 
\oplus \mathbb{H}_\epsilon$, the $\epsilon$-quaternionic structures in each factor
are of the same type which leads to a Weyl-compatible action of the Schur group
which respects the decomposition into semi-spinor modules. Indeed, using 
 \eqref{bstargammastar} we can see explicitly that the 
Weyl-compatible $\epsilon$-complex structures $J^{(\epsilon)}_+$ and $J^{(\epsilon)}_-$ become proportional when restricted to Weyl spinor modules
\begin{align}
    J^{(\epsilon)(\alpha)}_+ (\lambda_\pm) = \begin{cases} \alpha^* B^*_+ \lambda^*_\pm = i \alpha^* B^* _- \lambda^*_\pm = J^{(\epsilon)(i \alpha)}_- (\lambda_\pm)\;,  \qquad &D = 2, 6, 10 \\ \alpha^* B^*_+ \lambda^*_\mp = \mp \alpha^* B^* _- \lambda^*_\pm = J^{(\epsilon)(\mp \alpha)}_- (\lambda_\pm) \;, \qquad &D = 4, 8, 12  \end{cases} 
\end{align}
If the Schur algebra is the simple algebra
$\mathbb{C}(2)$, there are two $\epsilon$-quaternionic structures of opposite type, which
are related by the generator $\gamma_*$, but this time they map the Weyl spinor modules
to each other.

Let us note the implications for the construction of supersymmetry algebras, 
in particular for the existence of chiral supersymmetry algebras
where supersymmetry acts differently on left-handed and right-handed supercharges,
including the option to have a different number of left- and right-handed supercharges. 
In Weyl-compatible signatures the Weyl spinor modules carry an
$\epsilon$-quaternionic 
structure and are therefore self-conjugate, i.e 
$\bar{\mathbb{S}}_\pm = \mathbb{S}_\pm$. Since the Schur algebra
is $\mathbb{H}_\epsilon \oplus \mathbb{H}_\epsilon$, it 
follows that 
the real spinor module is reducible $S=S_+ + S_-$, and 
that the real semi-spinor modules are inequivalent $S_+\not\cong S_-$. 
Therefore the extended spinor module is of the form 
$\mathbb{S}_+ \otimes \mathbb{C}^{K_+}
\oplus \mathbb{S}_- \otimes \mathbb{C}^{K_-}$. To be able to define 
chiral real supersymmetry algebras, we also need to be in an orthogonal
dimension where the complex bilinear form can be restricted to
a fixed chirality $\mathbb{S}_\pm \otimes \mathbb{C}^{K_\pm} $,
as discussed in Section \ref{secbilformweyl}.
Thus chiral 
supersymmetry algebras 
can be constructed when the dimension is orthogonal ($D=2,6,10, \ldots$) 
and the signature is
Weyl-compatible, $\epsilon_+ = \epsilon_-$, where $\epsilon_\pm$ is the type
of $J_\pm$. 
Isotropic bilinear forms require an equal number of spinors of both chiralities, and these can naturally be combined into Dirac spinors. In this case we will require, even in Weyl compatible signatures, that 
the same reality condition is imposed on both chiralities, so that supercharges and fermions
can be combined into Dirac spinors. The extended spinor module is then $\mathbb{S} \otimes \mathbb{C}^K$. In Weyl incompatible signatures the real spinor module is either irreducible, or it is reducible and the real 
semi-spinors are equivalent, since in this case 
the Schur algebra of $S$ is $\mathbb{R}(2) \subset \mathbb{C}(2)$.
Therefore the extended spinor module takes again the form
$\mathbb{S} \otimes \mathbb{C}^K$.

\subsection{$\epsilon$-quaternionic structures on the auxiliary space $\mathbb{C}^{K}$} \label{subsecLform}

Next, we need to define $\epsilon$-quaternionic structures on $\mathbb{C}^K$. 
Given a complex $K\times K$ matrix $L=(L_{ij})$ which satisfies $L^* L = \epsilon \mathbbm{1}$, 
we  
can define an $\epsilon$-quaternionic structure
\begin{align}
    j^{(\epsilon)} : z^i \rightarrow (z^j)^*  L_{j i}  \;,
\end{align}
where the indices $i,j,\ldots$ comply with the NW-SE convention.  Spin invariance 
is trivially realized, since the spin group does not act on $\mathbb{C}^K$. 
Since we want to pick a real form of a complex supersymmetry algebra, 
we choose reality conditions on $\mathbb{C}^K$ such that it picks 
a real form of the isometry group $G_{\mathbb{C}^K}$ of the complex bilinear form $M$.
We will see later that this has the effect of reducing
the complex R-symmetry group
$G_R^\mathbb{C}$ to one of its real forms.

Let $x\in \mathfrak{g}_\mathbb{C}$ be an element of the Lie algebra of the 
group $G_{\mathbb{C}^K}$
which acts on $\mathbb{C}^K$ in the fundamental 
representation. The complex linear action $z \rightarrow L z$ of $L$ on 
$\mathbb{C}^K$ defines an involution 
\begin{equation}
\label{phi_L}
\phi_L: x \rightarrow L x L^{-1} \;,\;\;\phi_L^2=\mathbbm{1} \
\end{equation} 
of $\mathfrak{g}_\mathbb{C}$, provided that we require that $L$ is real, $L=L^*$, 
so that 
$L^* L = \epsilon \mathbbm{1}$ implies $L^2 = \epsilon \mathbbm{1}$ and $\phi_L^2 =\mathbbm{1}$.
We require in addition that $\phi_L$ is compatible with 
the Lie algebra structure, that is an involutive Lie algebra automorphism.
Given an involutive automorphism
\eqref{phi_L}, we can define a real structure 
\begin{equation}
\tau_L: x \rightarrow L \bar{x} L^{-1}
\end{equation}
on the complex Lie algebra $\mathfrak{g}_\mathbb{C}$.
The real points (fixed points) 
$(\mathfrak{g}_\mathbb{C})^{\phi_L}$  of the action 
of $\phi_L$ define a real Lie algebra $\mathfrak{g}$ with 
complexification $\mathfrak{g}_\mathbb{C}$, called a real form
of $\mathfrak{g}_\mathbb{C}$. 
It is a standard
result that all real forms of a complex semisimple Lie algebra arise from
involutive automorphisms.\footnote{See for example \cite{Gilmore} for real forms of complex semisimple
Lie algebras.}
 Moreover, two involutive automorphisms define the same
real form if they are related through conjugation by an automorphism of $\mathfrak{g}_\mathbb{C}$. 

We have two cases to consider. 
\begin{enumerate}
\item
The bilinear form 
$M$ on $\mathbb{C}^K$ is symmetric, with standard choice $M=\delta$.  Then the Lie algebra of 
infinitesimal isometries is $\mathfrak{o}(K,\mathbb{C})$, and
up to conjugation all involutive automorphisms 
are given by the following choices for $L$:
\begin{align}
    \mathds{1} _K, \qquad I_{p,q} = \begin{pmatrix} \mathds{1}_p & 0 \\ 0 & -\mathds{1}_q \end{pmatrix},\qquad J_{K} = \begin{pmatrix} 0 & \mathds{1}_k \\ - \mathds{1}_k & 0 \end{pmatrix},  \qquad K = p + q = 2k 
    \;. \label{ols}
\end{align}
$L = \mathds{1}_K, I_{p,q}$ define a real structure $j^{(\epsilon)} = j^{(+1)}$, 
while $L = J_K$, where $K$ must be even, defines a quaternionic structure $j^{(\epsilon)} = j^{(-1)}$.
\item
The bilinear form 
$M$ is antisymmetric, with standard choice $M=J$, and the Lie algebra of infinitesimal isometries
is $\mathfrak{sp}(K,\mathbb{C})$. In this case 
$K = 2k$ is necessarily even, and 
we have the following inequivalent possibilities for $L$:
\begin{align}
&\mathds{1}_K, \qquad J_{K} = \begin{pmatrix} 0 & \mathds{1}_k \\ - \mathds{1}_k & 0 \end{pmatrix}, \qquad I_{1,1} = \begin{pmatrix} 1 & 0 \\ 0 & -1 \end{pmatrix} \quad \text{(only when $K=2$)},  \\
&\tilde{I}_{2r,2s} = \begin{pmatrix} \mathds{1}_{r} & 0 & 0 & 0 \\ 0 & -\mathds{1}_{s} & 0 & 0 \\ 0 & 0 & \mathds{1}_r & 0 \\ 0 & 0 & 0 &-\mathds{1}_{s} \end{pmatrix} = \begin{pmatrix} I_{r,s} & 0 \\ 0 & I_{r,s} \end{pmatrix}, \quad k = r + s\nonumber. \label{spls}
\end{align}
Note that $\tilde{I}_{2r, 2s}$ cannot be used when $K=2$, where $I_{1,1}$ takes it place, and that
in general conjugation with $I_{p,q}$ is not an involutive automorphism for $\mathfrak{sp}(K,\mathbb{C})$.
For 
$L = \mathds{1}_K, I_{1,1}, \tilde{I}_{2r, 2s}$ we obtain a real structure 
$j^{(\epsilon)} = j^{(+1)}$, and for $L = J_K$ we obtain a quaternionic structure 
$j^{(\epsilon)} = j^{(-1)}$.
\end{enumerate}
We will refer to the above choices of representatives for $L$ as `standard' or
`canonical' in the following.  
The complex R-symmetry group $G_R^\mathbb{C}$, that is the invariance group
of $\beta = C \otimes M$ agrees with  $G_\mathbb{C}$ if the dimension of
space-time is odd. In even dimensions $G_R^\mathbb{C}$ also depends 
on whether $\beta$ is orthogonal or isotropic. Real R-symmetry
groups will be discussed in detail in Section \ref{rsym}. The real forms
selected by the above automorphisms are listed in Table \ref{tableoddrealforms}.

\subsection{Real structures on $\mathbb{S} \otimes \mathbb{C}^{K}$ and $\mathbb{S}_+ \otimes \mathbb{C}^{K_+} \oplus \mathbb{S}_- \otimes \mathbb{C}^{K_-}$.}

By combining information about $\epsilon$-quaternionic structures on $\mathbb{S}$ and 
$\mathbb{C}^K$, we obtain $\mathrm{Spin}(t,s)$-invariant 
real structures on $\mathbb{S} \otimes \mathbb{C}^K$, and, in even dimensions
on $\mathbb{S}_+ \otimes \mathbb{C}^{K_+} \oplus \mathbb{S}_- \otimes \mathbb{C}^{K_-}$.

\subsubsection{Odd dimensions}

We can define a real structure $\rho$ on $\mathbb{S} \otimes \mathbb{C}^{K}$ either as the product of two real or of two quaternionic structures. 
\begin{align}
    \rho = J^{(\epsilon)} \otimes j^{(\epsilon)} : \lambda^i \rightarrow \alpha B^* (\lambda^j)^* L_{j i}.
\end{align}
Which option is available depends on the signature. In odd dimensions
we have only one $\mathrm{Spin}(t,s)$-invariant structure at our disposition, which is either a real
structure or a quaternionic structure. 
The restriction of any tensor-valued bilinear forms to this subspace is either real or purely imaginary (and therefore real after multiplication by  $i$).  We choose the phase $\alpha$ so that the vector-valued bilinear form is real.

When $K$ is odd, we cannot define quaternionic structures on $\mathbb{C}^K$ because a quaternionic structure requires an even number of complex dimensions. This does not impede defining a real structure on $\mathbb{S} \otimes \mathbb{C}^K$, because $K$ is only 
odd if $\mathbb{S}$ has a real structure, so that we need a real structure on $\mathbb{C}^K$
in order to impose a reality condition on extended spinors.
Similarly, in cases where
we only have access to a quaternionic structure on $\mathbb{S}$ the extended spinor modules are always of the form $\mathbb{S} \otimes \mathbb{C}^{2K}$ so once again we can always define a real structure on the product space because there is no impediment to defining a quaternionic structure on $\mathbb{C}^{2K}$. A corollary is that in signatures without a real structure on $\mathbb{S}$ we cannot have theories with an odd number of supersymmetries.

\subsubsection{Even Dimensions}

In even dimensions we have two $\mathrm{Spin}(t,s)$-invariant structures for each signature.
There are two cases to be distinguished. Either the two structures are Weyl-compatible,
then they are of the same type and map $\mathbb{S}_\pm$ to $\mathbb{S}_\pm$. 
Or they are Weyl-incompatible,
then they are of opposite type and map $\mathbb{S}_\pm$
to $\mathbb{S}_\mp$. 

Weyl-compatible $\epsilon$-quaternionic structures work in the same way as $\epsilon$-quaternionic
structures  in odd dimensions, except that we replace the Dirac spinor module with either Weyl-spinor module. 
Real structures can be defined on each Weyl spinor module individually:
\begin{align}
    \rho_{(\pm)} (\lambda^i_\pm) = \alpha B^* _{(\pm)} (\lambda^i_\pm)^* L_{j i}\;,
\end{align}
where $B^* B = L^2 = \epsilon$. The canonical choices of $L$ were listed above. The total real structure is then $\rho = \rho_+ + \rho_-$. For superalgebras with both chiralities present, we can have different real structures defined on each chirality.

Weyl-incompatible signatures link the two Weyl spinor modules. In terms of Weyl spinors, the real structure is of the form
\begin{align}
    \rho (\lambda^i_\pm) = \alpha B^* _{(\pm)} (\lambda^i_\mp)^* L_{j i}.
\end{align}
Which  choice we make for $B$ makes a difference, as $B^*_+ B_+ = - B^*_- B_-$, and therefore the 
choice of $L$ depends on the choice of $B$. The reality condition can be written as a reality condition on
$\mathbb{S} \otimes \mathbb{C}^K$:
\begin{align}
    \rho (\lambda^i) = \rho (\lambda^i_+) + \rho (\lambda^i_-) =  \alpha B^* _{(\pm)} (\lambda^j_-)^* L_{j i} + \alpha B^* _{(\pm)} (\lambda^j_+)^* L_{j i} = \alpha B^* _{(\pm)} (\lambda^j)^*L_{ji}.
\end{align}

\section{Real supersymmetry algebras} \label{secsuperalg}

At this point we have all the elements in place that we need to define a real supersymmetry
algebra. 
\begin{enumerate}
\item
The odd part $\mathfrak{g}_1^\mathbb{C}$ of the complex supersymmetry algebra
takes the form $\mathbb{S}\otimes\mathbb{C}^K$ for odd dimensions and in even-dimensional signatures with isotropic bilinear forms or with Weyl-incompatible reality conditions,
while it takes the form 
$\mathbb{S}_+ \otimes\mathbb{C}^{K_+} \oplus \mathbb{S}_- \otimes\mathbb{C}^{K_-}$ in 
even-dimensional signatures with orthogonal vector-valued bilinear forms and Weyl-compatible reality conditions.
\item
On $\mathfrak{g}_1^\mathbb{C}$ we have shown how to construct a
super-admissible complex
bilinear form $\beta$ (that is, forms with $\sigma_\beta \tau_\beta=1$). The associated complex
vector-valued bilinear form
\begin{equation}
\beta^{(1)} = \beta(\gamma^\mu \cdot, \cdot) \;,: \;\;\;
\mathfrak{g}_1^\mathbb{C} \times \mathfrak{g}_1^\mathbb{C} \rightarrow V_\mathbb{C}
\end{equation}
is symmetric and $\mathrm{Spin}(t+s,\mathbb{C})$-equivariant and defines 
a complex superbracket, which gives  $\mathfrak{so}(V_\mathbb{C}) \oplus V_\mathbb{C} 
\oplus \mathfrak{g}_1^\mathbb{C}$ the structure of a complex Poincar\'e Lie superalgebra.
\item
On $\mathfrak{g}_1^\mathbb{C}$ we have constructed $\mathrm{Spin}(t,s)$ invariant
real structures $\rho$, which allow us to impose a spin-invariant reality condition and thus
to obtain real forms $\mathfrak{g}_1$ 
of $\mathfrak{g}_1^\mathbb{C}$, that is $\mathrm{Spin}(t,s)$-modules 
whose complexification is  $\mathfrak{g}_1^\mathbb{C}$.
\end{enumerate}
The data that we have to specify in order to define a real supersymmetry algebra
are the complex form  $\mathfrak{g}_1^\mathbb{C}$ of its odd part, a 
super-admissible complex bilinear form $C\otimes M$ to define the complex superbracket, 
and the matrices $B$ and $L$ defining
a real structure $\rho$ on $\mathfrak{g}_1^\mathbb{C}$, which then selects a 
real form $\mathfrak{g}_1 = (\mathfrak{g}_1^\mathbb{C})^\rho$ of 
$\mathfrak{g}_1^\mathbb{C}$. 
The final step to be taken in this section is to verify that
if we restrict the complex vector-valued bilinear form $\beta^{(1)}$ 
to spinors satisfying the reality condition $\rho$, this defines a real, symmetric, 
$\mathrm{Spin}(t,s)$-equivariant vector-valued bilinear form
\begin{equation}
\beta^{(1)} = \beta(\gamma^\mu \cdot, \cdot) \;: \;\;\;
\mathfrak{g}_1 \times \mathfrak{g}_1 \rightarrow V \;
\end{equation}
on $\mathfrak{g}_1 = (\mathfrak{g}_1^\mathbb{C})^\rho$. 
The only property that we need to verify is that $\beta^{(1)}$ becomes real-valued
when restricted to spinors which satisfy the reality condition, as the other
properties hold by construction.

\subsection{Reality conditions and superbrackets}

Let us start with the simpler case where $\mathfrak{g}_1^\mathbb{C} = \mathbb{S} \otimes \mathbb{C}^K$,
equipped with the complex super-admissible bilinear form $\beta$. The real structure
$\rho$ defines the real subspace $\mathfrak{g}_1$ by 
\begin{align}
    \rho (\lambda^i) = \lambda^i \Leftrightarrow (\lambda^i)^* = \alpha B \lambda^j L_{j i}. \label{realbits}
\end{align}
From Section  \ref{complexification} we know that 
\begin{equation}
\mathfrak{g}_1 = 
 (\mathbb{S} \otimes \mathbb{C}^K)^\rho \cong S^{\oplus N},
\end{equation}
where $S$ is the unique irreducible real spinor module.

We need to verify that the vector-valued bilinear form becomes real-valued 
when restricted to $\mathfrak{g}_1$, 
\begin{align}
    ([C \otimes M] (\gamma^\mu \lambda, \chi))^* = [C \otimes M] (\gamma^\mu \lambda, \chi) \;,\;\;\;
    \mbox{for}\;\;\;\lambda, \chi \in (\mathfrak{g}_1^{\mathbb{C}})^\rho \cong \mathfrak{g}_1 \;.
\end{align}
At this point it is important to take into account that our reality conditions contain an
arbitrary phase factor $\alpha$. We compute
\begin{align}
    ([C \otimes M] (\gamma^\mu \lambda, \chi))^* &= 
    [(\gamma^\mu \lambda^i)^T C \chi^j M_{j i} ]^*\nonumber \\
    &= \alpha^2 (\gamma^\mu B \lambda^k  L_{k i})^T C^* B \chi^l L_{l j} M_{j i}  \\
    &= \alpha^2 \tau_B (-1)^t (\gamma^\mu \lambda^i)^T B^T C^* B \chi^j (L M L^T)_{j i}. \nonumber
\end{align}
Note that in even dimensions we can choose the matrix $B=B_\pm$, which defines the reality condition,
independently from our choice of  $C=C_\pm$, which defines the bilinear form. 
Therefore we have written $\tau_B$ to indicate that we refer to the sign in \eqref{gamma_star}.
To further evaluate the right hand side requires to fix the signature. However, it is
clear that we can always choose $\alpha$ such that the bilinear form is real-valued, since,
for all signatures, $B^T C^* B = \pm C$ and $L^T M L = \pm M$ \footnote{Here we use
the properties of the canonicial choices 
for $L$ specified in Section \ref{subsecLform}. Note
that $M$ is real and either symmetric or antisymmetric.} so that 
$\alpha^2 = \pm 1$ is sufficient to make the bilinear form real. 
This fixes $\alpha$ up to a sign, so that either $\alpha = \pm 1$ or $\alpha =\pm i$.
The sign is conventional, since changing the overall sign of the superbracket 
is an isomorphism of Poincar\'e Lie superalgebras, see for example 
\cite{Cortes:2019mfa}.

The second case is where $\mathfrak{g}_1^\mathbb{C} = \mathbb{S}_+ \otimes \mathbb{C}^{K_+} \oplus \mathbb{S}_- \otimes \mathbb{C}^{K_-}$. Here the bilinear form and real structure 
are sums of two terms, $\beta = \beta_+ + \beta_-$ and $\rho=\rho_+ + \rho_-$, respectively. 
According to Section \ref{complexification} the real form of $\mathfrak{g}_1^\mathbb{C}$ is 
\begin{equation}
(\mathbb{S}_+ \otimes \mathbb{C}^{K_+})^{\rho_+} \oplus (\mathbb{S}_- \otimes \mathbb{C}^{K_-})^{\rho_-} \cong S^{\oplus N_+}_+ \oplus S_-^{\oplus N_-},
\end{equation}
where $S_+\not\cong S_-$ are the real semi-spinor modules. Since we have two
inequivalent spin representations and an orthogonal superbracket,
we must require that the restrictions of the vector-valued bilinear form to
$S_+ \times S_+$ and $S_- \times S_-$ are both real valued. This fixes the
phases $\alpha_\pm$ of $\rho_\pm$ up to sign. While the overall sign of the
superbracket is conventional, the relative sign between $\alpha_+$ and $\alpha_-$
is not and distinguishes non-isomorphic supersymmetry algebras. 
In particular, we will see in Section \ref{Sec:typeII1,3}
that this distinguishes between the type-IIA and type-IIA* algebras in 
ten dimensions, which both have the same (discrete) R-symmetry group. 

Our convention is to fix a complex superbracket and then to determine
the phase $\alpha$ by imposing that the superbracket is real-valued
on fixed points of the chosen real structure. 
While we find this convenient,
there are other conventions in the literature, where  the phase $\alpha$
is fixed but arbitrary, and the reality of the superbracket is achieved by changing the
bilinear form $M$ by a phase factor if necessary. This is for example done in  
\cite{Hull:1998ym}.
In Section \ref{Sec:typeII1,3} we will see in an example how the two conventions
are mapped to one another.

\subsection{Real supersymmetry algebras with $\mathfrak{g}_1^\mathbb{C} = \mathbb{S} \otimes \mathbb{C}^K$}

In this section we take a closer look a supersymmetry algebras with $\mathfrak{g}_1^\mathbb{C} \cong \mathbb{S} \otimes \mathbb{C}^K$. All these algebras are defined by specifying
a complex vector-valued bilinear form $\beta=C\otimes M$,
\begin{equation}
\beta(\gamma^\mu \lambda, \chi) = (\gamma^\mu \lambda^i)^T C \chi^j \otimes M_{ji}
\end{equation}
and a reality condition 
\begin{equation}
(\lambda^i)^* = \alpha B \lambda^j L_{ji}\;.
\end{equation}
As mentioned in the introduction and explained in detail in Appendix \ref{secpoincare}
this is equivalent  to the anti-commutation relations 
\begin{align}
    \{Q^i _\alpha , Q^j _\beta \} = M^{i j} (\gamma^\mu C^{-1})_{\alpha \beta} P_\mu,
\end{align}
with supercharges $Q^i=(Q^i_\alpha)$ that satisfy the reality condition $\rho (Q^i) = Q^i$, i.e.
\begin{align}
    (Q^i)^* = \alpha B Q^j L_{j i}.
\end{align}
In odd dimensions the superbracket is unique, as there is a single choice for $C$ and this fixes the choice of $M$. Real supersymmetry algebras are then determined by the choice of $L$, subject to the
condition that $B$ and $L$ together define a real structure. Note that if $C$ is not
super-admissible, $M$ is antisymmetric so that the range of $i,j$ is necessarily even,
preventing the existence of supersymmetry algebras based on a single irreducible spinor
of supercharges.

In even dimensions we have to distinguish between orthogonal dimensions where the 
vector-valued bilinear
form preserves chirality, and isotropic dimensions where chirality is flipped. 
In the orthogonal case the bilinear forms $C_\pm$ are either both super-admissible, 
or both are not, and therefore require the same $M$ to define a super-admissible 
$C_\pm \otimes M$. One can write the superbracket in terms of chiral supercharges if desired,
\begin{align}
    \{Q^i _\alpha , Q^j _\beta \} = \{Q^i _{+\alpha} , Q^j _{+\beta} \} + \{Q^i _{-\alpha} , Q^j _{-\beta} \}\;,
\end{align}
but note that we are currently considering the case where the real semi-spinor modules are equivalent.

In isotropic dimensions one Majorana bilinear form is super-admissible, and one is anti-super-admissible. Let us denote the super-admissible bilinear form $C$, and the anti-super-admissible bilinear form $C'$ (either could be $C_\pm$ depending on dimension). There are two potentially non-equivalent 
superbrackets:
\begin{align}
    &\{Q^i _\alpha , Q^j _\beta \} = \delta^{i j} (\gamma^\mu C)^{-1}_{\alpha \beta} P_\mu \;,\\
    &\{Q^i _\alpha , Q^j _\beta \} =  J^{i j} (\gamma^\mu C')^{-1}_{\alpha \beta} P_\mu \;.
\end{align}
In Section \ref{Sec:IWC} we will show that the two bilinear forms in fact 
define isomorphic real supersymmetry
algebras if (any) Weyl-compatible reality condition is imposed. The relations between 
real algebras with Weyl-incompatible reality conditions will be discussed in Section \ref{Sec:IWI}.
If desired these algebras can be written in terms of chiral supercharges
\begin{align}
    \{Q^i _\alpha , Q^j _\beta \} = \{Q^i _{+\alpha} , Q^j _{-\beta} \} + \{Q^i _{-\alpha} , Q^j _{+\beta} \} = 2 \{Q^i _{+\alpha} , Q^j _{-\beta} \} .
\end{align}

\subsection{Real supersymmetry algebras with $\mathfrak{g}_1^\mathbb{C}=\mathbb{S}_+ \otimes \mathbb{C}^{K_+} \oplus \mathbb{S}_- \otimes \mathbb{C}^{K_-}$}

Orthogonal Weyl-compatible signatures are special, since we can restrict the superbracket
to real semi-spinor modules which are not isomorphic to each other. This allows to 
define chiral supersymmetry algebras where the positive and negative chirality
supercharges are neither related by the superbracket nor by the reality condition and thus are
completely independent.

The supersymmetry anti-commutators take the form 
\begin{equation}
\{
Q^i_{+,\alpha}, Q^j_{+, \beta} \} = M^{ij} (\gamma^\mu C)_{\alpha\beta} P_\mu \;,\;\;\;
\{ Q^{i'}_{-,\alpha}, Q^{j'}_{-, \beta} \} = M^{i'j'} (\gamma^\mu C')_{\alpha\beta} P_\mu \;.
\end{equation}
where $i,j=1, \ldots \mathcal{N}_+$ and $i', j' = 1 ,\ldots  \mathcal{N}_-$, and where $C,C'$
are the restrictions of the charge conjugation matrix to $\mathbb{S}_\pm$.

If the Majorana bilinear forms on the semi-spinor modules are super-admissible, 
then $M^{ij}, M'^{ij}$ are symmetric and  we can define a supersymmetry algebra using 
a single Weyl spinor module. Since $\mathcal{N}_\pm$ counts supercharges 
in multiples of Majorana-Weyl spinors, the smallest chiral supersymmetry algebras have
the form $(\mathcal{N}_+, \mathcal{N}_-)=(1,0), (0,1)$ if the supercharges
are Majorana-Weyl spinors and $(2,0), (0,2)$ if they are Weyl spinors. 
If the restrictions of the Majorana bilinear form to $\mathbb{S}_\pm$ 
are not super-admissible, we need to choose 
an antisymmetric bilinear form on $\mathbb{C}^K$ which is only non-degenerate
if $K$ is even. In this case the minimal chiral superalgebras involve two semi-spinor modules and take the form $(2,0), (0,2)$.\footnote{Note that since in Weyl compatible signatures 
$\mathbb{S}_\pm$ carries either a real or a
quaternionic structure, we can always impose a reality condition $\mathbb{S}_\pm \otimes_\mathbb{C}
\mathbb{C}^2$ to define a $(2,0)$ or $(0,2)$ algebra, that is, $(4,0)$ and $(0,4)$ algebras
are never minimal. In other words, whenever Majorana-Weyl spinors do not exist, we 
define `symplectic Majorana-Weyl spinors.'}

\section{R-symmetry groups} \label{rsym}

So far we have shown that we can construct, in a uniform way across dimensions
and signatures, real supersymmetry algebras with given odd part $\mathfrak{g}_1$ using the data 
$(C,M, B, L)$ defined on its complexification $\mathfrak{g}_1^\mathbb{C}$. 
These data are not independent, and, 
moreover, different sets may define isomorphic supersymmetry algebras.
As a first step towards classification, we will construct and classify the
R-symmetry groups that can occur in our construction. Supersymmetry algebras with 
different R-symmetry groups are clearly not isomorphic. In Section \ref{Sec:Isomorphisms}  
we will show
that all supersymmetry algebras which arise from our construction are indeed
classified up to isomorphism by their R-symmetry groups, together with a
choice of the relative sign between $\alpha_+$ and $\alpha_-$  for chiral supersymmetry algebras.

\subsection{Determination of R-symmetry groups}

We now turn to the determination of the real R-symmetry groups, that is the 
R-symmetry groups of the real supersymmetry algebras that we obtain by
imposing $\mathrm{Spin}(t,s)$-invariant reality conditions,
\begin{align}
G_R = \{ R \in G_R^\mathbb{C} \;, | 
     \rho (R \cdot) = R \rho (\cdot) \} \;.
\end{align}

{

The R-symmetry groups for all signatures with dimension $\leq 12$ will be calculated in the following 
sections. 
As we now have to include reality conditions into our considerations, we have 
to distinguish between Weyl-compatible and Weyl-incompatible signatures. As a 
result there are five cases in total: odd dimensions, and the four cases in even dimensions: Weyl-compatible 
orthogonal, Weyl-compatible isotropic, Weyl-incompatible orthogonal and Weyl-incompatible isotropic. 
The results will be summarized in various tables.

\subsubsection{Odd dimensions}

In odd dimensions the extended spinor module is $\mathbb{S}\otimes \mathbb{C}^K$ and the
complex R-symmetry group is $\mathrm{O}(K,\mathbb{C})$ or $\mathrm{Sp}(K,\mathbb{C})$.
We impose a reality condition of the form
\begin{align}
    (\lambda^i)^* = \alpha B \lambda^j L_{j i} \;,
\end{align}
where $B$ defines an $\epsilon$-quaternionic structure on $\mathbb{S}$ and where 
$(L_{ji}) \in \mathrm{GL}(K,\mathbb{C})$ is an involutive automorphism of 
$G_R^\mathbb{C}$. 

Invariance under an R-symmetry transformation leads to
\begin{align}
    (R\indices{^i _j})^* L_{k j} = R\indices{^j _k} L_{j i}  \Leftrightarrow R^* L^T = L^T R.
\end{align}

All canonical choices for $L$ have a definite symmetry, i.e. $L^T = \pm L$, such that the above equation can be written
\begin{align}
    R^* L = L R.
\end{align}

Lie algebra elements $r\indices{^i _j}$ obey the same equation
\begin{align}
    r^* L = L r \implies r = L^{-1} r^* L.
\end{align}
This is indeed the type of equation that defines a real form of the complex 
Lie algebra $\mathrm{O}(K,\mathbb{C})$ or $\mathrm{Sp}(K,\mathbb{C})$.
To determine which real forms can occur as a real R-symmetry in a given
signature, we need to know whether $L$ defines a real or a quaternionic
structure on $\mathbb{C}^K$. This information is summarized in Table 
\ref{tableoddrealforms}. By imposing that $\epsilon$ must have the
same value as for the $\epsilon$-quaternionic structure on $\mathbb{S}$, we
arrive at the  Table  \ref{tableoddR} of real forms in odd dimensions up to $D=11$.
\begin{table*}[th!]
\begin{center}
    \begin{tabu}{|[1.5pt]c|[1.5pt]c|[1.5pt]c|[1.5pt]c|[1.5pt]}
    \tabucline[1.5pt]{1-4} $G^\mathbb{C}_R$ & $G_R$ & $\epsilon$ & $L$ \\
    \tabucline[1.5pt]{1-4} $\mathrm{O}(K, \mathbb{C})$ & $\mathrm{O}(K)$ & $+1$ & $\delta$ \\
    \tabucline{2-4} & $\mathrm{O}(p,q)$ & $+1$ & $I_{p,q}$ \\
    \tabucline{2-4} & $\mathrm{SO}^*(K)$ & $-1$ & $J_K$ \\
    \tabucline[1.5pt]{1-4} $\mathrm{Sp}(K, \mathbb{C})$ & $\mathrm{Sp}(K, \mathbb{R})$ & $+1$ & $\delta$ \\
    \tabucline{2-4} & $\mathrm{USp}(2r,2s)$ & $+1$ & $\tilde{I}_{2r,2s}$ \\
    \tabucline{2-4} & $\mathrm{USp}(K)$ & $-1$ & $J_K$ \\    
    \tabucline[1.5pt]{1-4} \end{tabu}
    \end{center}
    \caption{In this table we list the 
    real forms of the complex Lie groups $\mathrm{O}(K,\mathbb{C})$ and $\mathrm{Sp}(K,\mathbb{C})$, which are the building blocks of the complex R-symmetry groups. Note that 
    $p + q = K$ and $2r + 2s = K$. The third column specifies whether the reality condition which defines the real form provides the auxiliary space $\mathbb{C}^K$ with a real structure ($\epsilon=1$) or with a 
    quaternionic structure ($\epsilon=-1$). This decides which reality conditions are available to define a real structure on $\mathfrak{g}_1^\mathbb{C} \cong \mathbb{S} \otimes \mathbb{C}^K$. In the fourth 
    column we list the standard or `canonical' choices for the matrix $L$ which defines the reality 
    condition.}
    \label{tableoddrealforms}
\end{table*}

\begin{table*}[h!]
\resizebox{\textwidth}{!}{%
\begin{tabu}{|[1.5pt] c|[1.5pt]c|c|c|c|c|c|c|[1.5pt]}
\tabucline[1.5pt]{1-8} $D$ & $(0, D)$ & $(1, D - 1)$ & $(2, D - 2)$ & $(3, D-3)$ & $(4, D-4)$ & $(5, D-5)$ & $(6, D- 6)$ \\
\tabucline[1.5pt]{1-8} 1 & $\mathrm{O}(p,q)$ & $\mathrm{O}(p,q)$ & & & & & \\
\hline 3 & $\mathrm{SO}^*(K)$ & $\mathrm{O}(p,q)$ & $\mathrm{O}(p,q)$ & $\mathrm{SO}^*(K)$ & & & \\
\hline 5 & $\mathrm{USp}(K)$ & $\mathrm{USp}(K)$ & $\mathrm{Sp}(K,\mathbb{R})$, $\mathrm{USp}(2r,2s)$ & $\mathrm{Sp}(K,\mathbb{R})$, $\mathrm{USp}(2r,2s)$ & $\mathrm{USp}(K)$ & $\mathrm{USp}(K)$ & \\
\hline 7 & $\mathrm{Sp}(K,\mathbb{R})$, $\mathrm{USp}(2r,2s)$ & $\mathrm{USp}(K)$ & $\mathrm{USp}(K)$ & $\mathrm{Sp}(K,\mathbb{R})$, $\mathrm{USp}(2r,2s)$ & $\mathrm{Sp}(K,\mathbb{R})$, $\mathrm{USp}(2r,2s)$ & $\mathrm{USp}(K)$ & $\mathrm{USp}(K)$ \\
\hline 9 & $\mathrm{O}(p,q)$ & $\mathrm{O}(p,q)$ & $\mathrm{SO}^*(K)$ & $\mathrm{SO}^*(K)$ & $\mathrm{O}(p,q)$ & $\mathrm{O}(p,q)$ & $\mathrm{SO}^*(K)$ \\
\hline 11 & $\mathrm{SO}^*(K)$ & $\mathrm{O}(p,q)$ & $\mathrm{O}(p,q)$ & $\mathrm{SO}^*(K)$ & $\mathrm{SO}^*(K)$ & $\mathrm{O}(p,q)$ & $\mathrm{O}(p,q)$ \\
 \tabucline[1.5pt]{1-8}
\end{tabu}}
\caption{This table lists the real R-symmetry groups which can occur in odd dimension with any signature, $p + q = K$.
\label{tableoddR}}
\end{table*}

These two tables make manifest how the R-symmetry groups of real 
supersymmetry algebras vary across dimensions and signatures. Firstly, in 
signatures where $\mathbb{S}$ carries a real structure, in other words where Majorana
spinors exist, there is a broader range of R-symmetry groups since apart from 
$\mathrm{O}(K)$ or $\mathrm{Sp}(K,\mathbb{R})$ there are further real forms
$\mathrm{O}(p,q)$ or $\mathrm{Sp}(2r,2s)$ which preserve real bilinear forms 
with indefinite signature. These correspond to `twisted' Majorana conditions which introduce
relative signs between the Majorana conditions imposed on different spinors.
This phenomenon is known for Lorentz signature
from  Hull's type-* supersymmetry algebras
and we will relate our formalism to the slightly different formalism used
in  \cite{Hull:1998ym}
later when looking into specific examples.
In signatures where $\mathbb{S}$ carries a quaternionic structure the R-symmetry group is
fixed to be  $\mathrm{SO}^*(K)$ if the Majorana bilinear form is super-admissible or 
$\mathrm{USp}(K)$ if it is not. The latter case corresponds, for $K=2$, to symplectic
Majorana spinors. 
For $K=1$ the only possible group is $\mathrm{O}(1) \cong \mathbb{Z}_2$, which is discrete.
According to our counting conventions for supersymmetries, where $\mathcal{N}=K$, 
so that we count in multiples of Majorana spinors, 
such algebras are $\mathcal{N}=1$ algebras. $\mathcal{N}=1$ 
algebras only exist in signatures where the entry in our Table \ref{tableoddR}
is $\mathrm{O}(p,q)$.  In all other cases the minimal value of $\mathcal{N}=K$ for a non-degenerate
supersymmetry algebra is $\mathcal{N}=2$, that is, the supercharges form a Dirac 
spinor. 
For example, in five dimensions
the smallest supersymmetry algebra is the $\mathcal{N}=2$ algebra, 
for all signatures. Note that while  Majorana spinors exist in 
signatures $(t,s)=(2,3), (3,2)$, there is no non-degenerate
superbracket which would allow to define an $\mathcal{N}=1$ 
supersymmetry algebra, as already observed in \cite{Gall:2018ogw}.

In 11 dimensions (and as well in 3 dimensions)
the R-symmetry
group alternates between $\mathrm{O}(p,q)$ and $\mathrm{SO}^*(K)$. 
The latter
requires $K$ even and is related to quaternionic structures on $\mathbb{S}$ and 
$\mathbb{C}^{K}$. According to Table \ref{Table:epsilon-structures}, Majorana spinors, and, hence,
$\mathcal{N}=K=1$ algebras exist in 11 dimensions for
signature (1,10), (2,9) and (5,6). These algebras are realized in M-, M*- and M'-theory,
 respectively \cite{Hull:1998ym}. In some cases R-symmetry groups can be rewritten
 using accidental isomorphism between Lie groups. In particular, 
for $K=2$ we note that $\mathrm{SO}^*(2) \cong \mathrm{SO}(2)$, $\mathrm{Sp}(2,\mathbb{R}) \cong 
\mathrm{SU}(1,1)$ and $\mathrm{USp}(2) \cong \mathrm{SU}(2)$.

\subsubsection{Orthogonal Weyl-compatible signatures}

In orthogonal Weyl-compatible signatures we can define superbrackets independently for
the two Weyl spinor modules, and for each of them the situation is the same as for
the spinor module in odd dimensions. 
The complex R-symmetry groups were found to be
\begin{align}
    G^\mathbb{C}_R = \begin{cases} \mathrm{O}(K_+, \mathbb{C}) \times \mathrm{O}(K_-, \mathbb{C})\qquad &D = 2, 10, \\
    \mathrm{Sp}(K_+, \mathbb{C}) \times \mathrm{Sp}(K_-, \mathbb{C}) \qquad &D = 6. \end{cases}
\end{align}

We can also impose reality conditions independently for each chirality. But since 
Weyl-compatibility implies that $J^{(\epsilon)}_+$ and $J^{(\epsilon)}_-$ are either
both real structures or both quaternionic structures, 
the allowed combinations are quite restricted. In Table \ref{tableorthwcrealforms} 
we have listed the possible R-symmetry groups for orthogonal Weyl-compatible signatures.
\begin{table*}[th!]
\begin{center}
    \begin{tabu}{|[1.5pt]c|[1.5pt]c|[1.5pt]c|[1.5pt]}
    \tabucline[1.5pt]{1-4} $G^\mathbb{C}_R$ & $G_R$ & $\epsilon$ \\
    \tabucline[1.5pt]{1-4} $\mathrm{O}(K_+, \mathbb{C}) \times \mathrm{O}(K_-, \mathbb{C})$ & $\mathrm{O}(p_+, q_+) \times \mathrm{O}(p_-, q_-)$ & $+1$\\
    \tabucline{2-4} & $\mathrm{SO}^*(K_+) \times \mathrm{SO}^*(K_-)$ & $-1$\\
    \tabucline[1.5pt]{1-4} $\mathrm{Sp}(K_+, \mathbb{C}) \times \mathrm{Sp}(K_-, \mathbb{C})$ & \tiny{($\mathrm{Sp}(K_+, \mathbb{R})$ or $\mathrm{USp}(2r_+, 2s_+)$) $\times$ ($\mathrm{Sp}(K_-, \mathbb{R})$ or $\mathrm{USp}(2r_-, 2s_-)$)} & $+1$ \\
    \tabucline{2-4} & $\mathrm{USp}(K_+) \times \mathrm{USp}(K_-)$ & $-1$ \\    
    \tabucline[1.5pt]{1-4} \end{tabu}
    \end{center}
    \caption{Here we list the real R-symmetry groups  which can
    occur in orthogonal Weyl-compatible signatures. Note that  $p_\pm + q_\pm = K_\pm$ and $2r_\pm  + 2s_\pm  = K_\pm $. The last column
    shows which type of $\epsilon$-quaternionic structure on $\mathbb{S}$ is correlated with a given
    real form of the complex R-symmetry group.}
    \label{tableorthwcrealforms}
\end{table*}

\subsubsection{Orthogonal Weyl-incompatible signatures \label{Sec:OWI}}

For orthogonal Weyl-incompatible signatures the complex vector-valued bilinear form preserves
chirality while the reality condition flips it. Therefore the 
extended spinor module is of the form $\mathbb{S} \otimes \mathbb{C}^K$, but R-symmetry transformations can still act differently on the Weyl spinor modules.

In matrix notation, invariance of the real superbracket requires
\begin{align}
    \rho \bigg( R \begin{pmatrix} \lambda^j_+ \\ \lambda^j_- \end{pmatrix} \bigg) = R \rho \begin{pmatrix} \lambda^j_+ \\ \lambda^j_- \end{pmatrix} \label{orthwc2}.
\end{align}
Since the $\epsilon$-quaternionic structure flips chirality, 
\begin{align}
     R^* \begin{pmatrix} 0 & L \\ L & 0 \end{pmatrix} = \begin{pmatrix} 0 & L \\ L & 0 \end{pmatrix} R \label{orthwc4}.
\end{align}
Using that the complex R-symmetry group acts diagonally, we obtain 
\begin{align}
    \begin{pmatrix} 0 & A^* L \\ B^* L & 0 \end{pmatrix} = \begin{pmatrix} 0 & L B \\ L A & 0 \end{pmatrix},
\end{align}
which leads to
\begin{align}
    B = L^{-1}  A^* L.
\end{align}
Thus, we see that
\begin{align}
    R = \begin{pmatrix} A & 0 \\ 0 & L^{-1} A^* L \end{pmatrix},
\end{align}
where $A \in \mathrm{O}(K, \mathbb{C})$ or $\mathrm{Sp}(K,\mathbb{C})$.

This is a reducible representation of $\mathrm{O}(K, \mathbb{C})$ or $\mathrm{Sp}(K,\mathbb{C})$,
given by the direct sum of the fundamental representation with a representation that is 
equivalent to the anti-fundamental representation (complex conjugate of the fundamental
representation). 
Using Table \ref{bilinears}, it follows that 
the real R-symmetry group for orthogonal Weyl-incompatible signatures is, 
$\mathrm{O}(K, \mathbb{C}) \subset \mathrm{O}(K,\mathbb{C}) \times \mathrm{O}(K,\mathrm{C})$ 
in dimensions $D=2, 10, \ldots$ (where both Majorana bilinear forms are super-admissible)
and $\mathrm{Sp}(K,\mathbb{C})\subset \mathrm{Sp}(K,\mathbb{C}) \times 
\mathrm{Sp}(K,\mathbb{C})$, in dimensions $D=6, \ldots$ (where both Majorana bilinear
forms are anti-super-admissible). Note that 
due to the Weyl-incompatible reality condition we are restricted to the case 
$K_+=K_-=K$, and that the complexifications of $\mathrm{O}(K,\mathrm{C})$ and
$\mathrm{Sp}(K,\mathbb{C})$, considered
as a real Lie groups, are  $\mathrm{O}(K,\mathbb{C}) \times \mathrm{O}(K,\mathbb{C})$
and $\mathrm{Sp}(K,\mathbb{C}) \times 
\mathrm{Sp}(K,\mathbb{C})$, respectively.

\subsubsection{Isotropic Weyl-compatible signature}

Previously we found that for isotropic signatures R-symmetry transformations take the form
\begin{align}
    R = \begin{pmatrix} A & 0 \\ 0 &  M^{-1} (A^T)^{-1} M \end{pmatrix} .\label{formofR}
\end{align}
To simplify the following calculations we pass to the Lie algebra by setting 
$A = e^a$ and $R = e^r$, so that \eqref{formofR} takes the form
\begin{align}
    r = \begin{pmatrix} a & 0 \\ 0 &  - M^{-1} a^T M \end{pmatrix} \label{isor}.
\end{align}

In isotropic signatures, the two Majorana bilinear forms have opposite superadmissibility 
so that we need to consider both  $M=\delta$ and $M=J$. We remark that we will
see below that  the R-symmetry group only depends on whether the $\epsilon$-quaternionic 
structure on $\mathbb{C}^K$ is a real or quaternionic structure, 
but is insensitive otherwise to the choice of $M$ and $L$. 

We now specialise to Weyl-compatible signatures, where the reality condition 
preserves chirality, so that $r$ must satisfy
\begin{align}
    r^* \begin{pmatrix} L & 0 \\ 0 & L \end{pmatrix} = \begin{pmatrix} L & 0 \\ 0 & L \end{pmatrix}  r.
\end{align}

For $r$ given in \eqref{isor} this implies
\begin{align}
    a = L^{-1} a^* L, \qquad M^{-1} a^T M = L^{-1} (M^{-1} a^T M)^* L .\label{reqa}
\end{align}

Rearranging the final equation in \eqref{reqa} we get
\begin{align}
    a = (M L M^{-1})^{-1} a^* (M L M^{-1}).
\end{align}
For $M = \delta$ we obviously have $M L M^{-1} = L$,  
and the same is true for  $M = J$ for all possible choices $\mathrm{Id}, \tilde{I}_{2r,2s}, J$ 
for $L$. Therefore the only condition is 
\begin{align}
    a = L^{-1} a^* L \label{reqafinal}.
\end{align}
In Weyl-compatible signatures the possible choices for $L$ are further restricted by
the fact that both $\epsilon$-quaternionic structures on $\mathbb{S}$ have the same type. 

\subsubsection*{Real Structures on $\mathbb{S}$}

We first consider the case where $\mathbb{S}$ carries two real structures. Then 
for $M = \delta$ we can have
$L=\mathrm{Id}, I_{p,q}$ and for  $M = J$ we can have
$L=\mathrm{Id}, \tilde{I}_{2r, 2s}$. 

For $L = \mathrm{Id}$ equation \eqref{reqafinal} directly tells us that $a \in \mathfrak{gl}(K, \mathbb{R})$, 
and we will now show that we obtain the same Lie algebra in all other cases. 
For $L = I_{p,q}$ we see that
\begin{align}
    a = I_{p,q}^{-1} a^* I_{p,q} \implies a = \begin{pmatrix} w & ix \\ iy & z \end{pmatrix}\;,
\end{align}
where $w$ is a $p \times p$ real matrix, $z$ is $q \times q$, $x$ is $p \times q$ and $y$ is $q \times p$. 
We see that $a$ is a $K \times K$ matrix depending on $K^2$ real numbers. Since the form
of $a$ is preserved under matrix multiplication and hence under commutators, matrices
of the form $a$ define a $K^2$-dimensional real Lie subalgebra of $\mathfrak{gl}(K,\mathbb{C})$ 
which on dimensional grounds must be isomorphic to $\mathfrak{gl}(K,\mathbb{R})$. 
As already mentioned we will give explicit maps between different reality conditions later. 

If $M = J$ we need to consider $L = \mathrm{Id}$ and $L= \tilde{I}_{2r,2s}$. For $L = \mathrm{Id}$ we 
see directly that $a \in \mathfrak{gl}(K, \mathbb{R})$. For $L= \tilde{I}_{2r,2s}$ we find that $a$ must have the form
\begin{align}
    a = \begin{pmatrix} W & X \\ Y & Z \end{pmatrix},
\end{align}
where $W,X,Y,Z$ are $\frac{K}{2}\times\frac{K}{2}$ matrices that obey $V = I_{r,s} V^* I_{r,s}$ for $V = W, X, Y, Z$. Therefore all four blocks are of the form
\begin{align}
    V = \begin{pmatrix} V_1 & i V_2 \\ i V_3 & V_4     \end{pmatrix}.
\end{align}
By the same reasoning as in the previous case, matrices of this structure form a real
Lie subalgebra of $\mathfrak{gl}(K,\mathbb{C})$ which is isomorphic to 
$\mathfrak{gl}(K,\mathbb{R})$ for dimensional reasons. 
This means that $r \in \mathfrak{gl}(K, \mathbb{R})$ in isotropic Weyl-compatible signatures with real structures, regardless of the choice of $M$ and $L$. The R-symmetry group is 
$\mathrm{GL}(K,\mathbb{R})$.

\subsubsection*{Quaternionic Structures on $\mathbb{S}$}

In signatures where $\mathbb{S}$ has two quaternionic structures the only choice for 
$L$ is  $L=J$. Matrices  $a \in \mathfrak{gl}(K,\mathbb{C})$ which satisfy \eqref{reqafinal} 
have the form 
\begin{align}
    a = \begin{pmatrix} x & y \\ -y^* & x^* \end{pmatrix}, \qquad x, y \in M_{\frac{K}{2}}(\mathbb{C}) \label{ainustar}.
\end{align}
This defines the Lie subalgebra $\mathfrak{u}^*(K) \cong \mathfrak{gl}(\frac{K}{2}, \mathbb{H})$ of 
$\mathfrak{gl}(K,\mathbb{C})$.

Upon exponentiation these matrices retain their form,
\begin{align}
    A = e^a = \begin{pmatrix} X & Y \\ -Y^* & X^* \end{pmatrix}, \qquad X, Y \in \mathrm{GL}\left(\frac{K}{2},\mathbb{C} \right) .
\end{align}
The matrix  $r$ is determined by $a$ and provides a reducible representation 
of $\mathfrak{u}^*(K)$. The R-symmetry group is $\mathrm{U}^*(K)$.

\subsubsection{Isotropic Weyl-incompatible signatures} \label{risoWI}

The last case are signatures where both the complex bilinear form and the
reality condition pair opposite chiralities. 
Computations are conveniently carried out at the group level. 
To commute with the reality condition, $R$ must obey
\begin{align}
    R^* \begin{pmatrix} 0 & L \\ L & 0 \end{pmatrix} = \begin{pmatrix} 0 & L \\ L & 0 \end{pmatrix}  R.
\end{align}
Substituting in the form \eqref{formofR} that R takes in isotropic dimensions, this becomes
\begin{align}
    \begin{pmatrix} 0 & A^* L \\ M^{-1} (A^\dagger)^{-1} M     L \end{pmatrix} =     \begin{pmatrix} 0 & L M^{-1} (A^T)^{-1} M \\ L A & 0 \end{pmatrix}.
\end{align}

This is two copies of the equation
\begin{align}
    A^\dagger (M L) A = (M L) \;,
\end{align}
which  defines the pseudo-unitary group $\mathrm{U}(p,q)$ where $(p,q)$ is the signature of the matrix $ML$. For some choices of $M$ and $L$ their product $ML$ will not be diagonal, so that 
matrices of the form $A$ provide a non-standard matrix realization of this group. Let us
consider all possible combinations in turn. 
For $M = \delta$ the signature depends entirely on $L$ and is $(K,0)$ for $L=\mathrm{Id}$, $(p,q)$ for $L = I_{p,q}$ and $(k,k)$ for $L = J$ (where $K = 2k)$.  If either $M=J$ or $L=J$ (but not both), 
the matrix $ML$ is antisymmetric
and therefore has purely imaginary eigenvalues.  It can be diagonalized by a unitary transformation,
and after pulling out a factor $i$ it defines a Hermitian form. The signature of this Hermitian form is 
$(k,k)$ for $L = \mathrm{Id}$ and $L = \tilde{I}_{2r,2s}$, and $(K,0)$ when $L = J$.
In the last two cases we have made a choice of overall sign, which is conventional since
$\mathrm{U}(p,q) \cong \mathrm{U}(q,p)$. 
Table \ref{tableRisotropic} lists the R-symmetry groups for isotropic signatures.

\begin{table*}[h!]
\centering
    \begin{tabu}{|[1.5pt]c|[1.5pt]c|[1.5pt]c|[1.5pt]c|[1.5pt]}
    \tabucline[1.5pt]{1-4}  & $M$ & $L$ & $G_R$ \\
    \tabucline[1.5pt]{1-4} WC & $\delta$ & $\delta$ & $\mathrm{GL}(K,\mathbb{R})$\\
    \tabucline{2-4} & $\delta$ & $I_{p,q}$ & $\mathrm{GL}(K,\mathbb{R})$ \\
    \tabucline{2-4} & $\delta$ & $J$ & $\mathrm{U}^*(K)$ \\
    \tabucline{2-4} & $J$ & $\delta$ & $\mathrm{GL}(K,\mathbb{R})$ \\
    \tabucline{2-4} & $J$ & $\tilde{I}_{2r,2s}$ & $\mathrm{GL}(K,\mathbb{R})$ \\
    \tabucline{2-4} & $J$ & $J$ & $\mathrm{U}^*(K)$ \\ \tabucline[1.5pt]{1-4}
   \end{tabu}   \begin{tabu}{|[1.5pt]c|[1.5pt]c|[1.5pt]c|[1.5pt]c|[1.5pt]}
   \tabucline[1.5pt]{1-4}  & $M$ & $L$ & $G_R$ \\
    \tabucline[1.5pt]{1-4} WI & $\delta$ & $\delta$ & $\mathrm{U}(K)$\\
    \tabucline{2-4} & $\delta$ & $I_{p,q}$ & $\mathrm{U}(p,q)$ \\
    \tabucline{2-4} & $\delta$ & $J$ & $\mathrm{U}(k,k)$ \\
    \tabucline{2-4} & $J$ & $\delta$ & $\mathrm{U}(k,k)$ \\
    \tabucline{2-4} & $J$ & $\tilde{I}_{2r,2s}$ & $\mathrm{U}(k,k)$ \\
    \tabucline{2-4} & $J$ & $J$ & $\mathrm{U}(K)$ \\
    \tabucline[1.5pt]{1-4}    \end{tabu}
    \caption{In isotropic dimensions the R-symmetry groups are 
    real forms of $\mathrm{GL}(K,\mathbb{C})$. Depending on whether the signature 
    is Weyl-compatible (WC) or Weyl incompatible (WI), this table lists the R-symmetry groups
    for all possible combinations of a bilinear form $M$ with a reality condition $L$. Note that $K = 2k = 2r + 2s = p + q$.}
\label{tableRisotropic}
\end{table*}

\subsection{Real supersymmetry algebras and their  R-symmetry groups \label{sect:R_even}}

In even dimensions R-symmetry groups vary much more than in odd dimensions, since
we have to distinguish between orthogonal and isotropic dimensions, and between Weyl-compatible
and Weyl-incompatible signatures. In this section we provide tables for easy access, which expose
patterns and provide information that one needs in applications. 
The first set of tables collects the supersymmetry algebras with a minimal number 
of supercharges. Recall that our convention is that $\mathcal{N}$, and $\mathcal{N}_\pm$
count supersymmetries in multiples of Majorana and Majorana-Weyl spinors, irrespective
of whether such spinors, or a supersymmetry algebra based on such spinors, exist in the
signature under consideration. In the next section we will show that the supersymmetry
algebras constructed by our method are classified, up to the relative sign between $\alpha_+$ and 
$\alpha_-$
for orthogonal Weyl-compatible algebras, by their R-symmetry group. 
Therefore our tables provide a classification of supersymmetry algebras and not
only of R-symmetry groups.

\subsubsection{$(1,0)$ or $(0,1)$ algebras}

In table \ref{TableNonehalf} we list  those signatures in even dimensions where 
the minimal superalgebras have $\sfrac{d_\mathbb{S}}{2}$ real supercharges,
where $d_\mathbb{S}$ is the complex dimension of the complex spinor module
$\mathbb{S}$. In this case the supercharges form a single Majorana-Weyl spinor. 
This is only possible in orthogonal Weyl-compatible signatures, 
which can be read off from Table \ref{bilinears},
which in addition
must admit Majorana-Weyl spinors, that is $t-s$ must be 0 modulo 8, or, equivalently
$(\epsilon_+,\epsilon_-)=(1,1)$ in Table \ref{Table:epsilon-structures}. 
We denote these algebras as $(\mathcal{N}_+, \mathcal{N}_-)=(1,0), (0,1)$.
The real R-symmetry group is the discrete group $O(1)= \mathbb{Z}_2$, the same
as for algebras based on a single Majorana spinor in odd dimensions.

\begin{table*}[t]
\resizebox{\textwidth}{!}{
\begin{tabu}{|[1.5pt] c|[1.5pt]c|c|c|c|c|c|c|[1.5pt]}
\tabucline[1.5pt]{1-8} $D$ & $(0, D)$ & $(1, D - 1)$ & $(2, D - 2)$ & $(3, D-3)$ & $(4, D-4)$ & $(5, D-5)$ & $(6, D- 6)$ \\
\tabucline[1.5pt]{1-8} 2 & -- & $\mathbb{Z}_2$ & -- & & & &  \\
\hline 4 & -- & -- & -- & -- & -- & & \\
\hline 6 & -- & -- & -- & -- & -- & -- & --  \\
\hline 8 & -- & -- & -- & -- & -- & -- & -- \\
\hline 10 & -- & $\mathbb{Z}_2$ & -- & -- & -- & $\mathbb{Z}_2$ & --  \\
\hline 12 & -- & -- & -- & -- & -- & -- & -- \\
 \tabucline[1.5pt]{1-8}
\end{tabu}}
\caption{In this table we list
chiral $(\mathcal{N}_+, \mathcal{N}_-)=(1,0),(0,1)$ supersymmetry algebras, where the supercharges
form a single Majorana-Weyl spinor ($\sfrac{d_\mathbb{S}}{2}$ real supercharges, where 
$d_\mathbb{S} = \dim_\mathbb{C} \mathbb{S}$).
Where such algebras exist, they are represented by their discrete R-symmetry group 
$O(1) \cong \mathbb{Z}_2$. A dash means no such algebra can be defined.}
\label{TableNonehalf}
\end{table*}

In 10 dimensions the only signatures that allow a supersymmetry algebra with a single
Majorana-Weyl supercharge are (1,9) and (5,5) (as well as (9,1)). In Lorentz signature these algebras are 
realized by type-I and heterotic string theories and the corresponding supergravity theories
(which are usually denoted $\mathcal{N}=1$). Note that in many signatures which admit
Majorana-Weyl spinors, no algebra based on a single Majorana-Weyl spinor exists. In
particular, there are no $(1,0)$ and $(0,1)$ algebras in isotropic dimensions 
$D=4,8,12,\ldots$. Moreover, even in orthogonal dimensions, signatures where 
Majorana-Weyl spinors exist, such as $(t,s)=(3,3)$, may not admit 
$(1,0)$ and $(0,1)$ algebras, because none of the two Majorana bilinear forms are
super-admissible. In this case the bilinear form on $\mathbb{C}^{K_\pm}$ 
is antisymmetric, which forces $K_\pm$ to be even and prevents one to define
supersymmetry algebras based on a single irreducible spinor module. The corresponding
orthogonal dimensions are $D=6, 14, \ldots,$ with complex R-symmetry 
groups $\mathrm{Sp}(K_+,\mathbb{C}) \times \mathrm{Sp}(K_-,\mathbb{C})$ in
Table \ref{bilinears}.

\subsubsection{$(0,2), (1,1), (2,0)$ and $\mathcal{N}=1$ algebras}

We now turn to even-dimensional signatures which admit supersymmetry algebras with 
$d_\mathbb{S}$ real supercharges, which form a Majorana spinor or a Weyl spinor. The possible R-symmetry
groups are listed in Table \ref{RevenN1}. Chiral $(2,0)$ or $(1,1)$ supersymmetry algebras only 
exist in orthogonal dimensions $D=2,6, 10, \ldots$, while supersymmetry algebras in 
isotropic dimension $D=4,8,12,\ldots$ are necessarily non-chiral and thus denoted $\mathcal{N}=1$.

In orthogonal dimensions, the Majorana bilinear forms are either both super-admissible ($D=2,10,\ldots$) or both are  anti-super-admissible ($D=6,\ldots$). In Weyl-compatible signatures the $\epsilon$-quaternionic
structures  $J^{(\epsilon)}_\pm$ on $\mathbb{S}_\pm$ are either both real or both 
quaternionic. 
For orthogonal Weyl-compatible signatures with a super-admissible Majorana bilinear forms and $J^{(\epsilon)}_\pm$ both giving real structures we can define a $(1,1)$ superalgebra with 
R-symmetry $\mathbb{Z}_2 \times \mathbb{Z}_2$ or $(2,0)$ superalgebras with R-symmetry group $\mathrm{O}(1,1)$ or $\mathrm{O}(2)$. If, however, we have super-admissible Majorana bilinear forms, but $J^{(\epsilon)}_\pm$ are quaternionic structures on $\mathbb{S}_\pm$, then we can only define a $(2,0)$ superalgebra with $\mathrm{SO}(2)$ R-symmetry.\footnote{The quaternionic structure fixes the orientiation of the auxiliary space, which reduces the R-symmetry group from $\mathrm{O}(2)$ to $\mathrm{SO}(2)$.} When the Majorana bilinear forms are anti-super-admissible we can only define a $(2,0)$ algebra, with R-symmetry group given by $\mathrm{SU}(2)$ if $J^{(\epsilon)_\pm}$ are quaternionic structures or $\mathrm{SU}(1,1)$ if they are real structures.

Orthogonal Weyl-incompatible signatures can only have a $(1,1)$ superalgebra which therefore needs a super-admissible Majorana bilinear form. The result is a $\mathbb{Z}_2$ R-symmetry group because 
the reality condition links  the R-symmetry transformations on the two Weyl spinor modules.

Consider finally the non-chiral $\mathcal{N}=1$ algebras in isotropic dimensions. 
Here the R-symmetry group is $\mathrm{SO}(1,1)$ in 
isotropic  Weyl-compatible signatures, and $\mathrm{U}(1)$ in 
isotropic Weyl-incompatible signatures.

\begin{table}[t]
\resizebox{\textwidth}{!}{%
\begin{tabu}{|[1.5pt] c|[1.5pt]c|c|c|c|c|c|c|[1.5pt]}
\tabucline[1.5pt]{1-8} $D$ & $(0, D)$ & $(1, D - 1)$ & $(2, D - 2)$ & $(3, D-3)$ & $(4, D-4)$ & $(5, D-5)$ & $(6, D- 6)$ \\
\tabucline[1.5pt]{1-8} 2 & $\mathbb{Z}_2$ & $\mathrm{O}(1,1), \mathrm{O}(2), \mathbb{Z}_2 \times \mathbb{Z}_2$ & $\mathbb{Z}_2$ & & & &  \\
\hline 4 & -- & $\mathrm{U}(1)$ & $\mathrm{SO}(1,1)$ & $\mathrm{U}(1)$ & -- & & \\
\hline 6 & -- & $\mathrm{SU}(2)$ & -- & $\mathrm{SU}(1,1)$ & -- & $\mathrm{SU}(2)$ & --  \\
\hline 8 & $\mathrm{SO}(1,1)$ & $\mathrm{U}(1)$ & -- & $\mathrm{U}(1)$ & $\mathrm{SO}(1,1)$ & $\mathrm{U}(1)$ & -- \\
\hline 10 & $\mathbb{Z}_2$ & $\mathrm{O}(1,1), \mathrm{O}(2), \mathbb{Z}_2 \times \mathbb{Z}_2$ & $\mathbb{Z}_2$ & $\mathrm{SO}(2)$ & $\mathbb{Z}_2$ & $\mathrm{O}(1,1), \mathrm{O}(2), \mathbb{Z}_2 \times \mathbb{Z}_2$ & $\mathbb{Z}_2$  \\
\hline 12 & -- & $\mathrm{U}(1)$ & $\mathrm{SO}(1,1)$ & $\mathrm{U}(1)$ & -- & $\mathrm{U}(1)$ & $\mathrm{SO}(1,1)$ \\
 \tabucline[1.5pt]{1-8}
\end{tabu}}
\caption{R-symmetry groups for $\mathcal{N}=1$ and $(\mathcal{N_+}, \mathcal{N}_-)=(2,0), (1,1),
(0,1)$ supersymmetry algebras with $d_\mathbb{S}$ real supercharges
in even dimensions. A dash indicates signatures where no such supersymmetry algebra exists.}
\label{RevenN1}
\end{table}

Let us consider a few examples for illustration. 
In signature (1,9) the supersymmetry algebra with $\mathbb{Z}_2 \times \mathbb{Z}_2$ R-symmetry is that of Type IIA or IIA* theories. These have the same R-symmetry group but opposite
relative signs $\alpha_+\ = \pm \alpha_-$, as is further discussed 
in Section \ref{Sec:typeII1,3}.
The supersymmetry algebra with $\mathrm{O}(2)$ R-symmetry is realized in type-IIB supergravity and string theory theory, and that with $\mathrm{O}(1,1)$ R-symmetry in type-IIB*. 

In signature $(1,5)$, which is orthogonal and Weyl compatible, the entry $\mathrm{SU}(2)$ 
represents the minimal supersymmetry algebra, which is 
a chiral $(2,0)$ algebra based on a single Weyl spinor, which for this signature
is the same as a real semi-spinor ($\mathbb{S}_\pm \cong S_\pm, S_+\not\cong S_-$). 
This algebra is usually denoted $(1,0)$ in the literature, but with our conventions 
the notation $(1,0)$ is reserved for algebras based on Majorana-Weyl spinors, which 
do not exist in this signature. In four dimensions we see the standard minimal
supersymmetry algebra in Lorentz signature $(1,3)$ or $(3,1)$ with R-symmetry group
$\mathrm{U}(1)$. The minimal supersymmetry algebra in neutral signature $(2,2)$ has
R-symmetry group $\mathrm{SO}(1,1)$. All these algebras are based on Majorana spinors.
Note that while Majorana-Weyl spinors exist in signature $(2,2)$, there is 
no $(\mathcal{N}_+, \mathcal{N}_-)=(1,0)$ supersymmetry algebra, since four is an isotropic
dimension and the restriction of the $\mathcal{N}=1$ superbracket to Majorana-Weyl
spinors is degenerate, as already observed in  \cite{Cortes:2019mfa}.

Any signatures without an entry in this table have a minimal superalgebra with $2 d_\mathbb{S}$ real supercharges, that is the supercharges form a  Dirac spinor. With our convention these are  (2,2) superalgebras
 in orthogonal dimensions, and $\mathcal{N}=2$ algebras 
 in isotropic dimensions.  For example, in spacetime signature $(t,s)=(0,4)$ 
 the minimal supersymmetry algebra is 
 the $\mathcal{N}=2$ algebra.

\subsubsection{General $(\mathcal{N}_+, \mathcal{N}_-)$}

We conclude this section by providing further tables for reference.
Table \ref{mastertableeven} lists the R-symmetry groups for general extended supersymmetry 
algebras in even dimensions and arbitrary signature. 
This is followed by a master table, Table\ref{mastertable}, which combines our
results on R-symmetry groups in even and odd dimensions.

\begin{table*}[th!]
\resizebox{\textwidth}{!}{%
\begin{tabu}{|[1.5pt] c|[1.5pt]c|c|c|c|c|c|c|[1.5pt]}
\tabucline[1.5pt]{1-8} $D$ & $(0, D)$ & $(1, D - 1)$ & $(2, D - 2)$ & $(3, D-3)$ & $(4, D-4)$ & $(5, D-5)$ & $(6, D- 6)$ \\
\tabucline[1.5pt]{1-8} 2 & $\mathrm{O}(\mathcal{N},\mathbb{C})$ & $\mathrm{O}(p_+, q_+) \times \mathrm{O}(p_-, q_-)$ & $\mathrm{O}(\mathcal{N},\mathbb{C})$ & & & &  \\
\hline 4 & $\mathrm{U}^*(\mathcal{N})$ & $\mathrm{U}(p,q)$ &  $\mathrm{GL}(\mathcal{N} \mathbb{R})$  & $\mathrm{U}(p,q)$ & $\mathrm{U}^*(\mathcal{N})$ & & \\
\hline 6 & $\mathrm{Sp}(\mathcal{N},\mathbb{C})$& $\mathrm{USp}(\mathcal{N}_+) \times \mathrm{USp}(\mathcal{N}_-)$ & $\mathrm{Sp}(\mathcal{N},\mathbb{C})$& X & $\mathrm{Sp}(\mathcal{N},\mathbb{C})$& $\mathrm{USp}(\mathcal{N}_+) \times \mathrm{USp}(\mathcal{N}_-)$ & $\mathrm{Sp}(\mathcal{N},\mathbb{C})$ \\
\hline 8 & $\mathrm{GL}(\mathcal{N}, \mathbb{R})$ & $\mathrm{U}(p,q)$ & $\mathrm{U}^*(\mathcal{N})$ & $\mathrm{U}(p,q)$ & $\mathrm{GL}(\mathcal{N}, \mathbb{R})$ & $\mathrm{U}(p,q)$ & $\mathrm{U}^*(\mathcal{N})$  \\
\hline 10 & $\mathrm{O}(\mathcal{N},\mathbb{C})$ & $\mathrm{O}(p_+, q_+) \times \mathrm{O}(p_-, q_-)$ & $\mathrm{O}(\mathcal{N},\mathbb{C})$  & $\mathrm{SO}^*(\mathcal{N}_+) \times \mathrm{SO}^*(\mathcal{N}_-)$ & $\mathrm{O}(\mathcal{N},\mathbb{C})$ & $\mathrm{O}(p_+, q_+) \times \mathrm{O}(p_-, q_-)$ & $\mathrm{O}(\mathcal{N},\mathbb{C})$\\
\hline 12 & $\mathrm{U}^*(\mathcal{N})$ & $\mathrm{U}(p,q)$ & $\mathrm{GL}(\mathcal{N}, \mathbb{R})$ & $\mathrm{U}(p,q)$ & $\mathrm{U}^*(\mathcal{N})$ & $\mathrm{U}(p,q)$ & $\mathrm{GL}(\mathcal{N}, \mathbb{R})$ \\
 \tabucline[1.5pt]{1-8}
\end{tabu}}
\caption{R-symmetry groups possible in even dimension with  any signature. 
Note that $p_+ + q_+ = \mathcal{N}_+$, $p_- + q_- = \mathcal{N}_-$ and $\mathcal{N}_+ + \mathcal{N}_- = 2 \mathcal{N}$. $X = (\mathrm{Sp}(\mathcal{N}_+, \mathbb{R})$ or $\mathrm{USp}(2r_+, 2s_+)$) $\times$ ($\mathrm{Sp}(\mathcal{N}_-, \mathbb{R})$ or $\mathrm{USp}(2r_-, 2s_-)$).}
\label{mastertableeven}
\end{table*}

\begin{sidewaystable*}
\caption*{Table of R-symmetry groups in all signatures up to $D= 12$}
\resizebox{\textwidth}{!}{%
\begin{tabu}{|[1.5pt] c|[1.5pt]c|c|c|c|c|c|c|[1.5pt]}
\tabucline[1.5pt]{1-8} $D$ & $(0, D)$ & $(1, D - 1)$ & $(2, D - 2)$ & $(3, D-3)$ & $(4, D-4)$ & $(5, D-5)$ & $(6, D- 6)$ \\
\tabucline[1.5pt]{1-8} 1 & $\mathrm{O}(p,q)$ & $\mathrm{O}(p,q)$ & & & & & \\
\hline 2 & $\mathrm{O}(\mathcal{N},\mathbb{C})$ & $\mathrm{O}(p_+, q_+) \times \mathrm{O}(p_-, q_-)$ & $\mathrm{O}(\mathcal{N},\mathbb{C})$ & & & &  \\
\hline 3 & $\mathrm{SO}^*(\mathcal{N})$ & $\mathrm{O}(p,q)$ & $\mathrm{O}(p,q)$ & $\mathrm{SO}^*(\mathcal{N})$ & & & \\
\hline 4 & $\mathrm{U}^*(\mathcal{N})$ & $\mathrm{U}(p,q)$ &  $\mathrm{GL}(\mathcal{N}, \mathbb{R})$  & $\mathrm{U}(p,q)$ & $\mathrm{U}^*(\mathcal{N})$ & & \\
\hline 5 & $\mathrm{USp}(\mathcal{N})$ & $\mathrm{USp}(\mathcal{N})$ & $\mathrm{Sp}(\mathcal{N},\mathbb{R})$, $\mathrm{USp}(2r,2s)$ & $\mathrm{Sp}(\mathcal{N},\mathbb{R})$, $\mathrm{USp}(2r,2s)$ & $\mathrm{USp}(\mathcal{N})$ & $\mathrm{USp}(N)$ & \\
\hline 6 & $\mathrm{Sp}(\mathcal{N},\mathbb{C})$& $\mathrm{USp}(\mathcal{N}_+) \times \mathrm{USp}(\mathcal{N}_-)$ & $\mathrm{Sp}(\mathcal{N},\mathbb{C})$& X & $\mathrm{Sp}(\mathcal{N},\mathbb{C})$& $\mathrm{USp}(\mathcal{N}_+) \times \mathrm{USp}(\mathcal{N}_-)$ & $\mathrm{Sp}(\mathcal{N},\mathbb{C})$ \\
\hline 7 & $\mathrm{Sp}(\mathcal{N},\mathbb{R})$, $\mathrm{USp}(2r,2s)$ & $\mathrm{USp}(\mathcal{N})$ & $\mathrm{USp}(N)$ & $\mathrm{Sp}(\mathcal{N},\mathbb{R})$, $\mathrm{USp}(2r,2s)$ & $\mathrm{Sp}(N,\mathbb{R})$, $\mathrm{USp}(2r,2s)$ & $\mathrm{USp}(\mathcal{N})$ & $\mathrm{USp}(\mathcal{N})$ \\
\hline 8 & $\mathrm{GL}(\mathcal{N}, \mathbb{R})$ & $\mathrm{U}(p,q)$ & $\mathrm{U}^*(\mathcal{N})$ & $\mathrm{U}(p,q)$ & $\mathrm{GL}(\mathcal{N}, \mathbb{R})$ & $\mathrm{U}(p,q)$ & $\mathrm{U}^*(\mathcal{N})$  \\
\hline 9 & $\mathrm{O}(p,q)$ & $\mathrm{O}(p,q)$ & $\mathrm{SO}^*(N)$ & $\mathrm{SO}^*(\mathcal{N})$ & $\mathrm{O}(p,q)$ & $\mathrm{O}(p,q)$ & $\mathrm{SO}^*(\mathcal{N})$ \\
\hline 10 & $\mathrm{O}(N,\mathbb{C})$ & $\mathrm{O}(p_+, q_+) \times \mathrm{O}(p_-, q_-)$ & $\mathrm{O}(\mathcal{N},\mathbb{C})$  & $\mathrm{SO}^*(\mathcal{N}_+) \times \mathrm{SO}^*(\mathcal{N}_-)$ & $\mathrm{O}(\mathcal{N},\mathbb{C})$ & $\mathrm{O}(p_+, q_+) \times \mathrm{O}(p_-, q_-)$ & $\mathrm{O}(\mathcal{N},\mathbb{C})$\\
\hline 11 & $\mathrm{SO}^*(\mathcal{N})$ & $\mathrm{O}(p,q)$ & $\mathrm{O}(p,q)$ & $\mathrm{SO}^*(\mathcal{N})$ & $\mathrm{SO}^*(\mathcal{N})$ & $\mathrm{O}(p,q)$ & $\mathrm{O}(p,q)$ \\
\hline 12 & $\mathrm{U}^*(\mathcal{N})$ & $\mathrm{U}(p,q)$ & $\mathrm{GL}(\mathcal{N}, \mathbb{R})$ & $\mathrm{U}(p,q)$ & $\mathrm{U}^*(\mathcal{N})$ & $\mathrm{U}(p,q)$ & $\mathrm{GL}(\mathcal{N}, \mathbb{R})$ \\
 \tabucline[1.5pt]{1-8}
\end{tabu}}
\caption{$p + q = \mathcal{N}$, $2r + 2s = \mathcal{N}$. $p_+ + q_+ = \mathcal{N}_+$, $p_- + q_- = \mathcal{N}_-$. X = ($\mathrm{Sp}(K_+, \mathbb{R})$ or $\mathrm{USp}(2r_+, 2s_+)$) $\times$ ($\mathrm{Sp}(K_-, \mathbb{R})$ or $\mathrm{USp}(2r_-, 2s_-)$). Supersymmetries are counted
in multiples $\mathcal{N}_\pm$ of Majorana-Weyl spinors and of Majorana spinors $\mathcal{N}$
irrespective of whether such spinors actually exist in a given signature. Therefore depending on the signature the minimal values of $\mathcal{N}_\pm, \mathcal{N}$ can be 2 rather than 1. \label{mastertable}}
\end{sidewaystable*}

\newpage

\section{Isomorphisms and classification \label{Sec:Isomorphisms}}

So far we have shown that we can construct a real supersymmetry algebra 
given the following data: a complex bilinear form $C\otimes M$ on 
the complex extended spinor module $\mathfrak{g}_1^\mathbb{C}=  \mathbb{S} \otimes \mathbb{C}^K$, 
a real structure $\rho$ on this space, which is determined by 
an $\epsilon$-quaternionic structure $B$ on the complex spinor module
$\mathbb{S}$, 
and a map $L$ on the auxiliary space $\mathbb{C}^K$, which determines a real form of the complex R-symmetry group. In orthgonal Weyl-compatible signatures we can choose these data independently
for the chiral sectors.
While algebras with distinct R-symmetry groups cannot be 
isomorphic, there is, to our knowledge, no rigorous statement asserting
the converse, that is, that supersymmetry algebras are classified by their 
R-symmetry groups.  Still, by inspection of our classification 
tables, we observe that
different sets of data $(C, M, B, L)$ on $\mathfrak{g}_1^\mathbb{C}$ often lead
to the same R-symmetry group, and previous experience tells one that 
in such cases one can often construct isomorphisms between the corresponding
supersymmetry algebras. See in particular \cite{Cortes:2019mfa} on which we
will now elaborate to show that, with one qualification applying to the 
orthogonal Weyl-compatible case, the supersymmetry algebras which we have
constructed in this paper are classified by their R-symmetry groups. 
We proceed by discussing each of the five cases in turn, and investigate
which isomorphisms need to exist in order to relate all supersymmetry algebras with the
same R-symmetry group to each other. Details of these isomorphisms 
are given in Appendix \ref{Sec:Details_Isomorphisms}.

\subsection{Odd dimensions}

In odd dimensions there is only one inequivalent choice for $C$, which fixes 
$M$. Thus complex supersymmetry algebras with a given odd part $\mathfrak{g}_1^\mathbb{C}$ 
are unique, and in particular classified by their R-symmetry group. For real supersymmetry
algebras we have to pick a reality condition, which is defined by a choice of $L$ which
is compatible with our choice of $B$. Since $B$ is unique, real supersymmetry algebras
are classified by the choice of $L$, that is by their R-symmetry group.

\subsection{Orthogonal, Weyl-compatible signatures}

In orthogonal dimensions, $C_\pm$ are either both super-admissible or both are not, 
which means that $M$ is fixed by the dimension. It remains to be seen
whether the choice of $C_\pm$ has an effect. As far as complex supersymmetry
algebras are concerned, showing that $C_\pm$ lead to isomorphic supersymmetry
algebras implies that these are classified by their R-symmetry group. For real
supersymmetry algebras we have to choose a $B$-matrix. In Weyl-compatible
signatures $B_\pm$ are either both real structures or both quaternionic 
structures, and therefore they combine with the same set of $L$'s to define a real structure.
To show that real supersymmetry algebras are classified by the choice of $L$, 
we need to show that we can replace $B_\pm$ by $B_\mp$ in the reality 
condition without changing the complex bilinear form. We also need to show that
we can replace $C_\pm$ by $C_\mp$ without changing the reality condition. 

To be precise, an additional complication
arises because
there are two independent phase factors $\alpha_\pm$ in the 
reality conditions 
for the two chiral sectors. As we discussed before,  imposing reality of 
the superbracket fixes these factors to be either $\pm 1$ or $\pm i$. 
While the overall sign of the pair $\alpha_\pm$ corresponds to an isomorphism
of supersymmetry algebras, 
the relative sign distinguishes non-isomorphic supersymmetry algebras, in particular 
the IIA and IIA$^*$ algebras in ten dimensions, compare Section \ref{Sec:typeII1,3}.
This  is a particular feature of the orthogonal Weyl-compatible case, where both the 
superbracket and the reality condition are compatible with chirality. 
Therefore the 
best statement we can aim for is that real supersymmetry algebras are classified
by their R-symmetry group together with a choice of the relative sign between 
$\alpha_+$ and $\alpha_-$. 

As shown in Appendix \ref{Sec:Details_Isomorphisms}
the R-transformation (\ref{RCmap}) maps 
the orthogonal bilinear forms $C_\pm \otimes M$ to one another, while preserving
the reality condition up to an irrelevant overall phase factor.
Moreover, (\ref{bgammastar}) implies that in 
orthogonal dimensions $B_\pm \gamma_* = \mp i B_\mp$ and therefore
the reality condition $(\lambda^i_\pm)^* = \alpha_\pm B_- \lambda^j_\pm L_{j i}$ is equivalent to the reality condition $(\lambda^i_\pm)^* = \pm i \alpha_\pm B_+ \lambda^j_\pm L_{j i}$. 
This shows that the choice of $B_\pm$ only matters to the extent that it is correlated with the 
relative sign between $\alpha_+$ and $\alpha_-$. In other words the real
supersymmetry algebra only depends on the choice of $L$, that is on its
real R-symmetry group, and on the relative sign between $\alpha_+$ and $\alpha_-$. 
For complex supersymmetry 
algebras the choice of reality conditions and thus of relative signs is irrelevant, 
and we note as a corrolary that complex supersymmetry algebras in 
orthogonal dimensions are classified by their R-symmetry groups.
 
\subsection{Orthogonal, Weyl-incompatible signatures \label{Sec:OWI2}}

Here we again  have two options $C_\pm$ for $C$, while $M$ is fixed by dimension. 
Concerning reality conditions, one of $B_\pm$ defines a real, the
other a quaternionic structure and therefore the choice of $B$
is determined by $L$ being a real or quaternionic structure. 
We have shown in Section \ref{Sec:OWI} that the R-symmetry group
is $\mathrm{O}(K,\mathbb{C})$ if $C_\pm$ are super-admissible, and $M$ is
symmetric, while it is
$\mathrm{Sp}(K,\mathbb{C})$ if $C_\pm$ are not super-admissible and $M$ 
is antisymmetric. Since the R-symmetry group is determined by the 
dimension, we need to show that all possible choices of $C,L$ 
are equivalent. In Appendix \ref{Sec:Details_Isomorphisms} we show that the
R-transformation (\ref{RCmap}) exchanges the orthogonal bilinear forms,
while preserving Weyl-incompatible reality conditions. 
Moreover, we show in Appendix  \ref{Sec:Details_Isomorphisms}
that we can use the transformation $S_{L^{-1}}$ defined in (\ref{S-transformation})
to map any reality condition to the one with $L=\mathrm{Id}$. For a given $M$, 
the canonical choices for $L$ were specified in Section \ref{subsecLform}. 
For these $L$-matrices, the bilinear form is invariant under $S_{L^{-1}}$, 
or, in the case $M=J$, $L=I_{1,1}$ anti-invariant, which can be compensated
by an overall phase in the reality condition.

\subsection{Isotropic, Weyl-compatible signatures \label{Sec:IWC}}

In isotropic dimensions $C_\pm$ have opposite super-admissibility 
properties. The choice of $C$ determines the choice of $M$,
which for complex superalgebras implies that they are classified
by their R-symmetry group. If the signature is Weyl-compatible, 
then $B_\pm$ either both define a real or both define a quaternionic structure,
and are thus compatible with the same set of $L$'s. 
According to our classification, there are two possible R-symmetry groups:
if $L$ (and thus $B_\pm$) defines a real structure, the R-symmetry
group is $\mathrm{GL}(K,\mathbb{R})$, if 
$L$ defines a quaternionic structure, then the R-symmetry group 
is $\mathrm{U}^*(K)$. Thus the R-symmetry group is determined
by the signature. According to Table \ref{tableRisotropic} there
are four combinations of $(M,L)$ with R-symmetry group
$\mathrm{GL}(K,\mathbb{R})$ and two combinations of $(M,L)$ 
with R-symmetry group $\mathrm{U}^*(K)$. 
To show that
real supersymmetry algebras are classified by their R-symmetry
groups the following statements are sufficient:
\begin{enumerate}
\item
For $L=\delta$ and $L=J$ 
the bilinear forms $C\otimes \delta$ and $C'\otimes J$,
define isomorphic supersymmetry algebras, 
where $C$ denotes the super-admissible $C$-matrix and $C'$ the
non-super admissible $C$-matrix. This can be done using the 
transformation $S_J$ defined in Appendix \ref{Sec:S-tranformations}, which exchanges the bilinear forms while
preserving $L=\delta$ and $L=J$.
\item
For $C\otimes \delta$, the reality conditions with $L=\delta$ and 
$L=I_{p,q}$ define isomorphic supersymmetry algebras, while
for $C'\otimes J$ the reality conditions with $L=\delta$ and 
$L = \tilde{I}_{2r,2s}$ define isomorphic supersymmetry algebras.
This can be done using the transformation $T$, 
defined in Appendix \ref{Sec:T-transformation}
which maps
the $L$'s as required, while preserving the corresponding bilinear
forms. 
\item
The $R$-transformation, defined in Appendix \ref{Sec:R-transformation}
exchanges $B_\pm$ without
changing the bilinear form. Therefore it does not matter which $B$ we 
use in the reality condition. 
\end{enumerate}

\subsection{Isotropic, Weyl-incompatible signatures\label{Sec:IWI}}

In this case $C_\pm$ have opposite super-admissibility and 
$B_\pm$ have opposite $\epsilon$-type. Therefore $(C,B)$ 
are determined by $(M,L)$. By inspection of Table \ref{tableRisotropic}
there are seven distinct combinations of $(M,L)$ which correspond
to only three distinct types of R-symmetry groups, $\mathrm{U}(K)$, 
$\mathrm{U}(p,q)$, with $p\not=q$,  and $\mathrm{U}(k,k)$. 
To show that real supersymmetry algebras are classified by their
R-symmetry groups, we need to show that
\begin{enumerate}
\item
R-symmetry group $U(K)$: real supersymmetry algebras 
defined using $(M,L)=(\delta, \delta), (J,J)$ are isomorphic.
\item
R-symmetry group $U(k,k)$: real supersymmetry algebas
defined using $(M,L)=(\delta,J), (\delta, I_{k,k}), (J,\delta), (J,\tilde{I}_{2r,2s})$ are
isomorphic.
\end{enumerate}

These statements can be proved using the transformations
$S_J, F,G$ as indicated in the following diagram:
\begin{equation}
\label{Fig:IWI}
\begin{tikzpicture}
  \matrix (m) [matrix of math nodes,row sep=3em,column sep=3em,minimum width=2em]
  {	
  (\delta, \delta) & (\delta, J) & (\delta, I_{k,k}) & (\delta, I_{p,q}), p\not=q \\
    (J, J) & (J, \delta) & (J, \tilde{I}_{2r,2s}) \\
    \mathrm{U}(K) & \mathrm{U}(k,k) & & \mathrm{U}(p,q), p\not=q \\ };
     \path[-stealth]
    (m-2-1) edge node [right] {$S_J$} (m-1-1)
    (m-1-1) edge node [left] {} (m-2-1)
	(m-2-2) edge node [left] {$S_J$} (m-1-2)
    (m-1-2) edge node [left] {} (m-2-2)
    (m-1-2) edge node [above] {$F$} (m-1-3)
    (m-1-3) edge node [above] {} (m-1-2)
    (m-2-2) edge node [above] {$G$} (m-2-3)
    (m-2-3) edge node [above] {} (m-2-2);
\end{tikzpicture}
\end{equation}

Details are given in Appendix \ref{Sec:IWI_details}.

\subsection{Remarks on the general classification problem}

In this paper we have constructed Poincar\'e Lie superalgebras with odd part $\mathfrak{g}_1$
by specifying a discrete set $(C,M,B,L)$ of data, which is subject to certain consistency 
conditions. Within this construction we have shown that the resulting supersymmetry algebras
are classified by their R-symmetry groups, together with a choice of the relative sign between 
$\alpha_+$ and $\alpha_-$ for
orthogonal Weyl-compatible algebras. However, this does not necessarily provide a 
classification of $\mathcal{N}$-extended Poincar\'e Lie superalgebras in arbitrary signature
up to isomorphism. The general classification problem can be stated as follows. 
Firstly, for a given $\mathfrak{g}_1$ one needs to find the space of all superbrackets,
that is of symmetric, vector-valued $\mathrm{Spin}(t,s)$-equivariant bilinear forms. 
This problem was solved in  \cite{Alekseevsky:1997}, where explicit bases in terms
of super-admissible bilinear forms have been constructed. Secondly, one needs 
a criterion which allows one to decide when two
symmetric, vector-valued $\mathrm{Spin}(t,s)$-equivariant bilinear forms define
isomorphic Poincar\'e Lie superalgebras. This question is answered by Theorem
1 of \cite{Cortes:2019mfa}, which specifies  necessary and sufficient conditions for two Poincar\'e Lie superalgebras to be isomorphic. Up to checking certain discrete transformations, the 
classification up to isomorphism amounts to identifying the orbits of the Schur group
on the space of symmetric vector-valued $\mathrm{Spin}(t,s)$-equivariant bilinear forms.
The third step is the classification of Schur group orbits for the spaces of superbrackets
known from \cite{Alekseevsky:1997} (together with checking for additional identifications
by elements of $\mathrm{Pin}(t,s) \backslash \mathrm{Spin}(t,s)$, see 
\cite{Cortes:2019mfa} for details). Due to the mod 8 periodicity of the table of Clifford 
algebras, which imprints itself on the data defining supersymmetry algebras, this is 
a finite problem. The classification was carried out for algebras based on the complex spinor
module $\mathbb{S}$ ($\mathcal{N}=2$) in five and four space-time dimensions 
in \cite{Gall:2018ogw,Cortes:2019mfa}. In this case the algebras in a given signature
are indeed classified by their R-symmetry groups, which is encouraging, and this may well
be true in general. However, there are potential subtleties which make 
the full classification problem somewhat involved. Non-degenerate superbrackets 
correspond to open orbits of the 
Schur group on the space of superbrackets, and the R-symmetry group is the
stabilizer group of the orbit. It may happen that two connected open orbits have the same
stabilizer, in which case one needs to carefully check whether there exists a
discrete transformation relating these two orbits which provides an isomorphism. Comparing 
 to the work presented in the paper this relates to the question whether the discrete
 data we use samples all orbits of the Schur group. A priori, there could be 
 open orbits with no representatives within our construction, though we have no 
 evidence that this is the case.

\section{Applications to type-II string theories in $D=10,9,4,3$} \label{secappphys}

\subsection{$\mathcal{N}=2$ supersymmetry in signature $(1,9)$ and type-II string theories
\label{Sec:typeII1,3}}

We start in signature $(1,9)$ which is orthogonal Weyl-compatible, with two
super-admissible bilinear forms $C_\pm$ and two real structures $B_\pm$
on $\mathbb{S}$. 
This signature admits two $(1,1)$ algebras with R-symmetry $\mathbb{Z}_2 \times
\mathbb{Z}_2$, which are distinguished by the relative sign $\alpha_- = \pm \alpha_+$.
We can choose either of the $C_\pm$ to define the Majorana bilinear form and, independently, 
either of the $B_\pm$ to define a reality condition, but all choices lead to equivalent
real superbrackets. For definiteness we choose $C_+, B_+$ in the following.
The complex vector-valued bilinear form is 
 \begin{equation}
 \label{type-IIA/IIA-star}
       (\Gamma^\mu \lambda)^T C_+ \chi = (\Gamma^\mu \lambda_+)^T C_+ \chi_+ 
       +  (\Gamma^\mu \lambda_-)^T C_+ \chi_- \;,
\end{equation}
where $\lambda_\pm, \chi_\pm \in \mathbb{S}_\pm$.
Following the conventions of Appendix \ref{App:Dim_Red}
we denote the ten-dimensional Dirac matrices by $\Gamma^\mu$. 
 To select the type-IIA or type-IIA$^*$
algebra, we impose the reality conditions
\begin{align}
\label{IIA_reality}
    &(\lambda_+)^* = \alpha B_+ \lambda_+, \qquad &&(\lambda_-)^* =  \pm \alpha B_+ \lambda_- \;.
\end{align}
The $+$ sign is the standard Majorana condition, while the $-$ sign corresponds to a 
twisted Majorana condition which selects the type-IIA$^*$ algebra. We remark that 
with our conventions the complex bilinear form is fixed, and type-IIA and type-IIA$^*$
are distinguished by their reality conditions. One could equivalently impose the same
reality condition, but define the type-IIA$^*$ theory using a modified complex superbracket.
This amounts to $\lambda_- \rightarrow i \lambda_-$, 
and we will come back to this option below. For reference, let us also write down how
the vector-valued bilinear form (\ref{type-IIA/IIA-star}) translates into anti-commutators
of supercharges:
\begin{equation}
\{ Q_{+,\alpha}, Q_{+, \beta}  \} = (\Gamma^\mu C_+)_{\alpha \beta} P_\mu \;,\;\;\;
\{ Q_{-,\alpha'}, Q_{-, \beta'}  \} = (\Gamma^\mu C_+)_{\alpha' \beta'} P_\mu \;,
\end{equation}
where the spinor indices $\alpha, \beta, \ldots$ and $\alpha', \beta',\ldots$ 
refer to $\mathbb{S}_\pm$. 
In the following we will prefer to work with vector-valued bilinear forms as this allows us
to suppress spinor indices. 

Signature $(1,9)$ also admits chiral $\mathcal{N}=2$ algebras, and for definiteness
we take $(\mathcal{N}_+, \mathcal{N}_-) = (2,0)$. The complex superbracket 
takes the form
 \begin{equation}
 (\Gamma^\mu \lambda_+^i)^T C_+ {\chi}_+ ^j M_{j i}, \;\;\; M_{i j} = \delta_{ij}  \nonumber \;,
\end{equation}
and the reality conditions are
\begin{align}
({\lambda}_+ ^i)^* 
= \alpha B_+ {\lambda}^j_+ L_{j i}, \qquad L_{ i j} = \begin{pmatrix}
        1 & 0 \\ 0 & \pm 1 
    \end{pmatrix} \;.
\end{align}
Depending on the choice of the sign in $L_{ij}$, we obtain the type-IIB algebra with
R-symmetry group $\mathrm{O}(2)$ or the type-IIB$^*$ algebra with R-symmetry
group $\mathrm{O}(1,1)$. 

As is well known, the type-IIA/IIB algebras are realized in the type-IIA/IIB string theories,
whose massless sectors are described by type-IIA/IIB supergravity, and these two
string theories are related by T-duality. Moreover, it was shown in \cite{Hull:1998vg}
that timelike T-duality maps type-IIA/IIB string theory to type-IIB$^*$/IIA$^*$ string theory. 
The corresponding type-II$^*$ supergravity theories differ from their type-II counter parts
by a sign flip of the kinetic terms for all Ramond-Ramond fields, as well as by factors
of $i$ in their fermionic terms.

Apart from showing immediately  the (potential) existence of type-II$^*$ theories, our
formalism also makes it straightforward to show how these algebras are related
to one another by T-duality.  Two superstring theories are related by T-duality 
if theory $A$ on the background $\mathbb{R}^{t',s'} \times S^1_R$ is identical to theory $B$
on the background $\mathbb{R}^{t',s'} \times S^1_{1/R}$, where the radius $R$ of the circle
is measured in string units, and $t'+s'=9$. To include timelike T-duality, we allow the 
circle to be timelike.  The ten-dimensional theories $A,B$ which arise in the two 
decompactification limits $R\rightarrow \infty$ and $R'=1/R\rightarrow \infty$ are then also 
said to be T-dual to each other. Note that this does not imply that they are equivalent
as theories on $\mathbb{R}^{t,s}$, $t+s=10$. 
In particular, type-IIA and type-IIB string theory are distinct as theories 
on $\mathbb{R}^{1,9}$, and in particular have supersymmetry algebras which 
are not isomorphic to each other. Thus at the level of supersymmetry algebras
T-duality is a map between supersymmetry algebras, but in general is not an isomorphism.
It relates ten-dimensional supersymmetry algebra which upon dimensional reduction 
`contract' to the same nine-dimensional algebra, a fact that we will explore further in 
Section \ref{Sec:9d_susy}.

In the following we use that in theories of closed strings T-duality 
can be viewed as a `chiral reflection,' which only acts on the (say) the 
right-moving degrees of freedom. In particular, 
on a ten-dimensional spinor $\lambda$, T-duality acts by 
\begin{align}
\mathcal{T}\;:\;\;
    \lambda_+ \rightarrow   \tilde{\lambda}_+ =  \lambda_+ \;,\;\;\;    \lambda_- \rightarrow \tilde{\lambda}_- = T \lambda_-, \quad T = \beta \Gamma_* \Gamma^{9/0}, \quad |\beta| = 1,
    \end{align}
with $\Gamma^9$ for spacelike and $\Gamma^0$ for timelike T-duality. We
have introduced an arbitrary phase $\beta$ which will be used later to interpolate between different conventions for supersymmetry algebras. 

We start from the type-IIA/IIA$^*$ algebras which are based on a spinor $\lambda$
subject to the reality condition (\ref{IIA_reality}). 
We note that 
T-duality maps $\mathbb{S}_+ \oplus \mathbb{S}_- \rightarrow
\mathbb{S}_+ \oplus \mathbb{S}_+$, and therefore the image of a non-chiral
type-IIA/IIA$^*$  supersymmetry algebra will be a chiral $(2,0)$ supersymmetry algebra. 

The matrix
$T$ is unitary for both the spacelike and the timelike case. 
To verify that it acts as a reflection in the 9-direction and in the
0-direction, respectively, we compute
\begin{equation}
T \Gamma^\mu T^{-1} = \left\{ \begin{array}{rl}
- \Gamma^\mu\;, & \mbox{for} \;\;\mu=9/0\;, \\
\Gamma^\mu\;, & \mbox{for} \;\;\mu\not= 9/0 \;,\\
\end{array} \right\} = \tilde{\Gamma}^\mu \;.
\end{equation}
Since T-duality maps the matrices $\Gamma^\mu$ to a new set $\tilde{\Gamma}^\mu$,
this gives rise to new matrices $A$ and $C$:\footnote{$A$ is the matrix defining 
the Dirac Hermitian sesquilinear form. In signature $(1,D-1)$ we take $A=\Gamma_0$.} 
\begin{equation}
A \rightarrow \tilde{A} = T^{\dagger, -1} A T^{-1}  =
{T} A {T}^\dagger \;,\;\;\;
C_+ \rightarrow \tilde{C}_+ =  {T}^{T,-1} C_+ {T}^{-1} \;.\;\;\; 
\end{equation}
We compute:\footnote{The sign in the relation for $C_+$ depends on 
that we have chosen $C_+$ rather than $C_-$, since we need to 
make use of the relation $\tau_+=-1$.}
\begin{equation}
\begin{array}{llll}
\text{spacelike T-duality:} & \tilde{C}_+ = \frac{1}{\beta^2} C_+\;, & \tilde{A} = A \;, & 
\tilde{B}_+ = (\tilde{C}_+ \tilde{A}^{-1})^T = \frac{1}{\beta^2} B_+ \;,\\
\text{timelike T-duality:} & \tilde{C}_+ = -\frac{1}{\beta^2} C_+\;, & \tilde{A} = - A  \;, & 
\tilde{B}_+ = (\tilde{C}_+ \tilde{A}^{-1})^T = \frac{1}{\beta^2} B_+ \;.\\
\end{array}
\end{equation}
Thus the difference between spacelike and timelike T-duality is a relative
sign between the transformation behaviour of $C_+$ and $B_+$, which 
induces a corresponding sign in the relation between the complex bilinear form
and the reality condition.

To see this explicitly, we express $\lambda_-, \chi_-$ in terms of the transformed
spinors $\tilde{\lambda}_-, \tilde{\chi}_-$ in the bilinear form
\begin{eqnarray*}
&& (\Gamma^\mu \lambda_-)^T C_+ \chi_- = 
(\Gamma^\mu T^{-1} \tilde{\lambda}_-)^T C_+ T^{-1} \tilde{\chi}_- 
=  (\tilde{\Gamma}^\mu \tilde{\lambda}_- )^T T^{-1,T} C_+ T^{-1} \tilde{\chi}_-  \\
&=& (\tilde{\Gamma}^\mu \tilde{\lambda}_- )^T  \tilde{C}_+ \tilde{\chi}_-   =
\pm \frac{1}{\beta^2} (\tilde{\Gamma}^\mu \tilde{\lambda}_- )^T  {C}_+ \tilde{\chi}_-  \;.
\end{eqnarray*}
The upper sign refers to spacelike, the lower sign to timelike T-duality. 

Since the transformed spinors both have positive chirality, we prefer to use the notation
$(\lambda^1_+, \lambda^2_+)$ instead of  $(\tilde{\lambda}_+, \tilde{\lambda}_-)$, and we also 
relabel $\tilde{\Gamma}^\mu \rightarrow \Gamma^\mu$, though this change needs
to be taken into account in the reality condition satisfied by $\lambda^2_+$, see below.
Combining both chiral sectors, the new complex vector-valued bilinear form is 
\begin{equation}
 (\Gamma^\mu {\lambda}^1_+)^T  C_+  {\chi}^1_+ \pm 
 \frac{1}{\beta^2} (\Gamma^\mu {\lambda}^2_+)^T  {C}_+  {\chi}^2_+ 
    = (\Gamma^\mu \lambda_+^i)^T C_+ {\chi}_+ ^j M_{j i}, \;\;\; M_{i j} = \begin{pmatrix}
        1 & 0 \\ 0 &  \pm \frac{1}{\beta^2} \nonumber
    \end{pmatrix} \;,
\end{equation}
with $+/-$ for spacelike/timelike T-duality. 
The reality condition (\ref{IIA_reality}) is mapped to the new reality condition
\begin{align}
(\tilde{\lambda}_+ ^i)^* = \alpha_{(i)}  \tilde{B}_+ \tilde{\lambda}^i_+ 
= \alpha B_+ \tilde{\lambda}^j_+ L_{j i}, \qquad L_{ i j} = \begin{pmatrix}
        1 & 0 \\ 0 & \epsilon \frac{1}{\beta^2}
    \end{pmatrix} \;,
\end{align}
where we used that $\tilde{B}_+ = B_+$ when acting
on $\lambda^1_+ = \tilde{\lambda}_+$, while $\tilde{B}_+ = \frac{1}{\beta^2} B_+$ 
when acting on $\lambda^2_+ = \tilde{\lambda}_-$. To account for  the phases 
in the original reality condition (\ref{IIA_reality}) we have set $\alpha_{(1)} = \alpha_+=\alpha $, $\alpha_{(2)} = \alpha_- = \epsilon \alpha$, 
with $\epsilon=1$ if we start with type-IIA and with $\epsilon=-1$ if we start with 
type-IIA$^*$. 

There are two natural choices for the phase $\beta$.
\begin{enumerate}
\item
We can choose $\beta$ such that the complex bilinear form has the canonical
form $M_{ij} =\delta_{ij}$ , which is the convention used in this paper. This requires 
$\beta=\pm 1$ for spacelike and $\beta=\pm i$ for
timelike T-duality. In this case the real supersymmetry algebras are distinguished
by their reality conditions:
\begin{equation}
(M_{ij} )= \left( \begin{array}{cc}
{1} & 0 \\
0 & {1} \\
\end{array} \right)  \;,\;\;\;
(L_{ij}) = \left( \begin{array}{cc}
{1} & 0 \\
0 & \pm {\epsilon} \\
\end{array} 
\right) \;.
\end{equation}
This shows that the type-IIA algebra, $\epsilon=1$, is mapped by 
spacelike T-duality 
to the type-IIB algebra with R-symmetry group $\mathrm{O}(2)$, and
by timelike T-duality to the type-IIB$^*$ algebra with R-symmetry group
$\mathrm{O}(1,1)$. Similarly, the type-IIA$^*$ algebra, $\epsilon=-1$,  is mapped
to the type-IIB$^*$ algebra by spacelike T-duality and to the type-IIB 
algebra under timelike T-duality. 
In this convention 
it is manifest that the four real algebras are different real forms of two complex 
algebras, because we have imposed that the complex superbrackets assume
a standard form, that is $M_{ij} =\delta_{ij}$. 
\item
We can choose $\beta$ such that the reality conditions is a standard Majorana
condition, $L_{ij} = \delta_{ij}$, which requires $\beta^2=\epsilon$. 
In this case the real supersymmetry algebras are distinguished by their bilinear form,
and for the IIB$^*$ theories the bilinear form is twisted by a relative minus sign:
\begin{equation}
(M_{ij} )= \left( \begin{array}{cc}
1 & 0 \\
0 & \pm \epsilon \\
\end{array} \right)  \;,\;\;\;
(L_{ij}) = \left( \begin{array}{cc}
1 & 0 \\
0 & 1 \\
\end{array} 
\right)
\end{equation}
This convention was used when the type-IIB* algebra was originally introduced as a twisted
version of the IIB algebra \cite{Hull:1998vg}.
\end{enumerate}

\subsection{$\mathcal{N}=2$ supersymmetry in general ten-dimensional signatures}

By inspection of Table \ref{RevenN1}, there are three types of signature
for ten-dimensional $\mathcal{N}=2$ algebras.
\begin{itemize}
\item
In signatures $(1,9)$, $(5,5)$ and $(9,1)$ there are three possible R-symmetry groups
$\mathbb{Z}_2 \times \mathbb{Z}_2, \mathrm{O}(2), \mathrm{O}(1,1)$. These 
are orthogonal Weyl-compatible signatures with two real structures on $\mathbb{S}$.
We have discussed the case $(1,9)$ before. 
Signature $(9,1)$ can be viewed 
as signature $(1,9)$ with a mostly minus convention for the metric.\footnote{While
the Clifford algebras are not the same, the resulting spin representations are
isomorphic and we have not found any indication that signatures $(1,9)$ and 
$(9,1)$ give rise to different physics.} Our classification shows that in neutral
signature $(5,5)$ algebras of type IIA/IIA$^*$/IIB/IIB$^*$ exist and are related by 
T-duality. The existence of the corresponding string theories is known from 
\cite{Hull:1998ym}.
\item
In signature $(0,10)$, $(2,8)$, $(4,6)$, $(6,4)$, $(8,2)$, $(10,0)$ the unique
R-symmetry group is $\mathbb{Z}_2$. These are orthogonal Weyl incompatible
signatures which only admit $(1,1)$ algebras. The existence of type-IIA 
string theories in these signatures was established in \cite{Hull:1998ym}.
\item
In signature $(3,7)$ and $(7,3)$ the unique R-symmetry group is $\mathrm{SO}(2)$. 
These are orthogonal, Weyl-compatible signatures with two quaternionic
structures on $\mathbb{S}$. This implies that the only $\mathcal{N}=2$
algebras are chiral $(2,0)$ (or $(0,2)$) algebras. The existence of 
type-IIB string theories in these signatures was established in \cite{Hull:1998ym}.
\end{itemize}
In summary, the network of ten-dimensional type-II string theories
described in  \cite{Hull:1998ym} exhausts all ten-dimensional $\mathcal{N}=2$
supersymmetry algebras. To relate type-II string theories across signatures,
one needs to combine spacelike and timelike T-duality with S-duality. 
T-duality changes the space-time signature if the string world-sheet 
has Euclidean signature. S-duality is needed to map type-IIB$^*$ string theory
to type-IIB' string theory. This exchanges the fundamental string with the E2-brane,
thus providing a fundamental string with Euclidean world-sheet. Since we 
only consider the supersymmetry algebras, which do not distinguish between 
type-IIB$^*$ and type-IIB', we can establish the relation between ten-dimensional
$\mathcal{N}=2$ algebras in different signatures by relating them to the same
nine-dimensional $\mathcal{N}=2$ algebra. 

\subsection{$\mathcal{N}=2$ supersymmetry in nine dimensions \label{Sec:9d_susy}}

`Pure', that is spacelike and timelike T-dualities, arise whenever after compactification
over space or over time, the limit $R\rightarrow 0$ corresponds to an alternative 
decompactification limit $R'=1/R \rightarrow \infty$. At the level of the supersymmetry
algebras this requires that the two ten-dimensional supersymmetry algebras 
give rise, by compactification, to the same nine-dimensional supersymmetry algebra. 
Since ten-dimensional theories in signatures $(t,s)$ and $(t-1,s+1)$ can reduce to the same
theory in signature $(t-1,s)$ by applying timelike and spacelike reduction, respectively, 
this opens up the possibility of `mixed' T-dualities, where one reduces over time
and decompactifies over space, or vice versa. Our formalism gives a uniform
description of superymmetry algebras across dimensions and signatures, and
allows to work out universal formulae for the spacelike and timelike
reduction of these algebras. We have collected the relevant formulae in Appendix
\ref{App:Dim_Red}. Using this machinery it is straightforward to work out 
which of the ten-dimensional $\mathcal{N}=2$ algebras reduce to the same
nine-dimensional algebra, and are thus related by a pure (signature preserving)
or mixed (signature changing) T-duality. 

\subsubsection{Nine-dimensional $\mathcal{N}=2$ supersymmetry algebras}

We start by surveying nine-dimensional $\mathcal{N}=2$ supersymmetry algebras.
In nine dimensions the Majorana bilinear form on $\mathbb{S}$ is super-admissible, so we use a 
symmetric bilinear form on $\mathbb{C}^K$, thus leading to R-symmetry groups that are real forms 
of $\mathrm{O}(K,\mathbb{C})$. 
The two Majorana bilinears in
 ten dimensions are based on the two ten-dimensional 
 charge conjugation matrices  $C^{(10)}_\pm$, which  have invariants $(\sigma, \tau) = (\mp 1, \mp 1)$.
  The nine-dimensional charge conjugation matrix $C^{(9)}$ has invariants $(\sigma, \tau) = (+1, +1)$. 
  In 10D we will use $C^{(10)}_-$ 
 in this section as this turns out to be convenient, and we know that both choices define the same real supersymmetry algebra. The nine-dimensional $\mathcal{N}=2$ algebras and their R-symmetry groups
 are listed in Table \ref{Table:9d}.

\begin{table*}[t]
\begin{tabu}{|[1.5pt] c|[1.5pt] c|c|c|c|c|[1.5pt]}
\tabucline[1.5pt]{1-8} Signature & $(0, 9)$ & $(1, 8)$ & $(2, 7)$ & $(3, 6)$ & $(4, 5)$ \\
\tabucline[1.5pt]{1-8} R-sym. Group & $\mathrm{O}(1,1)$ or $\mathrm{O}(2)$ & $\mathrm{O}(1,1)$ or $\mathrm{O}(2)$ & $\mathrm{SO}(2)$ & $\mathrm{SO}(2)$ & $\mathrm{O}(1,1)$ or $\mathrm{O}(2)$ \\
 \tabucline[1.5pt]{1-8}
\end{tabu}
\caption{$\mathcal{N}=2$ R-symmetry groups in nine dimensions, the R-symmetry group for $(s,t)$ is the same as $(t,s)$.  \label{Table:9d}}
\end{table*}

There are two types of nine-dimensional signatures.
\begin{itemize}
\item
In signatures $(0,9)$, $(1,8)$, $(4,5)$, and those signatures obtained by flipping $t\leftrightarrow s$,
the complex spinor module $\mathbb{S}$ has an invariant real structure, which allows one
to impose a standard Majorana condition with R-symmetry group $\mathrm{O}(2)$ or a
twisted Majorana condition with R-symmetry group $\mathrm{O}(1,1)$. 
\item
In signatures $(2,7)$, $(3,6)$  and those signatures obtained by flipping $t\leftrightarrow s$,
the complex spinor module $\mathbb{S}$ carries an invariant quaternionic structure, 
which allows one to impose a symplectic Majorana condition. In this case the 
R-symmetry group is $\mathrm{SO}(2)$.
\end{itemize}
Explicitly, the reality conditions are:
\begin{align}
    &(\lambda^i)^* = \alpha B^{(t,s)} \lambda^i \implies G_R = \mathrm{O}(2), \\
    &(\lambda^i)^* = \alpha B^{(t,s)} \lambda^j \eta_{j i} \implies G_R = \mathrm{O}(1,1), \\
    &(\lambda^i)^* = \alpha B^{(t,s)} \lambda^j \epsilon_{j i} \implies G_R = \mathrm{SO}(2).
\end{align}

\subsubsection{Reduction of Clifford algebras}

Next, we give the explicit relations between ten- and nine-dimensional quantities. We 
suppress most details, which are straightforward to work out using Appendix
\ref{App:Dim_Red}. The basis for the ten-dimensional Clifford algebra is $\Gamma_\mu$ within which we embed the nine-dimensional gamma matrices, $\gamma_\mu$, according to
\begin{align}
    \Gamma_\mu = \gamma_\mu \otimes \sigma_1, \quad \Gamma_{10} = \mathrm{Id}  \otimes \sigma_2 \quad \text{or} \quad \Gamma_0 = i \mathrm{Id} \otimes \sigma_2 \;.
\end{align}
The Majorana bilinear forms on the respective complex spinor modules are related by
\begin{align}
    C^{(10)}_- = C^{(9)} \otimes \sigma_1.
\end{align}
The 10D chiral projection matrix $\Gamma_* = (-i)^{t+\frac{D}{2}} \Gamma_1 ... \Gamma_{10}$ 
decomposes as
\begin{align}
    \Gamma_* = 1 \otimes \sigma_3.
\end{align}

In non-chiral type-IIA theories we have two spinors of opposite chirality, denoted $\lambda_\pm$, which  
decompose into nine-dimensional spinors $\psi^1$ and $\psi^2$ as
\begin{align}
    \lambda_+ = \psi^1 \otimes \begin{pmatrix} 1 \\ 0 \end{pmatrix}, \\
    \lambda_- = \psi^2 \otimes \begin{pmatrix} 0 \\ 1 \end{pmatrix}.
\end{align}
In chiral type-IIB theories we have two spinors of the same chirality, 
$\lambda^1_+$ and $\lambda^2_+$, which decompose into nine dimensional 
spinors, again denoted $\psi^1$ and $\psi^2$, as
\begin{align}
    \lambda^1_+ = \psi^1 \otimes \begin{pmatrix} 1 \\ 0 \end{pmatrix} ,\\
    \lambda^2_+ = \psi^2 \otimes \begin{pmatrix} 1 \\ 0 \end{pmatrix}.
\end{align}

\subsubsection{Reduction of reality conditions}

Next, we dimensionally reduce ten-dimensional reality conditions. For simplicity and without
loss of generality we will only consider $B^{(p,q)}_-$ (with $p + q = 10$).
  A spacelike reduction gives
\begin{align}
    B^{(t,s+1)}_- = B^{(t,s)} \otimes \sigma_1^{t+1} = \begin{cases} B^{(t,s)} \otimes \sigma_1 \quad 
    \text{for} \;\; t \;\; \text{even}, \\ B^{(t,s)} \otimes 1 \quad \text{for} \;\;t \;\;\text{odd},\end{cases}
\end{align}
while  a timelike reduction gives
\begin{align}
    B^{(t+1, s)}_- = B^{(t,s)} \otimes i \sigma_3 \sigma_1^{t} = \begin{cases} i B^{(t,s)} \otimes \sigma_3 \quad \text{for} \;\; t \;\; \text{even} ,\\ - B^{(t,s)} \otimes \sigma_2 \quad \text{for} \;\; t \;\;  \text{odd} .\end{cases}
\end{align}
The second factor in the tensor products captures the Weyl-compatibility of the ten-dimensional signature; we observe that when the ten-dimensional signature has an even number of timelike directions we have a factor $\sigma_1$ or $\sigma_2$ that exchanges chiralities, as in these signatures the reality condition  is Weyl-incompatible. When the ten-dimensional theory has an odd number of timelike directions the reality condition is Weyl-compatible, so we get $\mathrm{Id}$ or $\sigma_3$ which do not mix the two chiralities.

As an example we reduce the type-IIA algebra in signature $(0,10)$ 
(this is the unique $(\mathcal{N}_+,\mathcal{N}_-)=(1,1)$ algebra in this signature) to $(0,9)$. The $(0,10)$
parent theory  involves a single Majorana spinor that can be written in terms of a Weyl-incompatible reality condition as
\begin{align}
    (\lambda_\pm)^*= \alpha B^{(0,10)}_- \lambda_\mp.
\end{align}

Decomposing into nine-dimensional quantities, we see this reads
\begin{align}
    (\psi^1)^* \otimes \begin{pmatrix} 1 \\ 0 \end{pmatrix} = \alpha (B^{(0,9)} \otimes \sigma_1 ) (\psi^2 \otimes \begin{pmatrix} 0 \\ 1    \end{pmatrix}) = \alpha B^{(0,9)} \psi^2 \otimes \begin{pmatrix} 1 \\ 0 \end{pmatrix}, \\
    (\psi^2)^* \otimes \begin{pmatrix} 0 \\ 1 \end{pmatrix} = \alpha (B^{(0,9)} \otimes \sigma_1) (\psi^1 \otimes \begin{pmatrix} 1 \\ 0    \end{pmatrix}) = \alpha B^{(0,9)} \psi^1 \otimes \begin{pmatrix} 1 \\ 0 \end{pmatrix}.
\end{align}

Dropping the vector $\begin{pmatrix} 1 \\ 0 \end{pmatrix}$ we can simply write
\begin{align}
    (\psi^i)^* = \alpha B^{(0,9)} \psi^j \eta_{j i}.
\end{align}

This leads to a $(0,9)$ $\mathcal{N}=2$ superalgebra with an $\mathrm{O}(1,1)$ R-symmetry group.

\subsubsection{Reduction of vector-valued bilinear forms}

Next, we need to reduce the vector-valued bilinear form. 
A non-chiral, type-IIA vector-valued bilinear form reduces as
\begin{align}
    &(\Gamma^\mu \lambda_+)^T C^{(10)}_- \chi_+ +     (\Gamma^\mu \lambda_-)^T C^{(10)}_- \chi_- = 
    (\gamma^\mu \psi^i)^T C^{(9)} \phi^j \delta_{j i} \otimes 1 - i (\psi^i)^T C^{(9)} \phi^j \eta_{j i} \otimes 1     \;.
\end{align}
The second term is a scalar under the nine-dimensional Poincar\'e Lie algebra and therefore
gives rise to a central extension of the nine-dimensional supersymmetry algebra. This term
is only relevant for states which have 
momentum along the direction we reduce over, as this momentum component corresponds
to the central charge. We remark that the ten-dimensional algebra also admits 
BPS extensions (poly-vector extensions) corresponding to terms on the r.h.s. which
transform as antisymmetric tensors under the Lorentz group. While such terms are 
important since they encode the possible BPS branes of a theory, we have decided to
leave the inclusion of BPS extensions to future work. Neglecting the central term,
non-chiral, type IIA vector-valued forms therefore reduce as
\begin{equation}
(\Gamma^\mu \lambda_+)^T C^{(10)}_- \chi_+ +     (\Gamma^\mu \lambda_-)^T C^{(10)}_- \chi_- \rightarrow
 (\gamma^\mu \psi^i)^T C^{(9)} \phi^j \delta_{j i} \;.
 \end{equation}
 Similarly, chiral, type IIB vector valued forms reduce as
 \begin{align}
    &(\Gamma^\mu \lambda^i_+)^T C^{(10)}_- \chi^j_+ \delta_{j i} \rightarrow  (\gamma^\mu \psi^i)^T C^{(9)} \phi^j \delta_{j i} \;.
\end{align}

\subsubsection{Summary of relations between ten-dimensional and nine-dimensional
$\mathcal{N}=2$ supersymmetry algebras}

We will not include further details of computations, as the principle is by now clear and 
the various signatures only differ by sign factors. 
The results 
are summarized in the diagram in Table \ref{10d-9d}. This diagram shows in 
particular that by a sequence of timelike reductions and spacelike 
oxidations, that is by mixed T-dualities, one can connect all signatures
to one another. While in generic ten-dimensional signatures there 
is a unique $\mathcal{N}=2$ algebra which is chiral or non-chiral, there are
two chiral and two non-chiral algebras in Lorentz and in neutral signature,
which then are mutually related by pure T-dualities.


\begin{sidewaystable}[h!]

\begin{tikzpicture}
\node[draw=none] (name0) at (-2,2) {$(0,10)$};
\node[draw] (a0) at (-2,0) {IIA};
\node[draw=none] (name0) at (1.5,2) {$(1,9)$};
\node[draw] (a1) at (0,0) {IIA};
\node[draw] (astar1) at (1,0) {IIA*};  
\node[draw] (b1) at (2,0) {IIB};  
\node[draw] (bstar1) at (3,0) {IIB*}; 
\node[draw=none] (name0) at (5,2) {$(2,8)$}; 
\node[draw] (a2) at (5,0) {IIA};
\node[draw=none] (name0) at (7,2) {$(3,7)$};
\node[draw] (b3) at (7,0) {IIB};
\node[draw=none] (name0) at (9,2) {$(4,6)$};
\node[draw] (a4) at (9,0) {IIA};
\node[draw=none] (name0) at (12.5,2) {$(5,5)$};
\node[draw] (a5) at (11,0) {IIA};
\node[draw] (astar5) at (12,0) {IIA*};  
\node[draw] (b5) at (13,0) {IIB};  
\node[draw] (bstar5) at (14,0) {IIB*}; 
\node[draw=none] (name0) at (-1,-4) {$(0,9)$};
\node[draw] (20) at (-2,-3) {$\mathrm{O}(2)$};
\node[draw] (110) at (-0.5,-3) {$\mathrm{O}(1,1)$};
\node[draw=none] (name1) at (3,-4) {$(1,8)$};
\node[draw] (21) at (2,-3) {$\mathrm{O}(2)$};
\node[draw] (111) at (3.5,-3) {$\mathrm{O}(1,1)$};
\node[draw=none] (name2) at (6,-4) {$(2,7)$};
\node[draw] (22) at (6,-3) {$\mathrm{SO}(2)$};
\node[draw=none] (name3) at (8,-4) {$(3,6)$};
\node[draw] (23) at (8,-3) {$\mathrm{SO}(2)$};
\node[draw=none] (name4) at (11,-4) {$(4,5)$};
\node[draw] (24) at (10,-3) {$\mathrm{O}(2)$};
\node[draw] (114) at (11.5,-3) {$\mathrm{O}(1,1)$};
\node[draw=none] (name5) at (14.5,-4) {$(5,4)$}; 
\node[draw] (25) at (13.5,-3) {$\mathrm{O}(2)$};
\node[draw] (115) at (15,-3) {$\mathrm{O}(1,1)$};

\draw [draw=red,-triangle 90,fill=red] (a1) edge (110) (astar1) edge (20) (b1) edge (20) (bstar1) edge (110) (a2) edge (111) (b3) edge (22) (a4) edge (23) (a5) edge (114) (astar5) edge (24) (b5) edge (24) (bstar5) edge (114);

\draw [draw=blue,-triangle 90,fill=blue] (a0) edge (110) (a1) edge (21) (astar1) edge (111) (b1) edge (21) (bstar1) edge (111) (a2) edge (22) (b3) edge (23) (a4) edge (114) (a5) edge (25) (astar5) edge (115) (b5) edge (25) (bstar5) edge (115);
 
 \path[draw=green, line width=0.2mm, every node/.style={font=\sffamily\small}]
    (a0) edge [bend left = 55] node {} (bstar1)  (bstar1) edge [bend left = 55] node {} (a2)  (a2) edge [bend left = 45] node {} (b3)   (b3) edge [bend left = 55] node {} (a4)   (a4) edge [bend left = 55] node {} (bstar5);
     
 \path[draw=red, line width=0.2mm, every node/.style={font=\sffamily\small}]
    (a1) edge [bend left = 85] node {} (bstar1)  (astar1) edge [bend left = 100] node {} (b1)  (a5) edge [bend left = 85] node {} (bstar5)   (astar5) edge [bend left = 100] node {} (b5);
    
 \path[draw=blue, line width=0.2mm, every node/.style={font=\sffamily\small}]
    (a1) edge [bend left = 55] node {} (b1)  (a5) edge [bend left = 55] node {} (b5)
    (astar1) edge [bend left = 55] node {} (bstar1)  (astar5) edge [bend left = 55] node {} (bstar5);
\end{tikzpicture}

\caption{
A schematic overview of the T-dualities between ten-dimensional type-II supersymmetry algebras 
and their relation to nine-dimensional $\mathcal{N}=2$ algebras by dimensional reduction. 
A blue arrow corresponds to a spacelike reduction and red arrow corresponds to a timelike reduction. A blue curved line means that two theories are related by a spacelike T-duality, a red curved line means that two theories are related by a timelike T-duality and a green line indicates a mixed T-duality.
The ten-dimensional superalgebras are given their conventional names; the nine-dimensional superalgebras are named after their R-symmetry.}

\label{10d-9d}

\end{sidewaystable}

\clearpage

\subsection{Four-dimensional $\mathcal{N}=2$ supersymmetry algebras and 
type-II/II$^*$ Calabi-Yau compactifications} 

The compactification of type-IIA/IIB string theory in signature $(1,9)$
on a Calabi-Yau three-fold with Hodge numbers $h^{1,1}, h^{2,1}$ 
leads to an $\mathcal{N}=2$ supergravity theory 
in signature (1,3) with 
$n_V$ vector multiplets and $n_H$ hypermultiplets, where 
$(n_V,n_H) = (h^{1,1}, h^{2,1}+1)$ for type-IIA and 
$(n_V,n_H) = (h^{2,1}, h^{1,1}+1)$ for type-IIB \cite{Cecotti:1988qn,Ferrara:1989ik,Bodner:1990zm}. 
The four-dimensional
supersymmetry algebra is the standard $\mathcal{N}=2$ supersymmetry
algebra with R-symmetry group $G_R = \mathrm{U}(2) \cong \mathrm{U}(1) 
\times \mathrm{SU}(2)$. The target space geometries of vector and hypermultiplets
are projective special K\"ahler and quaternionic-K\"ahler, see 
\cite{LopesCardoso:2019mlj} for a review. As already shown in 
\cite{Cortes:2019mfa}, there is a second, inequivalent $\mathcal{N}=2$
algebra in signature $(1,3)$, where the supercharges obey a twisted 
Majorana condition and the R-symmetry group is $\mathrm{U}(1,1) \cong \mathrm{U}(1) \times 
\mathrm{SU}(1,1)$. The corresponding theory of rigid vector multiplets was constructed
explicitly in \cite{Cortes:2019mfa}, where it was shown that while the scalar geometry 
of the vector multiplets is still special K\"ahler, the modified 
supersymmetry transformations imply a relative sign between the scalar 
and vector kinetic terms. While hypermultiplets for the supersymmetry with
$\mathrm{U}(1,1)$ R-symmetry have not yet been constructed, sign flips
between the kinetic terms of the hypermultiplet scalars are expected, 
and the factor $\mathrm{SU}(1,1)$ suggests that the scalar geometry
is para-quaternionic K\"ahler rather than quaternionic K\"ahler. 
The reason is as follow: for the standard algebra with R-symmetry 
$\mathrm{U}(2)\cong \mathrm{U}(1) \times \mathrm{SU}(2)$
the vector multiplet and hypermultiplet scalars transform under 
$\mathrm{U}(1)$ and $\mathrm{SU}(2)$, respectively. The factor
$\mathrm{U}(1)$ acts on the vector multiplet scalar manifold 
as multiplication by the complex structure, that is, as 
`multiplication by $i$'. Similarly, the factor $\mathrm{SU}(2)$ acts
on the hypermultiplet scalar manifold as multiplication by the 
unit quaternions $\mathrm{SU}(2)\subset \mathbb{H}^*$. A change
of the R-symmetry group therefore indicates a change of the scalar geometry.
For example, in signature $(0,4)$ the R-symmetry group is 
$\mathrm{U}^*(2) \cong \mathrm{SO}(1,1) \times \mathrm{SU}(2)$, 
and the geometry of the vector multiplet scalars is 
para-K\"ahler, which differs from K\"ahler by replacing the 
complex structure $I, I^2=-\mathrm{Id}$ by a para-complex 
structure $E\not=\mathrm{Id}, E^2 = \mathrm{Id}$. 
The action 
of the factor $\mathrm{SO}(1,1) \subset \mathrm{U}^*(2)$ 
on the vector multiplet scalar manifold is generated by 
the action of the para-complex structure, see 
\cite{Cortes:2003zd} for details. Similarly, the replacement of the 
factor $\mathrm{SU}(2) \subset \mathrm{U}(2)$ by 
$\mathrm{SU}(1,1) \subset \mathrm{U}(1,1)$ indicates that
the scalar geometry of hypermultiplets in the twisted 
$\mathcal{N}=2$ theory in signature (1,3) is para-hyper-K\"ahler for
rigid supersymmetry and para-quaternion-K\"ahler for 
local supersymmetry. By inspection of the four-dimensional
R-symmetry groups we can determine the scalar geometries
of vector and hypermultiplets for all signatures and inequivalent
algebras, see Table \ref{Tab:4d_susy}.

For some of these cases the scalar geometry has been verified 
by explicit construction, while other cases will be the subject of 
future work.  $\mathcal{N}=2$ supergravity in signature (0,4) arises from the 
compactification of the Euclidean (signature (0,10)) type-IIA theory
on a Calabi-Yau threefold. The R-symmetry group $\mathrm{U}^*(2) \cong \mathrm{SO}(1,1) \times \mathrm{SU}(2)$ indicates that the vector multiplet geometry
is special para-K\"ahler while the hypermultiplet geometry is quaternionic-K\"ahler,
which was indeed found in \cite{Sabra:2015tsa} by explicit dimensional reduction.
In \cite{Third} it will be shown that theories realizing the twisted 
$\mathcal{N}=2$ algebra in signature (1,3) arise from the compactification of 
type-IIA$^*$/IIB$^*$ string theory on Calabi-Yau three-folds, and that
their hypermultiplet geometry is indeed para-quaternion K\"ahler. 
$\mathcal{N}=2$ supergravity in signature (2,2)
arises from the compactification of the signature $(2,8)$ type-IIA theory 
on a Calabi-Yau threefold. The R-symmetry group $\mathrm{GL}(2, \mathbb{R}) \cong
\mathbb{R}^{>0} \times \mathrm{SL}^\pm(2,\mathbb{R})$ indicates that the
vector multiplet geometry is special para-K\"ahler while the hypermultiplet geometry 
is para-quaternionic-K\"ahler.\footnote{$\mathrm{SL}^\pm(2,\mathbb{R}) \subset \mathrm{GL}(2, 
\mathrm{R})$ is the group of real $2\times 2$ matrices with determinant $\pm 1$.} 
This will be verified by dimensional reduction in \cite{Third}.

\begin{table}[t]
\begin{tabu}{|[1.5pt] l |[1.5pt] l | l | l |[1.5pt]} \tabucline[1.5pt]{1-4}
Signature & R-symmetry & VM geometry & HM geometry \\ \tabucline[1.5pt]{1-4}
$(0,4)$ & $\mathrm{U}(2)^* \cong \mathrm{SO}(1,1) \times \mathrm{SU}(2)$ & SPK & QK \\ \hline
$(1,3)$ & $\mathrm{U}(2) \cong \mathrm{U}(1) \times \mathrm{SU}(2)$ & SK & QK \\ \cline{2-4}
 & $\mathrm{U}(1,1)\cong \mathrm{U}(1) \times \mathrm{SU}(1,1)$ & SK & PQK \\  \hline
$(2,2)$ & $\mathrm{GL}(2,\mathbb{R} )\cong \mathrm{SO}(1,1) \times 
\mathrm{SL}^\pm(2,\mathbb{R})$ & SPK & PQK \\ \tabucline[1.5pt]{1-4}
\end{tabu}
\caption{Four-dimensional $\mathcal{N}=2$ supersymmetry algebras, their
R-symmetry groups and the scalar geometries of vector multiplets (VM) and
hypermultiplets (HM). SK = special K\"ahler, SPK = special para-K\"ahler, 
QK = quaternionic K\"ahler, PQK = para-quaternionic K\"ahler.
\label{Tab:4d_susy}}
\end{table}

\subsection{Three-dimensional $\mathcal{N}=4$ supersymmetry algebras and 
their hypermultiplet geometries}

Three-dimensional $\mathcal{N}=4$ supergravity theories can be realized
by spacelike and timelike reductions of four-dimensional $\mathcal{N}=2$
supergravity. In three dimensions vector multiplets can be dualized into hypermultiplets
so that the scalar manifold is the product of two hypermultiplet manifolds.
The R-symmetry groups in three dimensions are $\mathrm{SO}^*(4)\cong 
\mathrm{SL}(2,\mathbb{R}) \times \mathrm{SU}(2)$ in Euclidean signature (0,3)
and $\mathrm{O}(p,q)$, $p+q=4$ in Lorentz signature (1,2). For all cases 
with the exception of $\mathrm{O}(1,3)$ this allows us to identify the 
hypermultiplet geometries, see Table \ref{Tab:3d_susy}. The case
$\mathrm{O}(1,3)$ is special in that it is the only one which does not
arise by dimensional reduction. For the other cases we note
the embeddings $\mathrm{U}^*(2) \subset \mathrm{SO}^*(4)$, 
$\mathrm{U}(2) \subset \mathrm{O}(4)$, and $\mathrm{U}(1,1) \subset
\mathrm{O}(2,2)$ of the respective R-symmetry groups.

For some of these cases the scalar geometry has been verified by explicit
construction, while others will be subject to future work. Dimensional
reduction of vector multiplets from signatures (0,4) or (1,3) to signature (0,3) leads 
to hypermultiplets with a para-quaternionic K\"ahler
target space \cite{Cortes:2015wca}. 
By comparing to Table \ref{10d-9d} we see that the resulting 
pattern of spacelike and timelike reductions replicates the one for ten- and nine-dimensional
type-II theories, with the exception that in signature $(1,3)$ there are only two rather
than four non-isomorphic supersymmetry algebras. However, compactifications of
type-IIA and type-IIB on the same Calabi-Yau threefolds lead to different theories
since, as reviewed above, the roles of vector and hypermultiplets are reversed. 
The resulting pattern of dimensional reductions and oxidations, and the induced
T-dualities between the four-dimensional theories 
will be studied in detail in \cite{Third}.

\begin{table}[t]
\begin{tabu}{|[1.5pt] l |[1.5pt] l | l | l |[1.5pt]} \tabucline[1.5pt]{1-4}
Signature & R-symmetry & HM$_1$  geometry & HM$_2$ geometry \\ \tabucline[1.5pt]{1-4}
$(0,3)$ & $\mathrm{SO}^*(4) \cong \mathrm{SL}(2,\mathbb{R}) \times SU(2)$ &PQK & QK \\ \hline
$(1,2)$ & $\mathrm{O}(4)  \cong \mathrm{SU}(2) \times \mathrm{SU}(2)$ & QK & QK \\ \cline{2-4}
 & $\mathrm{O}(1,3)$ & $-$  &  $-$\\  \cline{2-4}
& $ \mathrm{O}(2,2) \cong \mathrm{SL}(2,\mathbb{R}) \times 
\mathrm{SL}(2,\mathbb{R})$ &   PQK & PQK \\ \tabucline[1.5pt]{1-4}
\end{tabu}
\caption{Three-dimensional $\mathcal{N}=4$ supersymmetry algebras, their
R-symmetry groups and the scalar geometries of the two hypermultiplet manifolds. 
\label{Tab:3d_susy}}
\end{table}

\section{Conclusion and outlook}

In this paper we have provided a construction of extended supersymmetry algebras
which works uniformly across dimensions and signatures. We have classified the 
resulting R-symmetry groups, which ultimately leads to Table \ref{mastertable}. The
resulting pattern of R-symmetry groups is modulated by the properties of the complex
bilinear form and of the reality conditions. In some signatures multiple real forms lead
to several non-isomorphic algebras based on the same spinor module. This includes
`type-*' Lorentzian signature algebras with non-compact R-symmetry groups 
such as $\mathrm{O}(p,q)$ in signatures (1,9) and (1,2), and 
$\mathrm{U}(k,l)$ in signature (1,3) which correspond to non-standard, `twisted'
Majorana conditions imposed on complex supercharges. Our formalism 
always starts with a standard complex superbracket, so that the `twisting' is encoded in the reality 
condition. This has the advantage that the twisting is tied to
selecting a real form of the complex R-symmetry group, which allows us to see 
`twisting' as part of the larger pattern of variation of R-symmetry group across dimensions
and signatures. 

While we have provided evidence that 
supersymmetry algebras are classified by their R-symmetry groups, together with 
the choice of one relative sign for chiral algebras, solving the classification problem 
completely has been left to 
future work. The most promising approach is to combine the formalism applied 
in this paper with the methods and results of \cite{Alekseevsky:1997}, as has been 
done for the special case of $D=4, \mathcal{N}=2$ supersymmetry in \cite{Cortes:2019mfa}.
BPS-charges can be included along the lines of \cite{Alekseevsky:2003vw}, which will
also allow to discuss the tensor-valued bilinear forms needed to describe fermionic
terms in the action \cite{Hull:1998ym,Gall:2018ogw}. Our formalism, which encodes 
the signature dependence completely in the reality condition, while the superbracket 
is fixed in a given dimension allows to obtain the supersymmetry transformations and
actions in a given signature by imposing reality conditions on their complex versions
\cite{Gall:2018ogw}. 

Apart from ab initio construction, dimensional reduction can be used to obtain 
supersymmetry transformations and actions. In Section \ref{secappphys} we have
discussed Calabi-Yau compactifications of type-II superstring theories 
to theories in signatures (0,4), (1,3) and (2,2) using their 
supersymmetry algebras and R-symmetry groups. The corresponding bosonic
actions which will be constructed explicitly in \cite{Third}. This will not only
allow us to verify the claims we have made regarding the target space geometries
of four- and three-dimensional hypermultiplets in arbitrary signature, but also 
prepare the ground for studying solutions of `exotic' supergravity and string theories
in dimension 4 and 10. It was observed in \cite{Gutowski:2020fzb} that there is a correspondence 
between the planar cosmological solutions of standard $D=4, \mathcal{N}=2$
supergravity with vector multiplets constructed in \cite{Gutowski:2019iyo},
and planar black hole solutions of 
its twisted variant (with R-symmetry $\mathrm{U}(1,1))$. As these include 
solutions of the STU-model (and its twisted version), they can be lifted
to 10 and 11 dimensions to solutions of type-II string theory and M-theory.
In fact, in the untwisted case, this leads to the same higher-dimensional
brane configuration as the STU black hole \cite{Gutowski:2019iyo}. 
The solutions of the twisted four-dimensional theory lift to brane configurations
in type-II$^*$ theories, which strongly suggests that the duality between 
black hole and cosmological solutions can be understood as a combination 
of time-like and space-like T-dualities. As shown in \cite{Gutowski:2020fzb}
the horizons of the dual cosmological/black hole solutions satisfy the first
law of thermodynamics and have the same semi-classical Euclidean partition 
function. Investigating this from the ten-dimensional point of view is likely 
to provide new insights into the solutions of type-II$^*$ theories. This is but one example
how one can further explore the network of type-II string theories described
in \cite{Hull:1998vg,Hull:1998ym}. The work of 
\cite{Dijkgraaf:2016lym,Blumenhagen:2020xpq} shows that there are many
field-theoretical as well as phenomenological aspects to be explored, including
signature change, de Sitter solutions, and brane world scenarios. 

Another direction to explore is what can be learned about the symmetries 
underlying string theories. Dimensional reduction exhibits hidden symmetry,
and maximal symmetry is reached when all directions, including time,
are compactified \cite{Moore:1993zc}. However, if string theories in general
signature are part of the full configuration space of string/M-theory, then 
dimensional reduction is not enough to exhibit the full symmetry of string 
theory. In order to cover all possible spacetime signatures one can use
complexification followed by exploring all possible reality conditions. 
For example in \cite{Bergshoeff:2000qu,Bergshoeff:2007cg}
a uniform description of $M$-theory and type-II theory was given 
based on the complex form of the ortho-symplectic Lie superalgebra $\mathfrak{osp}(32)$. 
The formalism presented in this paper may offer certain advantages, because
it does not rely on obtaining Poincar\'e Lie superalgebras as contractions, 
but instead works directly with 
complex and real Poincar\'e Lie superalgebras. Another strategy to explore the
hidden symmetries of string theory is
to use an extended spacetime, as is done in doubled and exceptional field theory.
In these formalisms exotic versions of type type-II string theory appear 
naturally, see for example \cite{Hohm:2019bba}, and the formalism presented
in this paper should be useful to investigate this further.

\subsection*{Acknowledgements}
This paper is substantially based on parts of the first authors' PhD thesis 
\cite{Thesis_Louis}, which is available at the University of Liverpools `Elements'  repository.
T.M. thanks Vicente Cort\'es for useful discussions.

\begin{appendix}
\section{Conventions, notation and some useful formulae} \label{secconvs}

In this appendix we collect information about our notations and conventions. In particular, 
we explain our conventions for spinor indices, which we suppress where
possible in the paper. We also explain in detail how to relate the formulation
of supersymmetry algebras in terms of vector-valued bilinear forms to the
usual one in terms of anti-commutators of supercharges.

\subsection{Spinor index conventions \label{app:spinor_indices}}

For spinor indices we use the same conventions as  \cite{Cortes:2003zd}.
Dirac spinors $\psi \in \mathbb{S}$ have lower indices, $\psi =(\psi_\alpha)$. Since $\gamma$-matrices represent endomorphisms on the spinor module $\mathbb{S}$, their index structure is $\gamma_\mu  =(\gamma_{\mu\alpha}^{\;\;\;\;\beta})$.

The matrices $A$ and $C$ relate the $\gamma$-matrices to their Hermitian conjugate
and to  the transposed matrices, respectively (\ref{A}), (\ref{C}). 
They have two upper spinor indices, 
 $A= (A^{\alpha \beta})$ and $C = (C^{\alpha \beta})$, and define a sesquilinear form 
(the Dirac bilinear form) and a complex bilinear form (the Majorana bilinear form)
on $\mathbb{S}$, respectively:
\begin{align}
	A(\lambda, \chi) = \lambda^*_\alpha A^{\alpha \beta} \chi_\beta, \quad 	C(\lambda, \chi) = \lambda_\alpha C^{\alpha \beta} \chi_\beta.
\end{align} 
The inverse matrices are $A^{-1} = (A^{-1}_{\alpha \beta})$ and $C^{-1} =(C^{-1}_{\alpha \beta})$.

When using Dirac spinors with a sesquilinear form $A$, spinor indices are raised and lowered using $A$ and $A^{-1}$, i.e. $\lambda^\alpha = A^{\alpha \beta} \lambda_\beta$ and $\lambda_\alpha = A^{-1}_{\alpha \beta} \lambda^\beta$. Upper index spinors defined this way are elements of complex-conjugate
dual (transposed) Dirac spinor module, which we can identify with $\mathbb{S}$ using the spin
equivariant  isomorphism provided by $A$. Similarly, we can raise and lower 
Dirac spinor indices using $C$ and its inverse:
$\lambda^\alpha  = C^{\alpha \beta} \lambda_\beta$ and $\lambda_\alpha = C^{-1}_{\alpha \beta} \lambda^\beta$. If upper index spinors are defined this way, they are elements of the 
dual (transposed) Dirac spinor module, which we can identify with $\mathbb{S}$ using
the spin equivariant isomorphism provided by $C$. In this paper we exclusively use 
$C$ and $C^{-1}$ to raise and lower Dirac spinor indices. 
We use a convention where the matrix $C$ is equal to its inverse, so that
we can denote the components of $C^{-1}$ simply by $C_{\alpha\beta}$.  The matrix $C$
is either symmetric or antisymmetric. This is encoded by the invariant $\sigma=\sigma_C$, which also
is the symmetry of the Majorana bilinear form,
\begin{equation}
C^{\beta \alpha} = \sigma C^{\alpha \beta} \;.
\end{equation} 
In the antisymmetric case some care is required when raising and lowering indices. 
For example
\begin{equation}
C^{\alpha \beta} C_{\alpha \gamma} C_{\beta \delta} = 
\sigma C_{\gamma \alpha} C^{\alpha \beta} C_{\beta \delta} =
\sigma C_{\gamma \delta} \;.
\end{equation}
Note that the convention we use for spinor indices is not of the `NW-SE type.' This is
different from our convention for `internal' indices on spinors, see below.
In the bulk of this paper we work with expressions where spinor indices are contracted 
and can be omitted. However, index conventions are relevant when comparing
our results to the literature. In particular, they are relevant for translating
between vector-valued forms and anti-commutators of supercharges, see below.

Equations like $A = \Pi_\tau \gamma_\tau$ (where $\tau$ runs over all timelike 
Lorentz indices), 
which equate the matrix of a bilinear form 
(two upper indices) to a matrix representing an endomorphism (one lower, one upper index)
are equations between matrices, not between maps. Put differently, once we have 
fixed a set of $\gamma$-matrices, we can make a choice for the matrix $A$,
but this choice is tied to our choice of $\gamma$-matrices.

\subsection{Index conventions for internal indices}

For indices on the auxiliary space $\mathbb{C}^K$ we will usually\footnote{The only exception is the 
`matrix notation' which we use when doubling the dimension of the auxiliary space to completely
disentangle the spin and R-symmetry groups for even-dimensional space-times, see 
Appendix \ref{secmatrixnot}.} 
use the NW-SE convention. This reflects that our formalism generalizes symplectic
Majorana spinors, where the R-symmetry group $\mathrm{SU}(2)$ acts on these indices. 
Adopting this as the universal convention, bilinear forms on $\mathbb{C}^K$ 
take the form
\begin{align}
    M(z, w) = z^i w^j M_{j i}, \qquad i, j = 1,...,K.  \nonumber
\end{align}
Raising and lowering the indices is done using $M_{i j}$ and its contragradient (transposed of
inverse)  $M^{i j}$ such that
\begin{align}
    z^i = M^{i j} z_j , \quad z_i = z^j M_{j i},\nonumber
\end{align}
Note that the NW-SE convention then implies
\begin{equation}
M^{ij} M_{kj} = \delta^i_k \;,
\end{equation}
so that for anti-symmetric bilinear forms the matrices 
$(M^{ij})$ and $(M_{ij})$ are not each others inverse, but the inverse multiplied by $-1$. 

Our construction requires the choice of a non-degenerate complex bilinear form, which is 
either symmetric or antisymmetric. Our preferred choices are the
symmetric bilinear form $\delta$ and antisymmetric bilinear form $J$, with Gram matrices 
\begin{align}
    \delta = \mathds{1}_N, \quad J = \begin{pmatrix} 0 & \mathds{1}_{\frac{K}{2}} \\ -\mathds{1}_{\frac{K}{2}} & 0 \end{pmatrix}.\nonumber
\end{align}
We refer to these bilinears and matrices as `canonical.'

\subsection{Poincar\'{e} Lie superalgebras with and without indices} \label{secpoincare}

In this section we explain how the standard, index-based notation for supersymmetry  is
related to the index-free
description (or sometimes only partly index-free description)
used in \cite{Alekseevsky:1997} and in this paper.

A Poincar\'{e} Lie superalgebra, $\mathfrak{g}$, is a $\mathbb{Z}_2$-graded Lie algebra
\begin{equation}
    \mathfrak{g} = \mathfrak{g}_0 + \mathfrak{g}_1 \;,\;\;\;
    \mathfrak{g}_0 = \mathfrak{so}(V) + V \;,
\end{equation}
where the even part $\mathfrak{g}_0$ is the Poincar\'{e} Lie algebra based on the vector space
$V=\mathbb{R}^{t,s}$, that is $\mathbb{R}^{t+s}$ equipped with a real symmetric
bilinear form with signature $(t,s)$. 
The Lie bracket on $\mathfrak{g}_0$ is
\begin{align}
    [A, B] = AB - BA, \qquad [A, v] = Av, \qquad [v_1, v_2] = 0,
\end{align}
where $A,B \in \mathfrak{so}(V)$ and $v, v_1, v_2 \in V$.

The odd part $\mathfrak{g}_1$ of $\mathfrak{g}$  is an arbitrary sum of irreducible spinor modules, that is
of  $\mathfrak{spin}(V) \cong \mathfrak{so}(V)$ modules obtained
by decomposing irreducible modules of the
real Clifford algebra $Cl(V)\cong Cl_{t,s}$ into irreducible modules of $\mathfrak{spin}(V) \subset
Cl_{t,s}$. The Lie algebra structure on $\mathfrak{g}_0$ is extended to a Lie superalgebra
structure by the spinorial action $\rho_S$ of  $\mathfrak{so}(V)$ on $\mathfrak{g}_1$ together
with a trivial action of $V$ on $\mathfrak{g}_1$, and a symmetric vector-valued bracket
$\mathfrak{g}_1 \times \mathfrak{g}_1 \rightarrow V$, which is $\mathfrak{so}(V)$ equivariant
(covariant). The additional non-trivial relations are:
\begin{align}
    [A, \lambda] = \rho_S(A) \lambda, \qquad [\lambda, \chi] = \Pi(\lambda, \chi) \in V \;,\;\;\;A\in \mathfrak{so}(V),\;\;
    \lambda, \chi \in \mathfrak{g}_1 \;,
\end{align}
where $\Pi: \mathfrak{g}_1 \times \mathfrak{g}_1 \rightarrow V$ is a real, symmetric, 
$\mathrm{Spin}(t,s)$-equivariant vector-valued bilinear form.\footnote{Note that the required super-Jacobi identity 
is implied by these conditions.}

It was shown in \cite{Alekseevsky:1997}
that the real, symmetric, 
$\mathrm{Spin}(t,s)$-equivariant vector-valued bilinear forms on a spinor module
form a vector space, which is spanned by vector-valued bilinear forms which 
can be constructed out of so called super-admissible scalar-valued bilinear forms
using Clifford multiplication. Given a bilinear form $\beta$, the associated 
vector-valued bilinear form $\Pi_\beta$ is defined by 
\begin{align}
\label{SusyVBil}
	\langle \Pi_\beta(\lambda, \chi), v \rangle = \beta(\gamma(v) \lambda, \chi)\;, \;\;\;
	\lambda, \chi  \in \mathfrak{g}_1, 
	v \in V \;.
\end{align}
Here $\langle\cdot, \cdot\rangle$ is the bilinear form on $V=\mathbb{R}^{t,s}$ and 
$\gamma(v) \lambda$ is the Clifford multiplication of the vector $v$ with the spinor $\lambda$. 

If we denote the generators of $V$ and $\mathfrak{so}(V)$ by $P_\mu$ and $M_{\mu \nu}$,
respectively, the relations of the Poincar\'e Lie algebra are  \cite{Freedman:2012zz}
\begin{align}
    &[P_\mu, P_\nu] = 0, \quad [M_{\mu \nu}, P_\rho] = i (\eta_{\mu \rho} P_\nu - \eta_{\nu \rho} P_\mu), \\
    &[M_{\mu \nu}, M_{\rho \sigma}] = i ( \eta_{\mu \rho} M_{\nu \sigma} - \eta_{\mu \sigma} M_{\nu \rho} - \eta_{\nu \rho} M_{\mu \sigma} + \eta_{\nu \sigma} M_{\mu \rho}) \nonumber,
\end{align}
where $\eta_{\mu \nu}$ is the Gram matrix of the bilinear form $\langle\cdot, \cdot \rangle$ on 
$\mathbb{R}^{t,s}$. The generators of the odd part $\mathfrak{g}_1$ are the supercharges
$Q_\alpha$. To translate \eqref{SusyVBil} into anti-commutation relations beween
supercharges, we expand the spinors and vectors in their respective bases
\begin{equation}
\lambda = \lambda^\alpha Q_\alpha \;,\;\;\;
\chi = \chi^\beta Q_\beta \;,\;\;
v = v^\mu  P_\mu \;.
\end{equation}
Clifford multiplication is an operation of $V$ on the spinor module which in 
terms of components is given by the action of $\gamma$-matrices on 
spinors, 
\begin{equation}
P^\mu Q_\alpha = \gamma^{\mu\;\;\beta}_{\;\;\alpha} Q_\beta \;.
\end{equation}
Therefore
\begin{equation}
\gamma \;: \;\; V \times \mathfrak{g}_1 \rightarrow \mathfrak{g}_1 \;,\;\;
(v,\lambda) \mapsto \gamma_v \lambda =  v_\mu P^\mu \lambda = 
v_\mu \gamma^{\mu\alpha}_{\;\;\;\;\;\beta} \lambda^\beta Q_\alpha \;,
\end{equation} 
where $(\gamma^{\mu \alpha}_{\;\;\;\;\beta}) = (\gamma^{\mu})^T = \tau C \gamma^\mu C^{-1}$ is
the transposed of $\gamma^\mu$.\footnote{Note that for $\tau=-1$ the matrices $\gamma^\mu$
and $(\gamma^\mu)^T$ are not  related by raising/lowering indices using $C$ (extra sign).}
Therefore
\begin{equation}
\beta(\gamma(v) \lambda , \chi ) = v_\mu \gamma^{\mu \alpha}_{\;\;\;\;\;\gamma} \lambda^\gamma \chi^\beta
\beta(Q_\alpha, Q_\beta) \;.
\end{equation}
A non-degenerate bilinear form can be used to identify a module with its dual 
(also called transposed module). Since $\beta$ is required to be an admissible bilinear form
this isomorphism is spin-equivariant, and induces a map which maps the
$\gamma$-matrices to their transposed \cite{Cortes:2003zd}. In other words the 
Gram matrix $\beta(Q_\alpha, Q_\beta)$ 
of the bilinear form $\beta$ can be interpreted as a 
charge conjugation matrix $C$. With our index conventions 
the Gram matrix $\beta(Q_\alpha, Q_\beta)$ 
is the inverse charge conjugation matrix $C^{-1}=(C_{\alpha \beta})$. Comparing 
to \eqref{SusyVBil} we conclude that
\begin{equation}
\Pi_\beta (\lambda, \chi) = 
[\lambda^\beta \chi^\gamma  (\gamma^\mu)^\alpha_{\;\;\gamma} C_{\alpha \beta}] P_\mu =
\sigma_C [\lambda^\beta \chi^\gamma  C_{\beta \alpha}
(\gamma^\mu)^\alpha_{\;\;\gamma}] P_\mu  =
\sigma_C \tau_C [ \lambda^\beta \chi^\gamma 
(\gamma^\mu)_\beta^{\;\;\alpha} C_{\alpha \gamma}]P_\mu \;.
\end{equation}
Here we used $C^T = \sigma C$ and $(\gamma^\mu)^T = \tau  C \gamma^\mu C^{-1}$. 
Defining the superbracket using the vector-valued bilinear form,
$ \{\lambda, \chi \} = \Pi_\beta(\lambda, \chi)$, we obtain
\begin{equation}
\{ Q_\beta, Q_\gamma \} = \sigma \tau  (\gamma^\mu C^{-1})_{\beta \gamma} P_\mu \;.
\end{equation}
For this to be a symmetric bracket we need that
\begin{equation}
(\gamma^\mu C^{-1})^T = C^{-1,T} (\gamma^\mu)^T = \sigma  C^{-1} \tau   C \gamma^\mu C^{-1}
= \sigma \tau \gamma^\mu C^{-1}   =  \gamma^\mu C^{-1} \;,
\end{equation}
that is $\sigma \tau  = 1$. This shows explicitly how the symmetry of the superbracket is
related to the super-admissibility of the bilinear form $\beta$.

While we are ultimately interested in real supersymmetry algebras, all the above 
concepts and statement naturally extend to complex Poincar\'e Lie superalgebras. 
Of course, for complex bilinear forms the concept of signature loses its invariant 
meaning. 
We can use this to pass from a real algebra to its complexification, and from there to other
real forms. 

In general, the spinorial module $\mathfrak{g}_1$ can be reducible. For sums of 
inequivalent modules the contributions to the supersymmetry anti-commutator
just add up `incoherently', while  for sums of equivalent 
modules they can `mix.' This is what gives rise to the larger R-symmetry groups of
extended supersymmetry algebras.
Let us consider the case of $\mathcal{N}$ copies of an irreducible
module $S$, where $\mathfrak{g}_1 = S \oplus \cdots \oplus S = \mathcal{N} S = S\otimes_{\mathbb{K}}
 \mathbb{K}^\mathcal{N}$, where $\mathbb{K}$ is $\mathbb{R}$ or $\mathbb{C}$, depending
 on whether $S$ is a real or complex module. 
 Denoting irreducible spinor indices by $\alpha, \beta, \ldots$ and labeling copies
 by $i,j=1,\ldots \mathcal{N}$, the supercharges are $Q_{i \alpha}$ and spinors
 expand as $\lambda = \lambda^{i\alpha} Q_{i \alpha}$. The bilinear form 
 on $S\otimes_{\mathbb{K}}  \mathbb{K}^\mathcal{N}$ is $\beta = C \otimes M$, where $C$ is the
 bilinear form defined by the charge conjugation matrix on the irreducible module $S$, 
 and where $M$ is a symmetric or antisymmetric bilinear form on the multiplicity space
 $\mathbb{K}^\mathcal{N}$. 
 By a similar computation to the one above we find
 \begin{equation}
 \beta( \gamma^\mu \lambda, \chi) = \sigma \tau \lambda^{i\alpha} \chi^{j\beta} 
 \gamma^{\mu\;\;\gamma}_{\;\;\alpha} C_{\gamma \beta} M_{ji} =
 \sigma \tau \sigma_M \lambda^{i\alpha} \chi^{j\beta} 
 (\gamma^\mu C^{-1})_{\alpha \beta} M_{ij} \;,
 \end{equation}
 where $\sigma_M$ is the symmetry of $M$, that is $M^T =\sigma_M M$. 
 Therefore 
 \begin{equation}
 \{ Q_{i\alpha}, Q_{j\beta} \} = \sigma \tau \sigma_M (\gamma^\mu C^{-1})_{\alpha \beta} M_{ij}  
 P_\mu \;.
 \end{equation}
 The bracket is symmetric if 
\begin{equation}
(\gamma^\mu C^{-1})_{\alpha \beta} M_{ij} = 
(\gamma^\mu C^{-1})_{\beta \alpha } M_{ji} \;,
\end{equation}
 which requires $\sigma \tau \sigma_M=1$. Since the $\gamma$-matrices do not act
 on $\mathbb{K}^\mathcal{N}$, it follows that $\sigma \tau \sigma_M = \sigma_\beta \tau_\beta$, 
 that is the bracket is symmetric if 
  the bilinear form $\beta=C \otimes M$ is super-admissible. 
 
 To summarize, given a super-admissible bilinear form $\beta = C\otimes M$, 
 \begin{equation}
 \beta(\lambda, \chi) = \lambda^{i\alpha} C_{\alpha \beta} \chi^{j\beta} M_{ji}
 \end{equation}
 the corresponding anti-commutation relations are 
 \begin{equation}
 \label{Qrelations}
 \{ Q_{i\alpha}, Q_{j\beta} \} = (\gamma^\mu C^{-1})_{\alpha \beta} M_{ij}  P_\mu \;.
\end{equation}
  For a given charge conjugation matrix $C$ one has either $\sigma \tau=1$ or 
  $\sigma \tau =-1$. Therefore a charge conjugation matrix does not always
  define a supersymmetry algebra.
  However, by taking an even number 
  of copies and pairing $C$ with an antisymmetric bilinear form on the multiplicity 
  space we can obtain a super-admissible bilinear form on the 
  extended spinor module. Our approach is to always double the spinor module and to use a 
  super-admissible
  bilinear form on the doubled space to define a complex algebra with relations 
  \eqref{Qrelations}. This uses that the doubling can be interpreted as a complexification.
  Real algebras are obtained by choosing a real form 
  through imposing a reality condition of the form
  \begin{equation}
  (\lambda^i)^* = \alpha B \lambda^j L_{ji} 
  \end{equation}
  where $\alpha$ is a phase and where $B \otimes L$ defines a real structure $\rho$
  on $\mathfrak{g}_1 \oplus \mathfrak{g}_1 \cong \mathfrak{g}_1 \otimes_{\mathbb{R}} \mathbb{C}
  \cong \mathfrak{g}_1^\mathbb{C}$.

\subsection{Useful formulae relating to chirality \label{Sect_Useful_Formulae}}

The following formula are used extensively throughout this paper and are provided here for easy 
reference. They can be found in, or straightforwardly obtained from \cite{VanProeyen:1999ni,Cortes:2019mfa}. 
\begin{align}
	C_\pm \gamma_* &= \begin{cases} \pm i C_\mp\;, \quad &D = 2, 6, 10\;, \\ C_\mp\;, \quad &
	D = 4, 8, 12\;, \end{cases} \label{cgammastar} \\
	\gamma_* C_\pm &= \begin{cases} \pm i C_\mp\;, \quad &D = 2, 6, 10\;, \\ - C_\mp\;, \quad 
	&D = 4, 8, 12\;, \end{cases} \label{gammastarc} \\
	B_\pm \gamma_* &= \begin{cases} \pm i \sigma_+ \sigma_- B_\mp = \mp i B_\mp\;,  \quad 
	&D = 2, 6, 10\;, \\ \sigma_+ \sigma_- B_\mp = B_\mp \;, \quad &D = 4, 8, 12 \;,  \end{cases}  \label{bgammastar} \\
	\gamma_* B_\pm  &= \begin{cases} (-1)^t \mp i \sigma_+ \sigma_- B_\mp = \pm (-1)^t  i B_\mp \;, \quad &D = 2, 6, 10\;,  \\ (-1)^t \sigma_+ \sigma_- B_\mp = (-1)^t B_\mp \;, \quad &D = 4, 8, 12 \;,  \end{cases} \label{gammastarb} \\
	B^*_\pm \gamma_* &= \begin{cases} \pm i B^* _\mp \;,  \quad &D = 2, 6, 10\;,  \\ B^* _\mp \;,  \quad &D = 4, 8, 12 \;.  \end{cases} \label{bstargammastar}
\end{align}
For completeness we note that, as shown in the main text, the symmetry $\sigma_\pm$ of $C_\pm$ 
satisfies $\sigma_+ = - \sigma_-$ in  the Weyl compatible dimensions
$D = 2, 6, 10$,  and $\sigma_+ = \sigma_-$ in the Weyl incompatible 
dimensions $D = 4, 8, 12$.

\subsection{Proof that a signature flip $(t,s) \leftrightarrow (s,t)$ exchanges
$B_+ \leftrightarrow B_-$.} \label{appproofsigflip}

This is the proof of a statement that we used in the main part of this paper.
Consider the $(t,s)$ signature $\gamma$-matrices, which obey
\begin{align}
    (\gamma_i)^2 = \begin{cases}
        -1\;, \qquad i \leq t \;, \\ +1\;, \qquad i > t\;.
    \end{cases}
\end{align}

We define the $(s,t)$ signature $\gamma$-matrices as $\gamma'_m = i \gamma_{(D - m + 1)}$ (where $D = t + s$) such that they correctly obey
\begin{align}
    (\gamma'_i)^2 = \begin{cases}
        -1 \;,\qquad i \leq s \;,\\ +1 \;, \qquad i > s \;.
    \end{cases}
\end{align}

Both theories have the same charge conjugation matrices, $C_+$ and $C_-$ and $A$-matrices
\begin{align}
    &A^{(t,s)} = \gamma_1 ... \gamma_t\;, \\
    &A^{(s,t)} = \gamma'_1 ... \gamma'_s = i^s \gamma_{D} ... \gamma_{t+1} \;. \nonumber
\end{align}

We then see that, using $C_+ = k C_- \gamma_*$ ($k$ is the constant from \eqref{cgammastar})
\begin{align}
    B^{(t,s)} _+ & = (C_+ (A^{(t,s)})^{-1})^T \\
    & = (k C_- \gamma_* (A^{(t,s)})^{-1})^T  \;.\nonumber
\end{align}

Using our definitions for $A^{(t,s)}$ we find 
\begin{align}
    \gamma_* (A^{(t,s)})^{-1} & = (-i)^t \gamma_1 ... \gamma_{D} (-1)^t \gamma_t ... \gamma_1 \nonumber \\
    & = (-1)^{st} (-i)^t \gamma_{t+1} ... \gamma_{D} \\
    & = (-1)^{st} (-i)^t (-i)^s \gamma'_{s} ... \gamma'_{1} \nonumber \\
    & = (-1)^{st} (-i)^{D}(A^{(s,t)})^{-1}  = (-1)^{st + \frac{D}{2}}(A^{(s,t)})^{-1}  \;, \nonumber
\end{align}
such that
\begin{align}
    &B^{(t,s)} _+ = (k C_-  (-1)^{st + \frac{D}{2}} (A^{(s,t)})^{-1}) = k  (-1)^{st + \frac{D}{2}} B^{(s,t)} _- \\
    &\implies (B^{(t,s)} _+)^* B^{(t,s)} _+ = (B^{(s,t)} _-)^* B^{(s,t)} _- \;. \nonumber
\end{align}

\section{Real (semi-)spinors and Majorana spinors \label{App:Real_vs_Majorana}}

In this appendix we
summarize the relation between complex and real spinors and semi-spinors
as defined in the mathematics literature and Dirac, Weyl, Majorana and 
Majorana-Weyl spinors as defined in the physics literature. The complex and real
spinor module
$\mathbb{S}$ and $S$ are defined by restricting irreducible 
representations of the complex and real Clifford algebras
$\mathbb{C}l_D$ and $Cl_{t,s}$ to the real spin group $\mathrm{Spin}(t,s)$. 
As $\mathrm{Spin}(t,s)$-modules 
$\mathbb{S}$ and $S$ can be isomorphic or non-isomorphic. Complex spinors
$\psi \in \mathbb{S}$ are Dirac spinors, while real spinors $\psi \in S$ are
not always Majorana spinors. 
The following
cases can occur, depending on dimension and signature.
\begin{enumerate}
\item 
$\mathbb{S}\not\cong S$, 
the complex and real spinor module are not isomorphic. 
Then $\mathbb{S}$ is the complexification of $S$, $\mathbb{S}=S \otimes_{\mathbb{R}}
\mathbb{C}$ and $S$ is a real subset of $\mathbb{S}$ fixed under the
action of an invariant real structure $\rho$, $S=\mathbb{S}^\rho$.  The elements of $S$ are
Majorana spinors. $S$ can be irreducible or reducible.
\begin{enumerate}
\item
$S$ is irreducible.  Then Majorana spinors are the unique irreducible spinor representation. 
If, in addition, the dimension is even, Majorana spinors are equivalent
(as real $\mathrm{Spin}(t,s)$ representations) to Weyl spinors:
$S\cong \mathbb{S}_+ \cong \mathbb{S}_-$. This is, for example, the case
in signature $(1,3)$. 
\item
$S$ is reducible.  Then real spinors decompose into real semi-spinors, $S=S_+ + S_-$.
Given that $\mathbb{S} = S \otimes_\mathbb{R} \mathbb{C}$, 
the real semi-spinor modules must be non-isomorphic, since
otherwise their complexifications $S_\pm \otimes_{\mathbb{R}} \mathbb{C}$ 
would be isomorphic as complex modules, which is not true.
The real  semi-spinors $S_+\not\cong S_-$ are Majorana-Weyl spinors. 
As is well known, Majorana-Weyl spinors exist if and only if $t-s$ is 
0 modulo 8. In Table \ref{Table:epsilon-structures} these are the entries
where $(\epsilon_+,\epsilon_-)=(1,1)$. 
\end{enumerate}
\item
$\mathbb{S} \cong S$, the complex and real spinor module are isomorphic.
Here we have three subcases:
\begin{enumerate}
\item
$S$ is irreducible. Then Majorana spinors don't exist ($\mathbb{S}$ does not
admit an invariant real structure)
and Dirac spinors are the
unique irreducible spinor representation. This is realized in signature $(1,4)$. 
\item
$S = S_+ + S_-$,  $S_+ \not\cong S_-$. 
$S$ is reducible and decomposes into non-isomorphic semi-spinor modules.
In this case
real semi-spinors are the same as complex semi-spinors, that is Weyl
spinors, $S_\pm \cong \mathbb{S}_\pm$.  
This happens in even dimensions in those signatures where no invariant real 
structures, and hence no Majorana spinors exist, for example in signature $(4,0)$ or $(0,4)$.
\item
$S=S_++ S_-$, $S_+ \cong S_-$.
$S$ is reducible, and decomposes into isomorphic semi-spinor modules. 
In this case $\mathbb{S}\cong S\cong S_\pm \otimes \mathbb{C}$ carries an invariant real structure. 
The elements of $S_\pm$ 
are Majorana spinors, and we are in a signature where no Majorana-Weyl
spinors exist. There are two subcases:
\begin{enumerate}
\item
If the dimension is even, then complex and real
semi-spinors coincide and all types semi-spinor modules are isomorphic as
real modules: $\mathbb{S}_+ \cong \mathbb{S}_- \cong S_+ \cong S_-$. 
This happens, for example, in signature $(3,1)$. 
\item
If the the dimension  is odd, no Weyl spinors exist, and we have 
$\mathbb{S} = S = S_\pm \otimes_{\mathbb{R}} \mathbb{C}$. 
This is realized, for example, in signature $(3,2)$. 
\end{enumerate}
\end{enumerate}
\end{enumerate}
More details for the explicit examples we have mentioned can be 
found in \cite{Gall:2018ogw,Cortes:2019mfa}.

\section{Details on the complexification of spinor modules \label{App:Details_Complexification}}

In this appendix we provide details on the complexification of the odd parts
$\mathfrak{g}_1 = S^{\oplus  N} \cong S \otimes \mathbb{R}^N$, 
$\mathfrak{g}_1 =  S_+^{\oplus N}  \cong S_+ \otimes \mathbb{R}^{N}$
and $\mathfrak{g}_1 = S^{\oplus N_+} _+ \oplus  S_-^{\oplus N_-}  \cong
S_+ \otimes \mathbb{R}^{N_+} \oplus 
S_- \otimes \mathbb{R}^{N_-}$ of real supersymmetry algebras.

We start with the simpler case where $D$ is odd and no Weyl spinors exist. 
The unique real irreducible spinor representation is either $S$ or $S_+ \cong S_-$. 
There are two cases.
\begin{enumerate}
\item
$S$ is irreducible. Then
\[
\mathfrak{g}_1 \otimes_{\mathbb{R}} \mathbb{C} =
\left( S \otimes_{\mathbb{R}} \mathbb{C} \right)^{\oplus N} \cong
\left( S \otimes_{\mathbb{R}} \mathbb{C} \right) \otimes_\mathbb{C} \mathbb{C}^N \;.
\]
There are two subcases.
\begin{enumerate}
\item
$S\cong \mathbb{S}$. The smallest spinor representation is given by Dirac spinors,
and 
\[
\mathfrak{g}_1 \otimes_{\mathbb{R}} \mathbb{C} =
\left( \mathbb{S} \otimes_{\mathbb{R}} \mathbb{C} \right)^{\oplus N} \cong
\left( \mathbb{S} \otimes_{\mathbb{R}} \mathbb{C} \right) \otimes_{\mathbb{C}} \mathbb{C}^N \cong
\mathbb{S} \otimes_{\mathbb{C}} \mathbb{C}^{2N} \;.
\]
This case is characterized by $\mathbb{S}$ not admitting an invariant real structure.
\item
$S\otimes_\mathbb{R} \mathbb{C} \cong \mathbb{S}$. The smallest spinor representation
is given by Majorana spinors, and 
\[
\mathfrak{g}_1 \otimes_{\mathbb{R}} \mathbb{C} =
\left( S \otimes_{\mathbb{R}} \mathbb{C} \right)^{\oplus N} \cong
\mathbb{S}  \otimes_{\mathbb{C}} \mathbb{C}^N \;.
\]
This case is characterized by $\mathbb{S}$ admitting an invariant real structure.
\end{enumerate}
\item
$S$ is reducible and $S_+ \cong S_-$. 
Then 
$\mathbb{S} \cong S \cong S_\pm \otimes_{\mathbb{R}} \mathbb{C}$. The
unique irreducible spinor representation given by Majorana spinors
$S_+ \cong S_-$, and
\[
\mathfrak{g}_1 \otimes_{\mathbb{R}} \mathbb{C} = 
(S_+ \otimes_{\mathbb{R}} \mathbb{C})^{\oplus N} \cong \mathbb{S} \otimes_{\mathbb{C}} 
\mathbb{C}^N \;.
\]
In this case $\mathbb{S}$ admits an invariant real structure.
\end{enumerate}
If $D$ is even, then there is one additional case to consider, namely when $S$ is
reducible and the real semi-spinor modules are non-isomorphic, $S_+ \not{\cong} S_-$.
In this case the minimal spinors are either Weyl spinors 
or Majorana-Weyl spinors. Since there are two inequivalent irreducible real 
spinor representations, the general form of the odd part of the supersymmetry 
algebra is $\mathfrak{g}_1 = S_+^{\oplus N_+} \oplus S_-^{\oplus N_-}$. The additional third case is
\begin{itemize}
\item[3.]
$S=S_+ + S_-$, $S_+\not{\cong} S_-$. There are two sub-cases:
\begin{enumerate}
\item
$\mathbb{S}_\pm \cong S_\pm$. In this case $\mathbb{S}_\pm$ do not admit invariant
real structures and the inequivalent minimal spinors are Weyl spinors.
\item
$\mathbb{S}_\pm \cong S_\pm \otimes_{\mathbb{R}} \mathbb{C}$. In this case
$\mathbb{S}_\pm$ admit invariant real strucutures, and the inequivalent minimal spinors are
Majorana-Weyl spinors.
\end{enumerate}
In both cases
the complexification procedes analogous to case 1, with $\mathbb{S}$ 
replaced by $\mathbb{S}_+$ and $\mathbb{S}_-$.
\end{itemize}

\section{Matrix notation for Weyl Spinors} \label{secmatrixnot}

\subsection{Explanation of the matrix notation}

In this section, we will describe a notation which was introduced in \cite{Cortes:2019mfa},
and makes calculations easier when dealing with Weyl spinors in even dimensions. 
Using the natural embedding $\mathbb{S}_\pm \subset \mathbb{S}$ we combine the two Weyl spinor modules into a single `doubled-again' spinor module
\begin{align}
(\lambda^I) = 
     \begin{pmatrix} \lambda^i_+, \lambda^{\hat{i}}_- \end{pmatrix} = \begin{pmatrix}    \lambda^1_+, ..., \lambda^{K_+}_+, \lambda^1_-, ...,\lambda^{K_-}_-\end{pmatrix} \in \mathbb{S}^{\oplus K_+}_+ \oplus \mathbb{S}^{\oplus K_-}_- \subset \mathbb{S}^{\oplus (K_+ + K_-)} = \mathbb{S} \otimes \mathbb{C}^{K_+ + K_-} 
\end{align}
Using a mixture of matrix notation and index notation, 
orthogonal and isotropic bilinear forms take the form 
\begin{align}
    &\begin{pmatrix} \bar{\lambda}^i_+,\bar{\lambda}^{\hat{i}} _- \end{pmatrix} \begin{pmatrix} M_{j i} & 0 \\ 0 & M' _{\hat{j} \hat{i}} \end{pmatrix} \begin{pmatrix} \chi^j_+ \\ \chi^{\hat{j}}_- \end{pmatrix} \qquad \text{(Orthogonal)}, \\
    &\begin{pmatrix} \bar{\lambda}^i_+,\bar{\lambda}^i_- \end{pmatrix} \begin{pmatrix} 0 & M_{j i} \\ M_{j i} & 0 \end{pmatrix} \begin{pmatrix} \chi^j_+ \\ \chi^j_- \end{pmatrix} \qquad \text{(Isotropic)}.
\end{align}
In the orthogonal case
$M$ is the bilinear form on the $\mathbb{C}^{K_+}$ factor, $M'$ the one on the $\mathbb{C}^{K_-}$ factor, $i, j = 1,..., K_+$ and $\hat{i}, \hat{j} = 1,...,K_-$. For isotropic signatures, necessarily $M = M'$ and $K_+ = K_-$.

In addition, we have a real structure, $\rho$, which is either Weyl-compatible or Weyl-incompatible. For a Weyl-compatible reality condition, we write this as
\begin{align}
    &\rho (\lambda^i_+) = \alpha^* B^* (\lambda^j_+)^* L_{j i}, \qquad \rho (\lambda^{\hat{i}} _-) = \beta^* B'^* (\lambda^{\hat{j}} _-)^* L'_{\hat{j} \hat{i}} \\
 &\rightarrow \rho \begin{pmatrix} \lambda^i_+ \\ \lambda^{\hat{i}} _- \end{pmatrix} =  \begin{pmatrix} \alpha^* B^* L_{j i} & 0 \\ 0 & \beta^* B'^* L'_{\hat{j} \hat{i}} \end{pmatrix} \begin{pmatrix} \lambda^j_+ \\ \lambda^{\hat{j}} _- \end{pmatrix}^* \label{comprealcond}.
\end{align}

Here $B$ and $B'$ can refer to either $B_\pm$.

Weyl-incompatible reality conditions are written as
\begin{align}
    \rho (\lambda^i_\pm) = \alpha^* B^* (\lambda^j_\mp)^* L_{j i} \rightarrow \rho\begin{pmatrix} \lambda^i_+ \\ \lambda^i _- \end{pmatrix} = \alpha^* B^* \begin{pmatrix} 0 & L_{j i} \\ L_{j i} & 0 \end{pmatrix} \begin{pmatrix} \lambda^j_+ \\ \lambda^j_- \end{pmatrix}^*.
\end{align}

Often it is possible and convenient to suppress the indices $i, j$, and write expressions in terms of 
vectors-of-vectors and block matrices
\begin{align}
    &\begin{pmatrix} \underline{\bar{\lambda}}_+,\underline{\bar{\lambda}}_- \end{pmatrix} \begin{pmatrix} M & 0 \\ 0 & M' \end{pmatrix} \begin{pmatrix} \underline{\chi}_+ \\ \underline{\chi}_- \end{pmatrix} \qquad \text{(Orthogonal)}, \\
    &\begin{pmatrix} \underline{\bar{\lambda}}_+,\underline{\bar{\lambda}}_- \end{pmatrix} \begin{pmatrix} 0 & M \\ M & 0 \end{pmatrix} \begin{pmatrix} \underline{\chi}_+ \\ \underline{\chi}_- \end{pmatrix} \qquad \text{(Isotropic)}.
\end{align}

When we do this, we will change the indices such that normal matrix multiplication makes sense for the resulting expressions. This induces a sign if $M=J$ is the anti-symmetric bilinear form with components
$J_{i j} =-J_{ji}$: 
\begin{align}
    \begin{pmatrix} \bar{\lambda}^i_+,\bar{\lambda}^i_- \end{pmatrix} \begin{pmatrix} 0 & J_{j i} \\ J_{j i} & 0 \end{pmatrix} \begin{pmatrix} \chi^j_+ \\ \chi^j_- \end{pmatrix} = \begin{pmatrix} \underline{\bar{\lambda}}_+,\underline{\bar{\lambda}}_- \end{pmatrix} \begin{pmatrix} 0 & -J \\ -J & 0 \end{pmatrix} \begin{pmatrix} \underline{\chi}_+ \\ \underline{\chi}_- \end{pmatrix}.
\end{align}

For a final example, \eqref{comprealcond} is rewritten as
\begin{align}
    \rho\begin{pmatrix} \underline{\lambda}_+ \\ \underline{\lambda}_- \end{pmatrix} =  B^* \begin{pmatrix} \alpha^*L & 0 \\ 0 & \beta^* L' \end{pmatrix} \begin{pmatrix} \underline{\lambda}_+ \\ \underline{\lambda}_- \end{pmatrix}.
\end{align}

When using this notation the Schur group and consequently the R-symmetry group acts by linear transformations on the expanded internal space $\mathbb{C}^{K_+ + K_-}$, but not on the spinor indices. 
This disentangling of spinor and internal indices with respect to the action of the Schur group is the main advantage of this notation. Without the additional doubling, the Schur group can act
non-trivially on spinors indices in even dimensions by acting differently on spinors depending
on their chirality. For example, in signature $(1,3)$  positive and negative chirality spinors
carry opposite charge under $\mathrm{U}(1) \subset G_R = \mathrm{U}(2)$, see \cite{Cortes:2003zd}. 
By doubling the auxiliary space, any chiral action of the 
Schur group is encoded in the larger matrix acting on the doubled space. Afterwards, the effects on each Weyl spinor module can be reconstructed and rewritten in terms of $\mathrm{Id}$ and $\gamma_*$ acting on the original spinor module.

\subsection{Action of R-symmetry transformations on the complex spinor module $\mathbb{S}$}

In even dimensions R-symmetry transformations act non-trivially on the complex spinor module
$\mathbb{S}$, though only through a relative sign between complex semi-spinors, since
R-symmetry transformations by definition commute with the Lie algebra of the spin group. We have
shown that the actions of the spin group and R-symmetry group can still be disentangled by
doubling the internal space. In the following section we briefly describe how our results
translate to a more conventional notation, where we employ Dirac spinors and do not
double the internal space. R-symmetry transformations are then given by products
of actions on the internal space with an action of $\mathbbm{1}, \gamma_*$ on 
Dirac spinors.

The chirality matrix $\gamma_*$ acts on the Weyl spinors $\lambda_\pm \in \mathbb{S}_\pm$ as 
$\gamma_* \lambda_\pm = \pm \lambda_\pm$. We choose a basis of $\mathbb{S}$ where
\begin{align}
    \gamma_* = \begin{pmatrix} \mathbbm{1}  & 0 \\ 0 & -\mathbbm{1} \;,
     \end{pmatrix},
\end{align}
where $\mathbbm{1}$ is the identity matrix acting on complex semi-spinors  $\mathbb{S}_\pm$.
In matrix notation, $\gamma_* \lambda^i _\pm = \pm \lambda^i_\pm$ acts on $(\lambda^i_+, \lambda^i_-)$ as the matrix
\begin{align}
    \begin{pmatrix}    \mathds{1}_{K} & 0 \\ 0 & -\mathds{1}_{K} \end{pmatrix} = \gamma_* \otimes \mathds{1}_K
\end{align}

\subsubsection{Orthogonal Bilinear Forms}

In orthogonal Weyl-compatible signatures the R-symmetry transformations act independently on each Weyl spinor module and are specified by the action on the internal $\mathbb{C}^{K_\pm}$ factor. This is because the Weyl spinor modules are complex irreducible modules (so we can apply Schur's lemma exactly like in odd dimensions) and the reality condition is defined on a Weyl spinor module alone. 
If $K_+ \neq K_-$ we cannot combine Weyl spinors into Dirac spinors. 

When $K_+ = K_-$, we can combine the Weyl spinors into Dirac spinors. Then R-symmetry
transformations act as follows. We know that a general element of the R-symmetry 
Lie algebra has the form
\begin{align}
    r = \begin{pmatrix} a & 0 \\ 0 & b \end{pmatrix}\;,
\end{align}
where $a,b$ are Lie algebra elements for the factor of the R-symmetry which acts on 
spinors of given chirality. 
We can rewrite this as 
\begin{align}
    r = \begin{pmatrix} c & 0 \\ 0 & c \end{pmatrix} + \begin{pmatrix} d & 0 \\ 0 & -d \end{pmatrix} = (\mathrm{Id} \otimes c) + (\gamma_* \otimes d)\;,
\end{align}
for $c = \frac{1}{2}(a + b)$ and $d = \frac{1}{2}(a - b)$.

We can see that, at most, R-symmetry generators act as identity or $\gamma_*$ on the $\mathbb{S}$ factor. In this case, the re-writing is somewhat artificial since the Lie algebra elements $a,b$ are 
independent. But in the remaining cases, where we will do something similar,  the two transformations will depend on one another.

For a Weyl-incompatible orthogonal signature, the reality condition links the two chiralities, and 
we found that they take the form
\begin{align}
    r = \begin{pmatrix} a & 0 \\ 0 & L a^* L^{-1} \end{pmatrix},
\end{align}
where
$a$ acts entirely on $\mathbb{S}_+$ while the corresponding transformation on $\mathbb{S}_-$ is
$L a^* L^{-1}$.
This can be recast into transformations that act on the entire spinor module $\mathbb{S} = \mathbb{S}_+ + \mathbb{S}_-$. Using that conjugation by $L$ and complex conjugation are involutions, we can 
rewrite $a$ and $L a^* L^{-1}$ in terms of 
\begin{align}
    a_\pm = \frac{1}{2} (a \pm L a^* L^{-1})\;,
\end{align}
so that
\begin{align}
    r = \begin{pmatrix} a_+ & 0 \\ 0 & a_+ \end{pmatrix} + \begin{pmatrix} a_- & 0 \\ 0 & -a_-\end{pmatrix} = (\mathds{1} \otimes a_+) + (\gamma_* \otimes a_-).
\end{align}

This is slightly different from the previous case with Weyl-compatible signatures because $a_+$ and $a_-$ are functions of $a$ alone. However we see that similarly  the generators of the R-symmetry group can be written in a way where they act either as $\mathrm{Id}$ or $\gamma_*$ on $\mathbb{S}$. 

\subsubsection{Isotropic Bilinear Forms}

For an isotropic vector-valued bilinear form, $\gamma_*$ generates a real one-parameter 
subgroup of the R-symmetry group. 
Consider the transformation
\begin{align}
    \lambda^i \rightarrow e^{\omega \gamma_*} \lambda^i = e^\omega \lambda^i_+ + e^{-\omega} \lambda^i_- \label{gammastargen}\;,
\end{align}
where $\omega \in \mathbb{C}$. 
A complex vector-valued bilinear form is invariant under this transformation:
\begin{align}
    \beta(\gamma^\mu \lambda, \chi) = &(\gamma^\mu \lambda^i_+)^T C \chi^j_- M_{j i} + (\gamma^\mu \lambda^i_-)^T C  \chi^j_+ M_{j i} \\
    \rightarrow & (\gamma^\mu  e^\omega \lambda^i_+)^T C e^{-\omega}\chi^j_- M_{j i} + (\gamma^\mu  e^\omega \lambda^i_-)^T C e^{-\omega} \chi^j_+ M_{j i} = \beta(\gamma^\mu \lambda, \chi) .
\end{align}

In matrix notation the transformation in \eqref{gammastargen} is
\begin{align}
    \begin{pmatrix} \underline{\lambda}_+ \\ \underline{\lambda}_- \end{pmatrix} \rightarrow  \exp (\omega \gamma_* \otimes \mathds{1}_K) \begin{pmatrix} \underline{\lambda}_+ \\ \underline{\lambda}_- \end{pmatrix}.
\end{align}
Imposing that \eqref{gammastargen} commutes with the reality condition in isotropic dimensions
forces $\omega$ to be real in Weyl-compatible signatures and 
to be imaginary in Weyl-incompatible signatures. The corresponding one-dimensional subgroups
$\mathrm{SO}(1,1)$ and $\mathrm{U}(1)$ of the R-symmetry group are often generated
when performing a  dimensional reduction from odd to even dimensions. For example,
the reduction of a supersymmetry algebra based on a single Dirac spinor from five to
four dimensions increases the R-symmetry group from $\mathrm{SU}(2)$ to 
$\mathrm{U}(2) \cong_{\mathrm{local}} \mathrm{SU}(2) \times \mathrm{U}(1)$ 
for the reduction $(t,s)=(1,4) \rightarrow (1,3)$ and to 
$\mathrm{U}^*(2) \cong_\mathrm{local} \mathrm{SU}(2) \times 
\mathrm{SO}(1,1)$ for the reduction $(t,s)=(1,4) \rightarrow (0,4)$
\cite{Cortes:2003zd}. 
General R-symmetry elements act simultaneously as $\mathrm{Id}$ or $\gamma_*$ on $\mathbb{S}$
and by a non-trivial transformation on the $\mathbb{C}^K$ factor, therefore the R-symmetry
groups are only locally isomorphic to direct products.

The general form  of an R-symmetry transformation in isotropic signatures is
\begin{align}
    r = \begin{pmatrix} a & 0 \\ 0 &  -M^{-1} a^T M \end{pmatrix} \;.
\end{align}
Since conjugation by $M$ and transposition are both involutions,  that we can split $a$ into eigen-matrices under  the combination of these two operations
\begin{align}
    a = a_{+} + a_{-}, \qquad \text{with} \qquad a_{\pm} = \frac{1}{2} (a \pm M^{-1} a^T M).
\end{align}

Therefore we can write
\begin{align}
    r = \begin{pmatrix} a_+ & 0 \\ 0 & -a_+\end{pmatrix} + \begin{pmatrix} a_- & 0 \\ 0 & a_-\end{pmatrix} = (\gamma_* \otimes a_+) + (\mathds{1} \otimes a_-) \label{decompr2}.
\end{align}

From this, we conclude that in isotropic signatures the only possible non-trivial 
action of the R-symmetry generators
on $\mathbb{S}$ is through multiplication by $\gamma_*$.

\section{Details of Isomorphisms \label{Sec:Details_Isomorphisms}}

In this appendix we provide the details of the proof that real supersymmetry
algebras constructed using the data $(C,M,B,L)$ on $\mathfrak{g}_1^\mathbb{C}$
are classified by their R-symmetry group (together with a choice of the relative sign 
between $\alpha_+$ and 
$\alpha_-$ for orthogonal Weyl-compatible signatures). This involves showing
that certain sets of data can be swapped without changing the complex superbracket
and the reality condition. For this purpose it is useful to introduce certain 
transformations, denoted $R, S_L, T$ which allow one to establish the required
isomorphisms for orthogonal and isotropic Weyl-compatible signatures. 
The isotropic Weyl incompatible case is a bit more involved and is therefore
treated separately. 

\subsection{The R-transformation \label{Sec:R-transformation}}

We define an invertible map on $\mathfrak{g}_1^\mathbb{C}$ by 
\begin{align}
  R\;:\;\;  \lambda^i \mapsto \Psi^i = \frac{1}{\sqrt{2}} (1 + i \gamma_*) \lambda^i. \label{RCmap}
  \end{align}
  For reference, the inverse is given by 
  \[
 \lambda^i = \frac{1}{\sqrt{2}} (1 - i \gamma_*) \Psi^i  \;.
  \]
  
  \subsubsection{Orthogonal dimensions}
  
We claim that  in orthogonal dimensions the R-transformation  
exchanges the bilinear forms $C_+\otimes M$ and
 $C_- \otimes M$ and preserves the reality condition up to an overall phase factor. 

To show this consider the chiral projections of the spinors,

\begin{align}
    \Psi^i_\pm  = \frac{1}{\sqrt{2}} (1 + i \gamma_*) \lambda^i_\pm = \frac{1}{\sqrt{2}} (1 \pm i) \lambda^i_\pm \;.
\end{align}
Recall that for an orthogonal bilinear form, if $C_+ \otimes M$ is super-admissible then so is $C_- \otimes M$. Using \eqref{cgammastar} and \eqref{bgammastar} we find these two super-admissible bilinear forms are related by
\begin{align}
    (C_+ \otimes M) (\gamma^\mu \underline{\lambda}_\pm, \underline{\chi}_\pm) = \pm i (C_- \otimes M) (\gamma^\mu \underline{\lambda}_\pm, \underline{\chi}_\pm).
\end{align}

The R-transformation removes this factor of $\pm i$:
\begin{align}
    (C_+ \otimes M) (\gamma^\mu \underline{\lambda}_\pm, \underline{\chi}_\pm) = (C_- \otimes M) (\gamma^\mu \underline{\Psi}_\pm, \underline{\Omega}_\pm) .
\end{align} 

Additionally, we find that the R-transformation does not change the reality condition (up to modifying $\alpha$ by a factor of $i$). For definiteness, we take the reality condition to be defined using $B_-$. 
 First let us consider Weyl-compatible reality conditions. Given $\lambda^i$ with reality condition
\begin{align}
    (\lambda^i_\pm)^* = \alpha_\pm B_- \lambda^j_\pm L_{j i},
\end{align}

we find
\begin{align}
    (\Psi^i_\pm)^* =-  i \alpha_\pm B_- \Psi^j_\pm L_{j i},
\end{align}
and we see the reality condition is unchanged,  up to an overall phase.

Next we look at orthogonal Weyl-incompatible signatures, where we find that the
reality condition is invariant.
\begin{align}
	    (\lambda^i_\pm)^* = \alpha B_- \lambda^j_\mp L_{j i} \implies (\Psi^i_\pm)^* = \alpha B_- \Psi^j_\mp L_{j i} \;
\end{align}

\subsubsection{Isotropic dimensions}

In isotropic dimensions, we work with Dirac spinors, called $\lambda^i$ and $\Psi^i$. 
For a reality condition of the form $(\lambda^i)^* = \alpha B 
\lambda^j L_{j i}$ we find, using \eqref{bgammastar} and \eqref{gammastarb}:
\begin{align}
    (\lambda^i)^* = \alpha B_- \lambda^j L_{j i} \implies (\Psi^i)^* = (-1)^{t+1} i \alpha B_+ \Psi^j L_{j i}.
\end{align}
Since the overall phase can be absorbed in $\alpha$, the
$R$-transformation can be used in isotropic dimensions to exchange $B_-$ and $B_+$.
In Weyl-incompatible signatures this is not a useful transformation, because $B_\pm$ 
have opposite $\epsilon$-type and the result is not one of our reality conditions. 
However in the Weyl-compatible case the $R$-transformation can be used to show
that reality conditions are independent of the choice $B_\pm$. For this to be an 
isomorphism of supersymmetry algebras, we also need that the bilinear form is invariant.
Using that in isotropic dimensions $\gamma_* C_\pm = C_\pm \gamma_*$, we see that an isotropic vector-valued bilinear form is unchanged by this transformation
\begin{align}
    &(\gamma^\mu \lambda^i)^T C_\pm \chi^j M_{j i} \rightarrow  \frac{1}{2} (\gamma^\mu (1+i \gamma_*) \Psi^i)^T C_\pm (1 + i \gamma_*) \Omega^j M_{j i}   \\
    =& (\gamma^\mu \Psi^i)^T C_\pm \Omega^j M_{j i} \nonumber.
\end{align}
In Section \ref{Sec:IWC} the $R$-transformation is used (together with the $S$-transformation
introduced in the next subsection) to show that isotropic Weyl-compatible supersymmetry 
algebras are classified by their R-symmetry group.

\subsection{The S-transformations \label{Sec:S-tranformations}}

We define a family of invertible maps
$S=S_s$ by 
\begin{align}
S_s\;: \;\;
    \lambda^i_+ \rightarrow \lambda^i_+, \quad \lambda^i_- \rightarrow \lambda^j_- s_{j i} \;,
\end{align}
where $(s_{ji})$ is an invertible matrix. 
In matrix notation this reads
\begin{align}
\label{S-transformation}
    \begin{pmatrix} \underline{\lambda}_+ \\ \underline{\lambda}_- \end{pmatrix} \rightarrow S_s \begin{pmatrix} \underline{\lambda}_+ \\ \underline{\lambda}_- \end{pmatrix} = \begin{pmatrix} \mathds{1} & 0 \\ 0 & s^T \end{pmatrix} \begin{pmatrix} \underline{\lambda}_+ \\ \underline{\lambda}_- \end{pmatrix}.
\end{align}

For further analysis we need to distinguish between orthogonal and isotropic dimensions.
\subsubsection{$S$-transformations and orthogonal bilinear forms}
In orthogonal dimensions the  bilinear forms are entirely chiral, and the $S$-transformation 
 leaves the vector-valued bilinear form $(C \otimes M) (\gamma^\mu \lambda_+, \chi_+)$ on positive chirality spinors 
invariant by definition. On the negative chirality spinors $S$ acts as follows:
\begin{align}
    (\gamma^\mu \lambda^i _-)^T C \chi^j_- M_{j i} = (\gamma^\mu \Psi^k _-)^T C \Omega^l_- M_{j i} s_{k i} s_{l j}.
\end{align}
To preserve $M$, that is $M_{j i} s_{k i} s_{l j} = M_{l k}$, the matrix $s$ needs to be orthogonal 
for $M=\delta$ and symplectic for $M = J$. This is the case in particular if 
$s \in  \{ \delta, I, J \}$ for $M=\delta$ and $s \in \{ \delta, \tilde{I}, J \}$ for 
$M=J$. Note that these are the standard forms of $L$ which are used to impose
reality conditions for a given $M\in \{ \delta, J\}$ to select real forms of $\mathrm{O}(K,\mathbb{C})$ or
$\mathrm{Sp}(K,\mathbb{C})$, respectively.\footnote{$L=I_{1,1}$ acts anti-isometrically for $M=J$. This
case needs to be treated separately when constructing isomorphisms in the 
orthogonal Weyl-incompatible case, see Appendix \ref{SL+WI}.  We don't need to use $S$-transformations for the orthogonal Weyl-compatible case.}
Since in orthogonal dimensions $M$ is fixed by the dimension, it 
is useful to note that the $S$-map does not change the bilinear form on $\mathbb{C}^K$. It does, 
however, change the reality condition, see below for the case $s=L$ which is 
relevant for the classification of orthogonal Weyl-incompatible supersymmetry algebras.

\subsubsection{$S$-transformations and isotropic bilinear forms}

In the isotropic case the choice $s=J$ is the only one which preserves the standard form 
$M\in \{\delta, J\}$ for the bilinear form on $\mathbb{C}^K$. The map $s_J$ allows one
to map $C_\pm \otimes \delta$ to $C_\mp \otimes J$, where we choose the upper 
or lower sign depending on which choice leads to a super-admissible bilinear form.

Writing out the isotropic vector-valued bilinear form $(C_\pm \otimes J)(\gamma^\mu \cdot, \cdot)$ explicitly gives:
\begin{align}
    (\gamma^\mu \lambda^i _+) C_\pm \chi_- ^j J_{j i} + (\gamma^\mu\lambda^i _-)^T C_\pm \chi_+ ^j J_{j i}.
\end{align}

Using \eqref{cgammastar} we can re-write this in terms of the other charge conjugation matrix
\begin{align}
    &(\gamma^\mu \lambda^i _+)^T C_\pm \chi_- ^j J_{j i} + (\gamma^\mu\lambda^i _-)^T C_\pm \chi_+ ^j J_{j i} = - (\gamma^\mu\lambda^i _+)^T C_\mp \chi_- ^j J_{j i} + (\gamma^\mu\lambda^i _-)^T C_\mp \chi_+ ^j J_{j i}. 
\end{align}
In our matrix notation, this equation is
\begin{align}
    \begin{pmatrix} (\gamma^\mu\lambda^i _+)^T, (\gamma^\mu\lambda^i _-)^T \end{pmatrix} C_\pm \begin{pmatrix} 0 & J_{j i} \\ J_{j i} & 0 \end{pmatrix} \begin{pmatrix} \chi^j_+ \\ \chi^j_- \end{pmatrix} = \begin{pmatrix} (\gamma^\mu \lambda^i _+)^T, (\gamma^\mu\lambda^i _-)^T \end{pmatrix} C_\mp \begin{pmatrix} 0 & - J_{j i} \\ J_{j i} & 0 \end{pmatrix} \begin{pmatrix} \chi^j_+ \\ \chi^j_- \end{pmatrix} .\label{jver}
\end{align}
The other super-admissible bilinear form $(C_\mp \otimes \delta)(\gamma^\mu \cdot, \cdot)$ is
\begin{align}
    (\gamma^\mu\Psi^i _+)^T C_\mp \Omega_- ^j \delta_{i j} + (\gamma^\mu\Psi^i _-)^T C_\mp \Omega_+ ^j \delta_{i j} =  \begin{pmatrix} (\gamma^\mu \Psi^i _+)^T, (\gamma^\mu \Psi^i _-)^T \end{pmatrix} C_\mp \begin{pmatrix} 0 & \delta_{j i} \\ \delta_{j i} & 0 \end{pmatrix} \begin{pmatrix} \Omega^j_+ \\ \Omega^j_- \end{pmatrix} \label{deltaver}.
\end{align}
We look for a transformation of the form 
\begin{align}
    &\begin{pmatrix} \underline{\lambda}_+ \\ \underline{\lambda}_- \end{pmatrix} = S \begin{pmatrix} \underline{\Psi}_+ \\ \underline{\Psi}_- \end{pmatrix} \;,
\end{align}
which maps \eqref{jver} and \eqref{deltaver} to on another. Then $S$ must satisfy
\begin{align}
    &S^T \begin{pmatrix}
        0 & J \\ - J & 0 
    \end{pmatrix} S = \begin{pmatrix}
        0 & \mathds{1} \\ \mathds{1} & 0 
    \end{pmatrix}\;,
\end{align}
which is solved by
\begin{align}
    S = \begin{pmatrix}
        \mathds{1} & 0 \\ 0 & - J
    \end{pmatrix}.
\end{align}
Note that this map is an isomorphism between the complex supersymmetry algebras based on the bilinear forms $C_\pm \otimes J$ and $C_\mp \otimes \delta$ in isotropic signatures. When using it
for real supersymmetry algebras we need to take into account how it acts on reality 
conditions. This requires us to distinguish between Weyl-compatible and 
Weyl-incompatible signatures, and is done below.

\subsubsection{$S_L$ transformations and Weyl-incompatible reality conditions \label{SL+WI}}

Consider the Weyl-incompatible reality condition 
\begin{align}
	(\lambda^i_\pm)^* = \alpha B \lambda^j_\mp L_{j i}
\end{align}

We can apply an $S_s$-transformation  with $s=L^{-1}$ 
to change the reality condition. Given that
\begin{align}
	\Psi^i_+ = \lambda^i_+, \quad \Psi^i_- = \lambda^j_- L^{-1}_{j i},
\end{align}
it is easy to see that
\begin{align}
	&(\Psi^i_+)^* = (\lambda^i_+)^* = \alpha B \lambda^j_- L_{j i} = \alpha B \Psi^i_-, \\
	&(\Psi^i_-)^* =  (\lambda^j_-)^* L^{-1}_{j i} = \alpha B \lambda^k_+ L_{k j} L^{-1} _{j i}  = \alpha B \Psi^i_+.
\end{align}
Thus we can map a reality condition defined by any $L$ to a reality condition 
with $L=\delta$, and by  using the inverse transformations we can map any choice of $L$ to
any other choice $L'$. In order to have isomorphic supersymmetry algebras, we need
that the superbracket is invariant. As mentioned above, this is the case if $L$ is
one of the canonical choices for the given $M$  listed in Section \ref{subsecLform}, except
for the combination $M=J$, $L=I_{1,1}$ where $L^T M L = -M$. In this case we use
$s_{iI_{1,1}}$ which preserves the superbracket and modifies the reality condition by 
an irrelevant overall phase factor. 
In Section \ref{Sec:OWI2} $S$-transformations are  used  to show that for orthogonal, 
Weyl-incompatible 
signatures supersymmetry algebras are classified by their R-symmetry group.

\subsubsection{$S_J$ transformations in the isotropic Weyl-compatible signatures}

Consider spinors $\lambda^i_\pm$ that obey a generic Weyl-compatible reality condition 
\begin{align}
    (\lambda^i_\pm)^* = \alpha B_{(\pm)} \lambda^j_\pm L_{j i}.
\end{align}
where the choice of $B_{(\pm)}$ is independent of the chirality of the spinor. 
Under $S_J$ $\lambda^i$ is related to 
\begin{align}
    \Psi^i_+ = \lambda^i_+, \quad \Psi^i_- = - \lambda^j_- J_{j i}.
\end{align}

Obviously $\Psi^i_+$ obeys the same reality condition as $\lambda^i_+$
\begin{align}
    (\Psi^i_+)^* = \alpha B_{(\pm)} \Psi^j_+ L_{j i}.
\end{align} 

Calculating the reality condition for $\Psi^i_-$ is not so trivial:
\begin{align}
    (\Psi^i _-)^* &= - (\alpha B_{(\pm)} \lambda^k _- L_{k j}) J_{j i} \\
    &= - \alpha B_{(\pm)} \Psi^l _- J_{l k} L_{k j} J_{j i} \nonumber.
\end{align}

For the two choices $L\in \{ \delta, J \}$ we find that $\lambda^i_\pm$ and $\Psi^i_\pm$ obey the same reality condition:
\begin{align} \label{lchoices}
    J_{l k} L_{k j} J_{j i} = \begin{cases} - \delta_{l i}, \qquad L_{i j} = \delta_{i j}, \\
         - J_{l i}, \qquad L_{i j} = J_{i j}.
 \end{cases}
\end{align}
This shows that $S_J$ exchanges the two bilinear forms
$C\otimes \delta$ and $C'\otimes J$, while preserving reality conditions
based on $L=\delta$ and $L=J$. This is used in Section \ref{Sec:IWC} to show
that isotropic Weyl-compatible supersymmetry algebras are classified by their R-symmetry group. 
To complete the argument, one also needs the $T$-transformation  introduced in the 
next section. 

\subsection{The T-transformation \label{Sec:T-transformation}}

This transformation is only used in isotropic, Weyl compatible signatures, where it
maps certain choices of $L$ to one another, while preserving the bilinear form.
There are two cases. In the first case the bilinear form is $C\otimes \delta$ and the
reality condition is
\begin{align}
	(\lambda^i_\pm)^* = \alpha B \lambda^j_{\pm} (I_{p,q})_{j i}	 \;.
\end{align}
In matrix notation this is
\begin{align}
	\begin{pmatrix} \underline{\lambda}_+ \\ \underline{\lambda}_- \end{pmatrix}^* = \alpha B \begin{pmatrix} I_{p,q} & 0 \\ 0 & I_{p,q} \end{pmatrix} \begin{pmatrix} \underline{\lambda}_+ \\ \underline{\lambda}_-	\end{pmatrix}.
\end{align}

One can show that if
\begin{align}
	\begin{pmatrix} \underline{\Psi}_+ \\ \underline{\Psi}_-	 \end{pmatrix} = \begin{pmatrix} \mathds{1}_p & 0 & 0 & 0 \\ 0 & i \mathds{1}_q & 0 & 0 \\ 0 & 0 & \mathds{1}_p & 0 \\ 0 & 0 & 0 & - i \mathds{1}_q \end{pmatrix} \begin{pmatrix} \underline{\lambda}_+ \\ \underline{\lambda}_-	,\end{pmatrix}
\end{align}

then $\Psi_\pm$ obey the reality condition
\begin{align}
	\begin{pmatrix} \underline{\Psi}_+ \\ \underline{\Psi}_- \end{pmatrix}^* = \alpha B \begin{pmatrix} \underline{\Psi}_+ \\ \underline{\Psi}_-	\end{pmatrix}
\end{align}
where $L=I_{p,q}$ has been replaced by $L=\delta$. We remark that this transformation 
is not particularly useful in orthogonal dimensions, because it changes the 
bilinear form, and the result takes a non-standard form. However we only need to 
use $T$ in isotropic dimensions, where these reality conditions are paired with the 
bilinear form $C \otimes \delta$ which is unchanged:
\begin{align}
	[C \otimes \delta](\underline{\lambda},\underline{\chi}) = [C \otimes \delta](\underline{\Psi},\underline{\Omega})
\end{align}

In the second case the bilinear form is $C'\otimes J$, and the reality conditions 
involve $L = \tilde{I}_{2r, 2s}$ and $L=\delta$. These can be mapped  using the $T$-transformation 
\begin{align}
	T = \begin{pmatrix} \mathds{1}_r & 0 & 0 & 0 & 0 & 0 & 0 & 0 \\ 0 & i \mathds{1}_s & 0 & 0 & 0 & 0 & 0 & 0 \\ 0 & 0 & \mathds{1}_r & 0 & 0 & 0 & 0 & 0 \\0 & 0 & 0 & i \mathds{1}_s & 0 & 0 & 0 & 0 \\ 0 & 0 & 0 & 0 & \mathds{1}_r & 0 & 0 & 0 \\ 0 & 0 & 0 & 0 & 0 & - i \mathds{1}_s & 0 & 0 \\  0 & 0 & 0 & 0 & 0 & 0 & \mathds{1}_r & 0 \\ 0 & 0 & 0 & 0 & 0 & 0 & 0 & - i \mathds{1}_s \end{pmatrix} \;,
\end{align}
which leaves invariant the bilinear form $C' \otimes J$. Together with the $S_J$-transformation,
the $T$-transformation allows to show that isotropic Weyl-compatible supersymmetry algebras
are classified by their R-symmetry group, see Section \ref{Sec:IWC}.

\subsection{Relations between isotropic Weyl-incompatible supersymmetry algebras \label{Sec:IWI_details}}

We start with a generic Weyl-incompatible reality condition
\begin{align}
    (\lambda^i_\pm)^* = \alpha B_{(\pm)} \lambda^j_\mp L_{j i}.
\end{align}
In Weyl-incompatible signatures $B_\pm$ have opposite $\epsilon$-type, and therefore
the choice of $B$ is not free, but fixed by $L$. Here $B_{(\pm)}$ refers to the choice
which satisfies $B^* B = L^2$ and therefore in combination with $L$ defines a real structure. 
$B_{(\mp)}$ refers to the other $B$-matrix.

\subsubsection{$S_J$-transformation: $(\delta, \delta) \rightarrow (J,J)$ and $(\delta, J) \rightarrow (J,\delta)$}

Under the $S_J$-transformation the transformed spinors, $\Psi^i_\pm$ obey
\begin{align}
    (\Psi^i_+)^* &= (\lambda^i_+)^* = \alpha B_{(\pm)} \lambda^j_- L_{j i} \label{psiplus} = \alpha B_{(\pm)} \Psi^k_- J_{k j} L_{j i} =  -\alpha B_{(\mp)} \Psi^k_- J_{k j} L_{j i} \;,
        \\ 
    (\Psi^i_-)^* &= - (\lambda^i_-)^* J_{j i} = - \alpha B_{(\pm)} \lambda^j_+ L_{k j} J_{j i} \label{psiminus} = - \alpha B_{(\pm)} \Psi^k_+ L_{k j} J_{j i}  = -\alpha B_{(\mp)} \Psi^k_+ L_{k j} J_{j i}  \;,
\end{align}
where in the last step we used \eqref{bgammastar} to re-write the reality condition in terms
of the other $B$-matrix. This is needed because $L=\delta$ and $L=J$ have opposite
$\epsilon$-type. 
Using that 
\begin{align}
	 &J_{k j} \delta_{j i} = J_{k i}, \qquad \delta_{k j} J_{j i} = J_{k i},\\
	 &J_{k j} J_{j i} = -\delta_{k i}, \qquad J_{k j} J_{j i} = -\delta_{k i}
\end{align}
we find 
\begin{align}
	&(\lambda^i)^* = \alpha B_{(\pm)} \lambda^i \Rightarrow 	(\Psi^i)^* = -\alpha B_{(\mp)} \Psi^j J_{j i}, \\
	&(\lambda^i)^* = \alpha B_{(\pm)} \lambda^j J_{j i} \Rightarrow (\Psi^i)^* = \alpha B_{(\mp)} \Psi^i. \nonumber
\end{align}
This shows that the algebras defined by the data
$(M, L) = (\delta, \delta)$ and $(J, J)$, are related by the $s_J$ transformation. These are 
in fact the standard Majorana and the symplectic Majorana condition, both with 
R-symmetry group $\mathrm{U}(K)$. This shows that the two ways of realizing this R-symmetry
group lead to isomorphic supersymmetry algebras, see Section \ref{Sec:IWI} and the figure in 
(\ref{Fig:IWI}). 
For the same reasons, the pairs $(M,L) = (\delta, J)$ and $(J, \delta)$ which have R-symmetry
group $\mathrm{U}(k,k)$ are also related to one another by $S_J$.
There are two other ways of obtaining a $\mathrm{U}(k,k)$ R-symmetry group: 
with $(M, L) = (\delta, I_{k,k})$ and $(J, \tilde{I}_{2r,2s})$, which will now be discussed.

\subsubsection{$F$-transformation: $ (\delta, J) \rightarrow (\delta, I_{k,k})$}

We start with an algebra with $(M, L) = (\delta, J)$, so the initial spinors have reality condition
\begin{align}
	&(\lambda^i_\pm)^* = \alpha B_{(\pm)} \lambda^j_\mp J_{j i}.
	\end{align}

To transform this to  an algebra with $(M,L) =(\delta, I_{k,k})$, we apply the 
transformation (using the `doubled again' matrix notation)
 $\Psi^I = \lambda^J F_{J I}$ where
\begin{align}
	F = \frac{1+i}{2} \begin{pmatrix} - i \mathds{1}_k & \mathds{1}_k & 0 & 0 \\ \mathds{1}_k & - i \mathds{1}_k & 0 & 0 \\ 0 & 0 & \mathds{1}_k & - i \mathds{1}_k \\ 0 & 0 & - i \mathds{1}_k  & \mathds{1}_k 
	\end{pmatrix} \;.
\end{align}
The transformed spinors obey the reality condition
\begin{align}
    (\Psi^i)^* = i \alpha B_{(\mp)} \Psi^j (I_{k,k})_{j i}.
\end{align}
Since $J$ and $I_{k,k}$ have opposite $\epsilon$-type we have used
\eqref{bgammastar} to change the $B$-matrix accordingly.

\subsubsection{$G$-transformation: $ (J, \tilde{I}_{2r,2s}) \rightarrow (J, \delta)$}

Let $\lambda^i$ be the spinors from a supersymmetry algebra with $(M, L) = (J, \tilde{I}_{2r,2s})$ with reality condition
\begin{align}
    (\lambda^i)^* = \alpha B_{(\pm)} \lambda^j (\tilde{I}_{2r,2s})_{j i}.
\end{align}

We define $\Psi^i = \lambda^j G_{j i}$ where
\begin{align}
    G = \begin{pmatrix} \mathds{1}_r & 0 & 0 & 0 \\ 0 & i \mathds{1}_s & 0 & 0 \\ 0 & 0 & \mathds{1}_r & 0 \\ 0 & 0 & 0 & -i \mathds{1}_s \end{pmatrix}.
\end{align}
Note that this matrix has size $(2r+2s)\times (2r+2s)$, that is, we are not using
the matrix notation. 

One can show that
\begin{align}
    (\Psi^i)^* = \alpha B_{(\pm)} \Psi^i,
\end{align}
so that $G$ maps a reality condition with $L=\tilde{I}_{2r,2s}$ to a standard Majorana
condition with $L=\delta$. Moreover, one can show that the vector-valued
bilinear form $(C\otimes J)(\gamma^\mu \cdot, \cdot)$ is invariant. Therefore $G$
defines an isomorphism between isotropic Weyl-incompatible algebras with 
$(M,L)=(J,\tilde{I}_{2r,2s})$ and $(J,\delta)$, as indicated in Figure \ref{Fig:IWI} in 
Section \ref{Sec:IWI}. This almost completes the classification of
isotropic Weyl incompatible supersymmetry algebras, except for one special case
that is not covered by the $G$-transformation. 

\subsubsection{Remaining case $(J,I_{1,1}) \rightarrow (J, \delta)$ }

The reality conditions $L=\tilde{I}_{2r,2s}$ only cover R-symmetry groups
$\mathrm{U}(k,k) = \mathrm{U}(r+s, r+s)$ with $k>1$, since $r,s\geq 1$. For $k=1$ we have
a different canonical representative, namely $L=I_{1,1}$.
However, this case was already covered in \cite{Cortes:2019mfa}, where
the isomorphisms between the isotropic, Weyl-incompatible 
supersymmetry algebras in signature $(1,3)$ with R-symmetry groups 
$\mathrm{U}(2)$ and $\mathrm{U}(1,1)$ have been worked out. If one performs
the same computation in any other signature, the only detail which can 
change is which of the two matrices $B_\pm$ defines a real, and which defines
a quaternionic structures. Since this does not change whether an isomorphism
exists or not, the results of \cite{Cortes:2019mfa} imply that all isotropic
Weyl incompatible supersymmetry algebras with R-symmetry group $\mathrm{U}(1,1)$ 
are isomorphic.\footnote{The relevant transformation is called $V$ in \cite{Cortes:2019mfa}. 
Note that there is a typo in the diagram which represents the isomorphisms. Also note
that the off-diagonal matrix $\eta$ used in \cite{Cortes:2019mfa} is related to $I_{1,1}$ by an additional
basis transformation.} 
This completes proving that isotropic, Weyl-incompatible supersymmetry algebra are classified
by their R-symmetry group, and as well all statements about classification of supersymmetry
algebras made in Section \ref{Sec:Isomorphisms}.

\section{Dimensional Reduction \label{App:Dim_Red}}

In this section we derive the general formulas which allow one to perform
spacelike and timelike dimensional reductions starting in an arbitrary 
signature.

\subsection{Odd to even dimensions}

In this section we describe how the reduction from odd to even dimensions works
in general. We use the following conventions:
the space-time indices of the higher dimensional theory are $M = 0, ..., D$. When 
performing  a spacelike reduction we remove the final direction (when going from $(D+1)$ to $D$ dimensions this is the $D$-th direction) and when performing a timelike reduction we remove the $0$-th direction. Therefore the lower dimensional space-time indices are are $\mu = 1, ..., D$ for a timelike reduction or $\mu = 0, ..., D-1$ for a spacelike reduction.

When we reduce from an odd to an even space-time dimension, 
the dimension of the Dirac spinor module does not decrease, which makes reductions from 
odd to even dimensions simpler than the second case. 
We relate the higher-dimensional spinors and $\gamma$-matrices to the lower ones as follows:
\begin{align}
    \lambda^i_{(D+1)} = \lambda^i_{(D)}, \quad \Gamma_M = \begin{cases} \{ \gamma_\mu, \gamma_{(D+1)} = \gamma_* \}\;, \quad \text{for spacelike reduction}, \\ \{\gamma_{0} = i \gamma_* , \gamma_\mu\}\;, \quad \text{for timelike reduction}. \end{cases}
\end{align}

The `extra'  $\gamma$-matrix of the higher-dimensional theory
is proportional to the chirality operator $\gamma_*$ of the lower-dimensional one.
We choose representations such that for a spacelike reduction $\Gamma_{(D+1)} = \gamma_*$ and for a timelike reduction $\Gamma_0 = i \gamma_*$. This is always possible.

The charge conjugation matrix of the $(D+1)$-dimensional theory is equal to one of the two charge conjugation matrices in the even $D$ dimensions. From Table \ref{bilinears} we can infer that
under reduction it becomes $C_+$, if the lower-dimensional theory is orthogonal,
and $C_-$ if it is isotropic. The corresponding bilinear form of the reduced theory is $C_+ \otimes M$ or $C_- \otimes M$ with the bilinear form $M$ inherited from the parent theory.

The reality condition is likewise inherited from the higher dimensional theory, though one needs to rewrite the $B$-matrix in terms of the lower dimensional $B$ matrices. When going from odd 
to even dimensions the dimensionally reduced $B$ matrices satisfy, for orthogonal dimensions
\begin{align}
    &B^{(t,s)} = (C (A^{(t,s)})^{-1})^T = (C_- (A^{(t,s-1)})^{-1})^T = B^{(t,s-1)} _+, \\
    &B^{(t,s)} = (-1)^{t}  (- i C_- (A^{(t-1,s)})^{-1} )^T = (-1)^{t} B^{(t-1,s)}_-, \nonumber
\end{align}
and for isotropic dimensions
\begin{align}
    &B^{(t,s)} = (C (A^{(t,s)})^{-1})^T = (C_- (A^{(t,s-1)})^{-1})^T = B^{(t,s-1)} _-, \\
    &B^{(t,s)} = (-1)^{t}  (- i C_+ (A^{(t-1,s)})^{-1} )^T = (-1)^{t+1} i B^{(t-1,s)}_+. \nonumber
\end{align}
At least one of the $\epsilon$-quaternionic structures in the reduced signature has the same $\epsilon$ as the $\epsilon$-quaternionic structure in the parent signature when going from odd dimensions to even dimensions, so that the daughter theories can have the same $L$.
More details which allow to derive the above relations can be found in \cite{Cortes:2019mfa}.

\subsection{Even to odd dimensions}

This case is a bit more complicated since the dimensionality of
Dirac spinors halves as we dimensionally reduce.  Morally speaking we can  equate the Weyl spinors of the parent theory with the Dirac spinors of the daughter theory. The charge conjugation matrices and 
$\gamma$-matrices reduce in size, too. Relating the two theories requires some care,
though fortunately there are not too many possibilities.

In the odd-dimensional daughter signature, we only have a single $C$, but it will always be related to 
the $C$-matrix of one of its two even-dimensional parents. By inspection of our tables, $C$ 
has the same invariants as $C_-$ if the parent theory is orthogonal, while it has the same
invariants as $C_+$ if the parent theory is isotropic. This explains why we can embed $C$ into
the higher dimensional $C_\pm$ by
\begin{align}
    C^{(D+1)}_- = C^{(D)} \otimes \sigma_1, \quad \text{D = 5, 9} \qquad \text{or} \qquad C^{(D+1)}_+ = C^{(D)} \otimes 1, \quad \text{D = 3, 7, 11} .
\end{align}

The bilinear form on the extended spinor module of the parent theory is $C^{(D+1)} \otimes M$ with whatever $C^{(D+1)}$ is in the above formula and where $M = \{ \delta, J \}$ is the correct choice to make the bilinear form super-admissible. If the standard bilinear form in the 
parent theory is different from what is obtained this way, we can use the maps constructed in Appendix \ref{Sec:Details_Isomorphisms} to bring the bilinear form to its canonical form.

Finally we have to choose an embedding of the $\gamma$-matrices. Those in the parent theory will be called $\Gamma_M$, with $M = 1, ..., D+1$ if we are reducing along a spacelike direction and $M = 0, ..., D$ if we are reducing along a timelike direction. The $\gamma$-matrices of the daughter theory are $\gamma_\mu$, with $\mu = 1,..., D$ always. We embed the $\gamma$-matrices as follows:
\begin{align}
    \Gamma_\mu = \gamma_\mu \otimes \sigma_1, \quad \Gamma_{(D+1)} = 1 \otimes \sigma_2 \quad \text{or} \quad \Gamma_0 = i 1 \otimes \sigma_2.
\end{align}

We remove either $\Gamma_{(D+1)}$ or $\Gamma_0$ depending on whether we 
reduce along a spacelike or timelike direction. We define $\Gamma_*$ according to the conventional
\begin{align}
    \Gamma_* = (-i)^{D/2 + t} \prod_{\mu} \Gamma_\mu
\end{align}

and choose the $\gamma_\mu$ such that
\begin{align}
    \Gamma_* = 1 \otimes \sigma_3.
\end{align}

Note that it is always possible, as the daughter theory is in odd dimensions.  Therefore there are two inequivalent representations of the Clifford algebra that are distinguished by the
sign of $\gamma_{(D)}$. We can therefore choose the representation such that 
the above relation holds.

For completeness we then have the other charge conjugation matrix, from \eqref{cgammastar} given as
\begin{align}
    C^{(D+1)}_+ = C^{(D)} \otimes \sigma_2 \quad \text{D = 5, 9} \qquad \text{or} \qquad C^{(D+1)}_- = C^{(D)} \otimes \sigma_3 \quad \text{D = 3, 7, 11} .
\end{align}

We can therefore decompose the $D+1$ dimensional spinors into $D$-dimensional spinors according to
\begin{align}
    &\lambda^i_+ = \psi^i \otimes \begin{pmatrix} 1 \\ 0 \end{pmatrix}, \quad \lambda^{\hat{i}}_- = \psi^{i + K_+} \otimes \begin{pmatrix} 0 \\ 1 \end{pmatrix} \;,
\end{align}

where $\lambda^i_+$ and $\lambda^{\hat{i}}_-$ are the spinors in $D+1$ dimensions, of which we have $K_+$ and $K_-$ respectively, and $\psi^i$ the spinors in $d$ dimensions, of which we now have $K_+ + K_-$. We may need to transform the $\psi^i$ quantities to put the bilinear form and reality condition into canonical forms.

We are now able to dimensionally reduce the vector-valued bilinear form. We have two cases, namely orthogonal and isotropic vector-valued bilinear forms. We begin with an orthogonal vector-valued bilinear form with $K_+$ positive and $K_-$ negative chirality spinors
\begin{align}
    &(\Gamma^M \lambda^i_+)^T C^{(D+1)} \chi^j_+ M_{j i} + (\Gamma^M \lambda^{\hat{i}}_-)^T C^{(D+1)} \chi^{\hat{j}}_- M'_{\hat{j} \hat{i}} \\
    = &(\gamma^\mu \psi^i)^T C^{(D)} \phi^j M_{j i} \otimes (\sigma_1 \begin{pmatrix} 1 \\ 0 \end{pmatrix})^T \sigma_1 \begin{pmatrix} 1 \\ 0 \end{pmatrix} + (\gamma^\mu \psi^{\tilde{i}})^T C^{(D)} \phi^{\tilde{j}} M'_{\tilde{j} \tilde{i}} \otimes (\sigma_1 \begin{pmatrix} 0 \\ 1 \end{pmatrix})^T \sigma_1 \begin{pmatrix} 0 \\ 1 \end{pmatrix} \\
    =& \bigg( (\gamma^\mu \psi^i)^T C^{(D)} \phi^j M_{j i} + (\gamma^\mu \psi^{\tilde{i}})^T C^{(D)} \phi^{\tilde{j}} M'_{\tilde{j} \tilde{i}}  \bigg)\otimes 1 \\
    =&    (\gamma^\mu \psi^i) C^{(D)} \phi^j \begin{pmatrix} M & 0 \\ 0 & M'    \end{pmatrix}_{j i} \otimes 1.
\end{align}

Here $i,j = 1,...,K$ and $\tilde{i}, \tilde{j} = K_+ + 1, ..., K_+ + K_-$ until the final line where we have combined the indices so that $i, j = 1,...,K_+ + K_-$. $M$ and $M'$ will be of the same form, either $\delta$ or $J$, but are $K_+ \times K_+$ and $K_- \times K_-$ matrices in the reduced theory, respectively.

Note if $M = \delta$ this is already correctly lined up so that the $D$-dimensional theory has the vector-valued bilinear form
\begin{align}
    (\gamma^\mu \psi^i) C^{(d)} \phi^j \delta_{j i}, \qquad i = 1,...,K_+ + K_-.
\end{align}

However, if $M = J$ we are not in the canonical form, in that
\begin{align}
    \begin{pmatrix} J_{K_+} & 0 \\ 0 & J_{K_-}    \end{pmatrix} \neq J_{K_+ + K_-}.
\end{align}
We then need a change of basis for $\psi^i$ to realign the spinors into a canonical form (this will also affect the reality condition).

For isotropic dimensions, where $K_+ = K_-$ and $M = M'$, we find the following
\begin{align}
    &(\Gamma^M \lambda^i_+)^T C^{(D+1)} \chi^j_- M_{j i} + (\Gamma^M \lambda^i_-)^T C^{(D+1)} \chi^j_+ M_{j i} \\
    = &(\gamma^\mu \psi^i)^T C^{(D)} \phi^{\tilde{j}} M_{\tilde{j} i} \otimes (\sigma_1 \begin{pmatrix} 1 \\ 0 \end{pmatrix})^T \begin{pmatrix} 1 \\ 0 \end{pmatrix} + (\gamma^\mu \psi^{\tilde{i}})^T C^{(D)} \phi^{j} M_{j \tilde{i}} \otimes (\sigma_1 \begin{pmatrix} 0 \\ 1 \end{pmatrix})^T \begin{pmatrix} 0 \\ 1 \end{pmatrix} \\
    = &(\gamma^\mu \psi^i)^T C^{(D)} \phi^j \begin{pmatrix} 0 & M \\ M & 0    \end{pmatrix} \otimes 1 .
\end{align}
Here $i,j = 1,...,K$ and $\tilde{i}, \tilde{j} = K+1, ..., 2K$ until the final line where we have combined the indices so that $i, j = 1,...,2K$. In the final expression $M$ represent the original $K \times K$ Gram matrices inherited from the parent theory. We will then need a basis transformation to obtain the  canonical form.

Next we consider the reduction of reality conditions. 
Due to different embeddings of $C$ into the parent theory, 
we have different factorisations of the $B$ matrices depending on whether the parent is orthogonal or isotropic. For a spacelike dimension from $(t, s+1)$ to $(t, s)$ we find
\begin{align}
    B^{(t,s+1)}_+ =& \begin{cases} - B^{(t,s)} \otimes \sigma_1^t \sigma_2 \qquad \text{for orthogonal parent}, \\  B^{(t,s)} \otimes \sigma_1 ^{t} \qquad \text{for isotropic parent}, \end{cases} \\
    B^{(t,s+1)}_- =& \begin{cases} B^{(t,s)} \otimes \sigma_1 ^{t+1} \qquad \text{for orthogonal parent}, \\ B^{(t,s)} \otimes \sigma_1^t \sigma_3 \qquad \text{for isotropic parent}    .\end{cases} 
\end{align}
Along a timelike direction, from $(t+1, s)$ to $(t,s)$ we find
\begin{align}
    B^{(t+1,s)}_+ =& \begin{cases} (-1)^{t+1} i B^{(t,s)} \otimes \sigma_1^t  \qquad \text{for orthogonal parent}, \\ i B^{(t,s)} \otimes \sigma_2 \sigma_1^t \qquad \text{for isotropic parent},    \end{cases} \\
    B^{(t+1,s)}_- =& \begin{cases} i B^{(t,s)} \otimes \sigma_2 \sigma_1^{t+1} \qquad \text{for 
    orthogonal parent}, \\ (-1)^{t+1} B^{(t,s)} \otimes \sigma_1^{t+1}\qquad \text{for isotropic parent}.    \end{cases} 
\end{align}
Note the presence of powers of $\sigma^1$ resulting from the decomposition 
of the matrix $A$, which is the product of all timelike $\gamma$-matrices.

\end{appendix}


\providecommand{\href}[2]{#2}\begingroup\raggedright\endgroup

\end{document}